\documentclass[twocolumn]{aastex63}
\usepackage{graphicx}
\usepackage{epsfig}
\usepackage{natbib}
\usepackage{multirow}
\usepackage{amsmath}
\usepackage{ifpdf}
\newcommand{\eg}{{\rm e.g.,\ }}
\newcommand{\cf}{{\rm cf.\ }}

\newcommand{\km}{{\rm\thinspace km}}
\newcommand{\g}{{\rm\thinspace g}}

\newcommand{\s}{{\rm\thinspace s}}

\newcommand{\kmps}{\hbox{$\km\s^{-1}\,$}}

\newcommand{\halpha}{H$\alpha$}
\newcommand{\lya}{Ly$\alpha$}

\newcommand{\G}{{\rm G}}
\newcommand{\K}{{\rm K}}
\newcommand{\hi}{H\thinspace{\sc i}}
\newcommand{\hii}{H\thinspace{\sc ii}}

\newcommand{\oiii}{[O\thinspace{\sc iii}]}

\newcommand{\oii}{[O\thinspace{\sc ii}]}
\newcommand{\oifb}{[O\thinspace{\sc i}]}
\newcommand{\oi}{O\thinspace{\sc i}}
\newcommand{\ov}{O\thinspace{\sc v}}

\newcommand{\neiii}{[Ne\thinspace{\sc iii}]}

\newcommand{\sii}{[S\thinspace{\sc ii}]}

\newcommand{\nhi}{$N_{\rm H\thinspace{\sc I}}$}

\newcommand{\heii}{He\thinspace{\sc ii}}
\newcommand{\hei}{He\thinspace{\sc i}}
\newcommand{\sithree}{Si\thinspace{\sc iii}}
\newcommand{\sitwo}{Si\thinspace{\sc ii}}
\newcommand{\siiv}{Si\thinspace{\sc iv}}
\newcommand{\cii}{C\thinspace{\sc ii}}
\newcommand{\ciif}{[C\thinspace{\sc ii}]}

\newcommand{\ciii}{C\thinspace{\sc iii}}

\newcommand{\civ}{C\thinspace{\sc iv}}

\newcommand{\nv}{N\thinspace{\sc v}}
\newcommand{\fesclyc}{$f_{\rm esc, LyC}$}
\newcommand{\fesclya}{$f_{{\rm esc, Ly}\alpha}$}
\newcommand{\vpeaks}{$\Delta v_{\rm Ly\alpha}$}
\newcommand{\fmin}{$F_{\rm min}$}
\newcommand{\fcov}{$f_{\rm cov}$}
\newcommand{\fcont}{$F_{\rm cont}$}
\newcommand{\vchar}{$v_{\rm char}$}
\newcommand{\vmax}{$v_{\rm max}$}
\newcommand{\vblue}{$v_{\rm blue}$}
\newcommand{\vred}{$v_{\rm red}$}
\newcommand{\rcor}{$\rho$}
\newcommand{\AV}{$A_{V}$}
\newcommand{\ewnet}{EW$_{\rm net}$}

\newcommand{\Msol}{\hbox{\thinspace M$_{\sun}$}}

\shorttitle{\lya\ Escape at High Ionization}

\begin{document}

\title{New Insights on \lya\ and Lyman Continuum Radiative Transfer in the Greenest Peas \footnote{Based on observations made with the NASA/ESA Hubble Space Telescope, obtained at the Space Telescope Science Institute, which is operated by the Association of Universities for Research in Astronomy, Inc., under NASA contract NAS 5-26555. These observations are associated with programs GO-14080.}}

\author{Anne E. Jaskot}
\altaffiliation{Hubble Fellow}
\affiliation{Astronomy Department, Williams College, Williamstown, MA 01267, USA.}
\affiliation{Department of Astronomy, University of Massachusetts, Amherst, MA 01003, USA.}
\author{Tara Dowd}
\affiliation{The Chandra X-Ray Center, Cambridge, MA 02138, USA.}
\affiliation{Department of Astronomy, University of Massachusetts, Amherst, MA 01003, USA.}
\author{M. S. Oey}
\affiliation{Department of Astronomy, University of Michigan, Ann Arbor, MI 48109, USA.}
\author{Claudia Scarlata}
\affiliation{Minnesota Institute for Astrophysics, University of Minnesota, Minneapolis, MN 55455, USA.}
\author{Jed McKinney}
\affiliation{Department of Astronomy, University of Massachusetts, Amherst, MA 01003, USA.}

\begin{abstract}
As some of the only Lyman continuum (LyC) emitters at $z\sim0$, Green Pea (GP) galaxies are possible analogs of the sources that reionized the universe. We present {\it HST} COS spectra of 13 of the most highly ionized GPs, with \oiii/\oii\ $=6-35$, and investigate correlations between \lya, galaxy properties, and low-ionization UV lines. Galaxies with high \oiii/\oii\ have higher H$\alpha$ equivalent widths (EWs), and high intrinsic \lya\ production may explain the prevalence of high \lya\ EWs among GPs. While \lya\ escape fraction is closely linked to low gas covering fractions, implying a clumpy gas geometry, narrow \lya\ velocity peak separation (\vpeaks) correlates with the ionization state, suggesting a density-bounded geometry. We therefore suggest that \vpeaks\ may trace the residual transparency of low-column-density pathways. Metallicity is associated with both \oiii/\oii\ and \vpeaks. This trend may result from catastrophic cooling around low-metallicity star clusters, which generates a compact geometry of dense clouds within a low-density inter-clump medium. We find that the relative strength of low-ionization UV emission to absorption correlates with \lya\ emission strength and is related to \lya\ profile shape.  However, as expected for optically thin objects, the GPs with the lowest \vpeaks\ show both weak low-ionization emission and weak absorption. The strengths of the low-ionization absorption and emission lines in a stacked spectrum do not correspond to any individual spectrum. Galaxies with high \oiii/\oii\ contain a high fraction of LyC emitter candidates, but \oiii/\oii\ alone is an insufficient diagnostic of LyC escape.

\end{abstract}

\section{Introduction}
\label{sec:intro}
The radiative transfer of ionizing, Lyman continuum (LyC), photons produced by star-forming regions has important implications for observational cosmology. Star-forming galaxies with escaping LyC radiation are the leading candidates responsible for the reionization of the universe at $z>6$ \citep[\eg][]{robertson15, finkelstein15}. Nevertheless, the LyC escape fraction (\fesclyc) from star-forming galaxies remains one of the most poorly constrained cosmological parameters \citep[\eg][]{fernandez11, robertson15, shull15}, and few galaxies at any redshift have confirmed direct detections of escaping LyC \citep[\eg][]{bergvall06, leitet13, borthakur14, leitherer16, mostardi15, vanzella16, fletcher18}. On the contrary, most galaxy samples show low LyC escape fractions (\fesclyc) $<10\%$ \citep[\eg][]{leitherer95, leitet13, rutkowski16, naidu18, tanvir19}. However, the recent detection of LyC radiation with \fesclyc $=2.5-72\%$ from 11 out of 11 targeted ``Green Pea" (GP) galaxies \citep{izotov16a,izotov16b,izotov18a,izotov18b} demonstrates the existence of a population of star-forming galaxies where LyC escape is not rare, but common. As such, these galaxies may reveal the physical conditions that enable LyC escape.

Characterized by strong \oiii~$\lambda$5007 emission and compact morphologies \citep{cardamone09}, the $z<0.3$ GP galaxies closely resemble $z>2$ galaxies. Like many high-redshift galaxies, the GPs show extreme nebular equivalent widths (EWs; \oiii~$\lambda$5007 EW $\sim100-2000$ \AA), high specific star formation rates ($10^{-9}-10^{-7}$ yr$^{-1}$), compact and clumpy morphologies, low metallicities ($12+\log({\rm O/H}) \lesssim 8.0$), and low dust content (E(B-V)$\lesssim0.3$; \citealt{cardamone09, amorin10, izotov11}). Notably, the GPs also exhibit extremely high ionization. Their \oiii~$\lambda$5007/\oii~$\lambda$3727 ratios are $\gtrsim10$ times higher than typical $z\sim0$ galaxies and are comparable to galaxies at $z>2$ \citep{nakajima14, khostovan16}. High \oiii/\oii\ ratios could indicate a high ionization parameter, and reduced \oii\ emission from density-bounded nebulae with LyC escape can also lead to high observed \oiii/\oii\ \citep[\eg][]{jaskot13, nakajima14}. The known sample of $z\sim0$ LyC-emitting galaxies shows a tentative correlation between \oiii/\oii\ and \fesclyc\ \citep{izotov18b}, and one of the strongest known LyC emitters (LCEs) at $z\sim3$ also shows high \oiii/\oii\ \citep{vanzella16}. However, the connection between high ionization and LyC optical depth is not yet clear.

\lya\ radiative transfer provides a powerful indirect probe of optical depth in the neutral interstellar medium (ISM). \lya\ photons produced in \hii\ regions by recombining hydrogen gas resonantly scatter as they traverse the galaxy. These scatterings alter the intrinsic \lya\ line profile, with \lya\ photons typically escaping after they have shifted out of resonance. High \hi\ optical depths increase the number of scatterings the \lya\ photons undergo. More scatterings increase the chance of absorption by dust, which reduces the overall \lya\ escape fraction (\fesclya) and broadens the observed \lya\ profiles. \lya\ radiative transfer models using homogeneous shell and clumpy gas geometries both predict that galaxies with low \hi\ optical depths should show systematically stronger and narrower \lya\ profiles \citep{verhamme15, dijkstra16}. 

Observational evidence also points to a connection between \lya\ and LyC. \lya\ emitters (LAEs) at $z\sim2$ show lower velocity offsets between \lya\ and the systemic redshift compared to Lyman Break Galaxies, which may indicate that lower \hi\ column densities (\nhi) are enhancing \lya\ escape \citep{shibuya14, hashimoto15}. As expected for galaxies with below-average \hi\ optical depths, almost all GPs show strong \lya\ emission \citep{henry15,yang17}, and the separations of the redshifted and blueshifted \lya\ emission peaks are consistent with low \hi\ optical depths facilitating \lya\ escape \citep[\eg][]{jaskot14,henry15}. Moreover, confirmed LyC-emitting GPs obey the expected trends between \lya\ and optical depth; the galaxies with stronger \fesclyc\ show stronger \lya\ escape fractions (\fesclya), higher \lya\ EWs, and narrower \lya\ emission peak separations \citep{verhamme17,izotov18b}. 

Nevertheless, although the GP population includes the strongest low-redshift LCEs known to date, not all highly ionized GPs exhibit the expected indirect signatures of LyC escape. In a case study of four GPs with extreme \oiii/\oii\ ratios (\oiii/\oii $>7$), \citet{jaskot14} find that the two highest \oiii/\oii\ ratio GPs in this sample are likely strong LCEs; their high \lya\ EWs ($>70$\AA), narrow separations of the blue-shifted and red-shifted \lya\ emission peaks ($<300$\kmps), and net \lya\ emission at systemic velocity suggest a low \nhi\ and consequent LyC escape. However, the \lya\ profiles of the other two GPs in this high \oiii/\oii\ sample are not consistent with LyC escape. One GP even shows weak, broad, double-peaked \lya\ emission within a deep \lya\ absorption trough, suggestive of a high line-of-sight \hi\ column density \citep{jaskot14}. 

Low-ionization state (LIS) UV absorption and emission lines appear closely linked to both \lya\ and optical depth and may trace the neutral gas geometry on and off the line of sight. The absorption of a UV photon in a  resonant transition (e.g., \sitwo~$\lambda$1260) will be followed by either emission of \sitwo~$\lambda$1260 and de-excitation back to the ground state or by emission of \sitwo* $\lambda$1265 and de-excitation to the first fine-structure level above ground. Since low-ionization species such as Si$^+$ and C$^+$ co-exist with neutral hydrogen gas, a high \hi\ gas column along the line of sight should correlate with deep low-ionization absorption \citep[\eg][]{heckman01}. At the same time, neutral gas off the line of sight will generate fluorescent emission from the non-resonant \sitwo* and \cii* transitions \citep[\eg][]{prochaska11,scarlata15}. 

The patterns of \sitwo, \sitwo*, \cii, and \cii* transitions in the four GPs studied by \citet{jaskot14} imply different gas geometries. The two LyC-emitting candidates show weak resonant absorption, consistent with a low line-of-sight optical depth, and clear \sitwo* and \cii* emission from low-ionization gas along other sight lines. Strong blue-shifted absorption in a third GP suggests an optically thick outflow, and in the fourth GP, strong resonant absorption and non-resonant emission near the systemic velocity may imply that the plane of the galaxy is aligned with the line of sight. Similar patterns of \sitwo\ and \sitwo* lines appear in stacked spectra of high-redshift galaxies; stronger \lya\ emitters show weaker low-ionization absorption and stronger fluorescent non-resonant emission \citep[\eg][]{shapley03, steidel18}. By considering information from both the low-ionization resonant and non-resonant transitions, we can investigate line-of-sight optical depth, outflows, neutral gas geometry, and the connection of these properties with \lya\ and LyC radiative transfer.

Together, the \lya\ and LIS lines can provide important insights regarding the gas geometries that allow \lya\ and LyC photons to escape galaxies. Models of LyC escape in the literature commonly consider two idealized geometries: a ``picket-fence" geometry or a density-bounded medium. In the picket-fence scenario, LyC photons escape through holes within higher column density material, whereas in a density-bounded medium, the emitting source is completely surrounded by low column density neutral gas. If \lya\ photons freely escape through holes, the \lya\ profile should show a peak at the systemic velocity, in contrast to the narrow, double-peaked profiles characteristic of radiative transfer through a low-column density shell \citep{behrens14, verhamme15, riverathorsen17b}. In both scenarios, the LIS absorption lines will not reach zero intensity. However, in a picket-fence geometry, the LIS absorption line ratios will be consistent with saturation \citep{heckman11}. Gas metallicity, velocity structure, and infilling by resonant emission can also affect LIS absorption line depths, such that weak LIS absorption lines by themselves do not guarantee LyC escape \citep[\eg][]{prochaska11, jones13, riverathorsen15, vasei16}.

To explore the relationships between \oiii/\oii, \lya\ escape, neutral gas geometry, and optical depth, we present {\it Hubble Space Telescope} ({\it HST}) Cosmic Origins Spectrograph (COS) observations of thirteen of the most highly ionized GP galaxies at low redshift, with \oiii/\oii\ $=6.6-34.9$. In combination with previous samples of GPs \citep{henry15,yang17}, we analyze the \lya\ and low-ionization line properties of galaxies with extreme \oiii/\oii\ ratios. We investigate the relationship between optical and UV spectral properties and \lya\ escape, and use the low-ionization absorption and emission lines to assess the GPs' neutral gas geometry. We assume a cosmology with $H_0=70$ km s$^{-1}$ Mpc$^{-1}$, $\Omega_m=0.3$, and $\Lambda_0$=0.7.

\section{Observations}
\label{sec:obs}
\subsection{Sample Selection and Optical Properties}
\label{sec:sdss}
To investigate the connection between extreme \oiii/\oii, optical depth, and \lya\ escape, we have obtained {\it HST} COS spectra of thirteen low-redshift, highly ionized GPs (GO-14080, PI Jaskot). We select this sample from the Sloan Digital Sky Survey (SDSS) Data Release 10 (DR10; \citealt{ahn14}) with observed \oiii~$\lambda$5007/\oii~$\lambda$3727$\geq7$ and a signal-to-noise (S/N) ratio $>3$ in each of these lines. We exclude galaxies that fall in the active galactic nuclei (AGN) region of the BPT diagram \citep{baldwin81}. We then prioritize galaxies whose {\it GALEX} far-UV (FUV) magnitudes allow COS spectra with a S/N $\geq$4 in the continuum to be obtained in five orbits or fewer. We also exclude galaxies where Milky Way or geocoronal lines block \lya\ or multiple LIS metal lines, and we include one fainter GP (J160810+352809), due to its excessively high \oiii/\oii\ ratio. 

\begin{figure*}
\epsscale{1}
\plotone{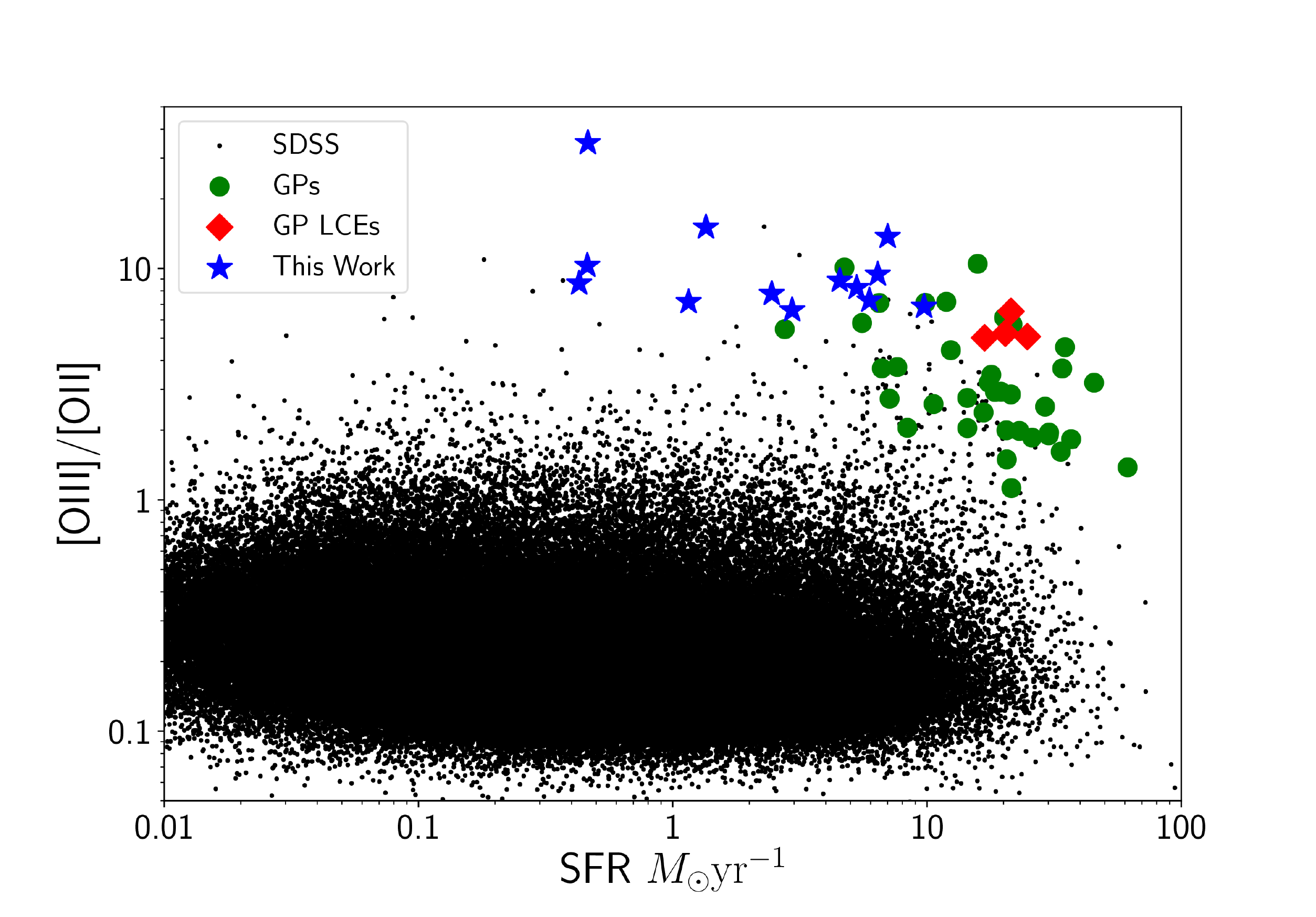}
\caption{The \oiii/\oii\ ratios and H$\alpha$-derived SFRs \protect\citep{kennicutt12} of the sample from GO-14080 (PI Jaskot; blue stars), compared with SDSS star-forming galaxies at $z<0.4$ (black points) and previous GP COS samples. Green circles show the compilation of GPs from \protect\citet{yang17}, and red diamonds show confirmed LCEs from \protect\citet{izotov16a,izotov16b}. The new GPs in this paper represent the most highly ionized star-forming galaxies in SDSS.} 
\label{fig:sample}
\end{figure*}

The resulting sample of thirteen GPs covers $z=0.027-0.124$ and has \oiii/\oii\ ratios ranging from 6.6-34.9 after dust correction. These \oiii/\oii\ ratios are comparable to or higher than the \oiii/\oii\ ratios of known and candidate LyC-emitting GPs \citep{izotov16a,izotov16b,izotov18a,izotov18b,jaskot14} and are similar to the \oiii/\oii\ ratios of $z>2$ LAEs \citep{nakajima14}. In Figure~\ref{fig:sample}, we compare our sample of extreme GPs with star-forming SDSS galaxies at $z<0.4$ and GPs previously observed with COS. Galaxies with such high \oiii/\oii\ ratios are exceedingly rare in the local universe, representing less than 0.01\% of all SDSS galaxies. One GP in this sample (J1608+3528) even has \oiii/\oii\ $=34.9$, the highest \oiii/\oii\ ratio among SDSS DR10 star-forming galaxies. Compared with previous GP samples, our sample also extends to lower star-formation rates (SFRs) and luminosities, where LyC escape may be more common \citep[\eg][]{wise09, paardekooper15, steidel18}.

\begin{figure*}
\epsscale{1}
\plotone{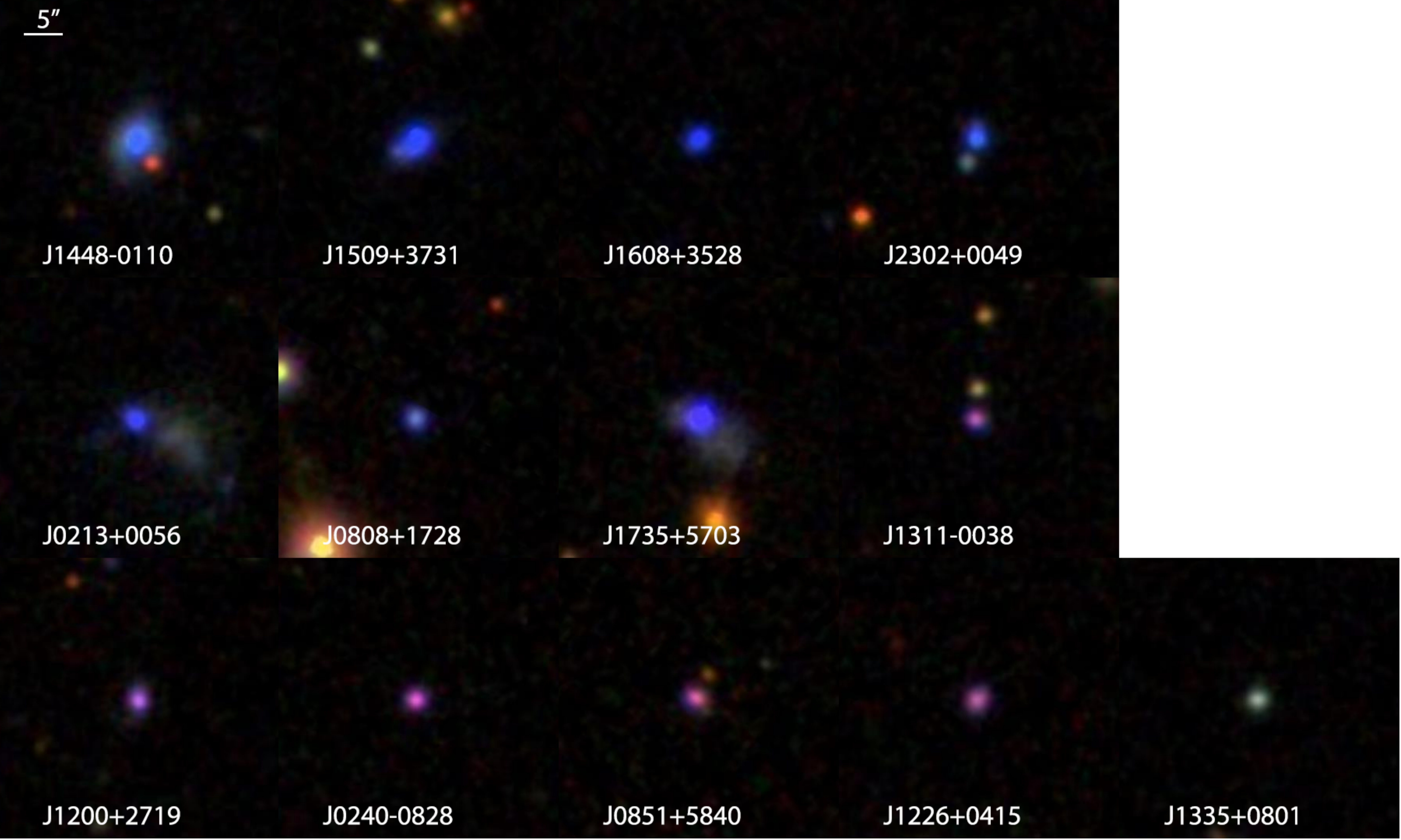}
\caption{SDSS $g,r,i$ images of the sample, ordered by increasing redshift from left to right. The scale bar at the upper left indicates 5\arcsec. Extended structure is visible in J1448-0110, J1509+3731, J0213+0056, and J1735+5703.} 
\label{fig:sdssmorph}
\end{figure*}

Although we did not explicitly select for compactness, all the galaxies in this sample show compact optical morphologies in SDSS with $r$-band half-light radii of $0.7-1.2$\arcsec\ or $0.6-1.7$ kpc. Figure~\ref{fig:sdssmorph} illustrates the morphologies of the sample, which are consistent with the morphologies of luminous compact galaxies \protect\citep{izotov11} and with the GP analog NGC 2366 \protect\citep{micheva17}. The brightest emission typically originates from a compact region with full width half maximum (FWHM ) $< 0.2$ kpc; in the lowest redshift galaxies ($z<0.05$), a lower surface brightness, extended component is sometimes visible. This lower surface brightness emission could be undetected or unresolved in SDSS images of GPs at higher redshift. 

We use the SDSS spectra to measure optical emission line fluxes and line ratios for both the sample of 13 extreme GPs and the compilation of COS-observed GPs from \citet{yang17}. We first fit the continuum with two fifth-order polynomials for $\lambda<5000$\AA\ and for $\lambda>5000$\AA. The SDSS spectroscopic pipeline uses a fifth-order continuum fit \citep[\eg][]{stoughton02}, and we find that a separate fit at the blue end of the spectrum better approximates the continuum for some galaxies. We exclude the 4620-4730\AA\ region from the continuum fits, since it may contain broad emission features from Wolf-Rayet stars. We also measure the median flux in 500 \AA\ intervals and iteratively exclude data points that deviate by more than 3$\sigma$ from the median; this process ensures that no spectral lines affect the continuum fits. After subtracting the fitted continuum from the spectra, we then fit the emission lines with Gaussian profiles. In some GPs' spectra, a single Gaussian does not capture the emission line wings. In these objects, we fit the emission lines with two Gaussian components at the same redshift. We require all lines to share the same redshift and Gaussian velocity widths. To estimate line flux and EW errors, we perturb each spectrum 100 times using the SDSS errors.

We correct all optical emission line fluxes for dust attenuation following \citet{izotov17a}. After correcting for Milky Way extinction using the \citet{fitzpatrick99} extinction law and the \citet{schlafly11} dust map, we correct for internal extinction using the observed H$\alpha$/H$\beta$ ratio and the \citet{cardelli89} extinction law. \citet{izotov17a} show that this law provides a more accurate fit to optical emission lines and UV and IR fluxes for extreme emission line galaxies such as the GPs. For GPs with H$\beta$ EWs $>150$\AA, we use $R_V=2.7$, and for all other GPs, we use $R_V=3.1$. We estimate the electron temperature using the \oiii~$\lambda\lambda$5007, 4959 to $\lambda$4363 flux ratio and the conversion in \citet{osterbrock06} and select the corresponding intrinsic H$\alpha$/H$\beta$ ratios by interpolating the values in \citet{storey95} for an electron density of 100 cm$^{-3}$. We compute the dust correction and temperature estimate iteratively. Because of the high EWs in the GPs, the observed Balmer line fluxes are insensitive to the value of underlying stellar absorption, and we therefore omit the stellar absorption correction. We note that the dust extinctions in this paper differ slightly from those in \citet{jaskot17}. Here, we use the H$\alpha$/H$\beta$ ratio only, whereas the previous work fit simultaneously to H$\alpha$/H$\beta$ and the weaker H$\gamma$ and H$\delta$ lines. Our results and conclusions do not change if we use the H$\gamma$/H$\beta$ and H$\delta$/H$\beta$ ratios instead of H$\alpha$/H$\beta$ to correct for dust extinction. Our error estimates include the errors in both the estimated electron temperature and the H$\alpha$/H$\beta$ line ratio. Finally, we exclude the H$\alpha$ fluxes of two GPs, J155926+084119 \citep{yang17} and J133304+624604 \citep{izotov16b}, whose H$\alpha$/H$\beta$ ratios are unphysically low ($<2$); for these two targets, the SDSS spectra in the H$\alpha$ region appear problematic due to strong night sky emission. We list properties derived from the SDSS spectra for the sample of 13 extreme GPs in Table~\ref{table:optical}.

\begin{table*}
\begin{center}
\caption{Optical Properties}
\begin{tabular}{lcccccccc}
\hline
\hline
Galaxy & $z$ & H$\alpha$ EW\tablenotemark{a} & \oiii/\oii\tablenotemark{b} & \oifb/H$\beta$\tablenotemark{c} & H$\alpha$/H$\beta$\tablenotemark{d} & $A_V$ & SFR\tablenotemark{e} & 12+log(O/H)\tablenotemark{f} \\
& & (\AA) & & & & & (\Msol\ yr$^{-1}$) & \\
\hline
J021307+005612 & 0.0399 & 1016$\pm$10 & 7.2$\pm$0.4 & 0.023$\pm$0.002 & 3.28$\pm$0.05 & 0.42$\pm$0.04 & 1.15$\pm$0.03 & 8.03$^{+0.08}_{-0.10}$ \\
J024052-082827 & 0.0822 &  1752$\pm$16 & 13.7$\pm$0.6 & 0.020$\pm$0.001 & 3.16$\pm$0.03 & 0.33$\pm$0.03 & 7.00$\pm$0.15 & 7.91$^{+0.09}_{-0.12}$   \\
J080841+172856 & 0.0442 & 424$\pm$5 &  10.3$\pm$0.9 & --- & 3.12$\pm$0.05 & 0.35$\pm$0.04 & 0.46$\pm$0.02 & 7.61$^{+0.13}_{-0.18}$ \\
J085116+584055 & 0.0919 & 1595$\pm$22 & 9.4$\pm$0.5 & 0.021$\pm$0.002 & 3.19$\pm$0.05 &0.36$\pm$0.04 & 6.40$\pm$0.19 & 7.87$^{+0.10}_{-0.14}$ \\
J120016+271959 & 0.0819 & 1057$\pm$10 & 8.9$\pm$0.5 & 0.022$\pm$0.001 & 3.01$\pm$0.03 & 0.19$\pm$0.03 & 4.54$\pm$0.10 & 8.06$^{+0.06}_{-0.07}$ \\
J122612+041536 & 0.0942 & 1060$\pm$13 & 8.3$\pm$0.5 & 0.023$\pm$0.002 & 3.11$\pm$ 0.05 & 0.28$\pm$0.04 & 5.29$\pm$0.16 & 8.00$^{+0.10}_{-0.12}$ \\
J131131-003844 & 0.0811 & 1106$\pm$10 & 6.6$\pm$0.2 & 0.022$\pm$0.001 & 3.21$\pm$0.03 & 0.36$\pm$0.02 & 2.95$\pm$0.05 & 7.98$^{+0.12}_{-0.17}$   \\
J133538+080149 & 0.1235 &  827$\pm$8 & 7.3$\pm$0.4 & 0.025$\pm$ 0.002 & 2.96$\pm$0.04 & 0.14$\pm$0.04 & 5.94$\pm$0.16 & 8.10$^{+0.22}_{-0.45}$  \\
J144805-011058 & 0.0274 &  805$\pm$6 & 7.8$\pm$0.3 & 0.021$\pm$0.001 & 3.23$\pm$0.03 & 0.36$\pm$0.03 & 2.45$\pm$0.05 & 8.11$^{+0.04}_{-0.05}$   \\
J150934+373146 & 0.0325 &  1411$\pm$16 & 15.1$\pm$0.9 & 0.015$\pm$0.001 & 3.05$\pm$0.04 & 0.24$\pm$0.04 & 1.35$\pm$0.04 & 7.88$^{+0.08}_{-0.10}$ \\
J160810+352809 & 0.0327 & 1472$\pm$27 &  34.9$\pm$3.5 & 0.009$\pm$0.002 & 3.21$\pm$0.06 & 0.39$\pm$0.05 & 0.46$\pm$0.02 & 7.83$^{+0.13}_{-0.19}$ \\
J173501+570309 & 0.0472 & 1442$\pm$9 & 6.8$\pm$0.3 & 0.022$\pm$0.001 & 3.29$\pm$0.03 & 0.41$\pm$0.02 & 9.74$\pm$0.15 & 8.11$^{+0.07}_{-0.08}$ \\
J230210+004939 & 0.0331 &  897$\pm$13 & 8.6$\pm$0.6 & 0.015$\pm$0.002 & 2.95$\pm$0.05 & 0.16$\pm$0.04 & 0.43$\pm$0.01 & 7.72$^{+0.07}_{-0.08}$ \\
\hline
\end{tabular}
\end{center}
\tablenotetext{a}{Rest-frame.}
\tablenotetext{b}{Extinction-corrected \oiii~$\lambda$5007/\oii~$\lambda$3727 ratio.}
\tablenotetext{c}{Extinction-corrected \oifb~$\lambda$6300/H$\beta$ ratio.}
\tablenotetext{d}{Corrected for Milky Way extinction.}
\tablenotetext{e}{Derived from H$\alpha$ following \citet{kennicutt12}.}
\tablenotetext{f}{Derived using the direct method, as described in the Appendix.}
\label{table:optical}
\end{table*}

\subsection{{\it HST} COS Observations}
\label{sec:cos}
We obtained {\it HST} COS spectra of the 13 targeted GPs with the G130M grating in Cycle 23 at lifetime position 3. All observations cover \lya\ and multiple low-ionization transitions, such as \sitwo~$\lambda$1190, \sitwo~$\lambda$1260, \oi~$\lambda$1302, and \cii~$\lambda$1334. We summarize the observations in Table~\ref{table:cosobs}. The observations were processed with CALCOS version 3.1 and downloaded from the {\it HST} MAST archive. We combined observations from separate visits with the IRAF task {\tt splice}. The target acquisition images were taken via near-UV (NUV) imaging with the primary science aperture and Mirror A.  

\begin{table*}
\begin{center}
\caption{COS Observations}
\begin{tabular}{lcccccccc}
\hline
\hline
& & & \multicolumn{3}{c}{Resolution} & \multicolumn{2}{c}{Flux Fraction} \\
Galaxy & Grating & Total Exposure & Segment A & Segment B & \lya & Inner Zone & Outer Zone \\
& & (s) & (\kmps) & (\kmps) & (\kmps) & & \\
\hline
J0213+0056  &  G130M-1300 & 4815 & 24 & 34 & 37 & 69-70\% & 97-98\% \\
J0240-0828 &  G130M-1309 & 8662 & 20 & 30 & 30 & 65-82\% & 98-100\% \\
J0808+1728  &  G130M-1309 & 8902 & 18 & 27 & 28 & 71-84\% & 99\% \\
J0851+5840  &  G130M-1318 & 12319 & 17 & 26 & 26 & 68-74\% & 96-99\% \\
J1200+2719  &  G130M-1309 & 4611 & 19 & 27 & 28 & 70-82\% & 98-99\% \\
J1226+0415 &  G130M-1318 & 11569 & 16 & 26 & 27 & 65-66\% & 95-96\% \\
J1311-0038  &  G130M-1309 & 8426 & 18 & 27 & 27 & 71-82\% & 98-99\% \\
J1335+0801  &  G130M-1327 & 9803 & 12 & 20 & --- & 74-76\% & 98-99\% \\
J1448-0110  &  G130M-1327 & 5021 & 13 & 20 & --- & 70-71\% & 97\% \\
J1509+3731 &  G130M-1327 & 8096 & 14 & 20 & 37 & 46-50\% & 90-93\% \\
J1608+3528 &  G130M-1327 & 21603 & 14 & 22 & 29 & 59-65\% & 96-98\% \\
J1735+5703  &  G130M-1309 & 5520 & 26 & 33 & 33 & 55-64\% & 95-96\% \\
J2302+0049  &  G130M-1327 & 10367 & 17 & 22 & 35 & 63-66\% & 97-99\% \\
\hline
\end{tabular}
\end{center}
\label{table:cosobs}
\end{table*}

The spectral resolution of COS observations depends on the spatial extent of the targets. Point sources result in $10-30$ \kmps\ resolution, while observations of more extended objects will have lower resolution. To determine the resolution of our observations, we collapse the two-dimensional spectrum of each object along the dispersion direction and measure the FWHM of the spatial profile in the A and B spectral segments. We exclude spectral regions containing \lya\ or geocoronal emission lines, since these wavelengths may have significantly more extended emission. The resulting FWHM of the collapsed profiles are only slightly wider than a point source profile and correspond to resolutions of $12-34$ \kmps\ (Table~\ref{table:cosobs}). Based on the FWHM of the cross-dispersion profiles in the \lya\ spectral region, the \lya\ lines have slightly lower spectral resolutions (26-37 \kmps; Table~\ref{table:cosobs}). We bin all spectra to match the expected resolution in the continuum for each segment.

The COS spectral extraction process is optimized for point sources, whereas the GPs are spatially extended. However, since most of the GPs' UV emission comes from a compact region (FWHM $<0.15$\arcsec), the default extraction method should not cause substantial flux calibration inaccuracies. The COS pipeline extracts pixels if they fall within two zones, containing the inner 80\% and 99\% of the flux from a point-source profile. The extraction excludes wavelength bins that have bad pixels within the inner zone, while bins that have bad pixels in the outer zone are preserved. 

For 12 of the targets, most flux ($95-100$\%) falls within the outer extraction boundaries and will be captured by the default extraction (Table~\ref{table:cosobs}). Of the total flux in these objects, a majority ($55-84$\%) falls within the inner extraction zone. The remaining GP, J1509+3731, consists of two UV-emitting knots. The weaker of the two knots is only partially captured by the extraction aperture, with its flux falling mostly in the outer zone. For this GP, $\sim$90\% of the total emission and $\sim$99\% of the emission from the dominant knot lie within the extraction boundaries. 

Flux missed by the extraction aperture will simply cause a reduction in the total flux. In contrast, the effect of bad pixels within the outer extraction zone can cause wavelength-dependent flux losses. The maximum flux loss expected at a given wavelength for a single grating position is 10\% for a point source \citep{romanduval16}. The use of all four FP-POS grating positions reduces the effect of flux loss by a factor of 4, so that the actual flux loss would be 2.5\% \citep{romanduval16}. Since some of our sample is more extended than a point source, flux losses at a particular wavelength would affect a higher fraction of the total flux. Our most extended targets (J1735+5703 and J1509+3731) have 42-49\% of their extracted flux within the outer zone, which is $2-2.5\times$ higher than the fraction of flux in the outer zone for a point source. Consequently, their maximum flux losses would be $2-2.5\times$ higher, corresponding to flux uncertainties of 5-6\%. The ratios of \lya\ flux within the inner zone to the outer zone are similar to these values, with similar maximum flux uncertainties of 6\%. We therefore add this uncertainty to each pixel in the final extracted spectra.

We estimate the continuum level near \lya\ with a linear fit to the rest-frame 1140-1290\AA\ region, excluding any absorption or emission lines. We conservatively set an error of 25\%\ on the continuum level. Measurements of \lya\ EW include all associated absorption and emission. To calculate the \lya\ escape fraction (\fesclya), we first measure the \lya\ emission component from the base of any \lya\ absorption trough; the uncertainties in \lya\ flux account for the difference between measuring the flux from the continuum level instead. We correct the \lya\ flux for Milky Way extinction using the \citet{fitzpatrick99} law and \citet{schlafly11} extinction maps. To estimate the intrinsic \lya\ flux, we multiply the extinction-corrected H$\alpha$ flux by the Case B \lya/\halpha\ ratio \citep{dopita03} corresponding to the galaxy's electron temperature and density (for our sample, \lya/\halpha\ $=8.24-8.96$). The uncertainties include the uncertainties in temperature, density, and H$\alpha$ flux. We also measure the separation of the blue and red \lya\ emission peaks (\vpeaks), which may trace the \hi\ column density \citep{verhamme15}. In low column-density media, \lya\ photons scatter less, resulting in narrower profiles with lower \vpeaks. We previously reported \vpeaks\ measurements for the sample in \citet{jaskot17}. We list these values and all other \lya\ measurements for the sample in Table~\ref{table:lya}. We also repeat these same measurements for the larger GP sample with COS observations \citep{yang17}. Two galaxies in the full sample (J1559+0841 and J1333+6246) do not have \fesclya\ measurements because their H$\alpha$ emission is affected by night sky lines in SDSS. 

\begin{table*}
\begin{center}
\caption{\lya\ Properties}
\begin{tabular}{lcccc}
\hline
\hline
Galaxy & \lya\ EW\tablenotemark{a} & \vpeaks & \fesclya & \fmin/\fcont\  \\
& (\AA) & (\kmps) & & \\
\hline
J0213+0056 & 42$\pm$4 & 397$\pm$47 & 0.12$\pm$0.01 & 0.7$\pm$0.3\\
J0240-0828 &  154$\pm$8 & 266$\pm$29 & 0.19$\pm$ 0.01 & 4.3$\pm$1.2\\
J0808+1728 & 31$\pm$2 & 156$\pm$37; 441$\pm$58\tablenotemark{b} & 0.36$\pm$0.03 & 3.3$\pm$0.9; 0.7$\pm$0.2\tablenotemark{b}\\
J0851+5840 &  26$\pm$2 & 361$\pm$25 & 0.04$\pm$0.01 & 0.4$\pm$0.4\\
J1200+2719 & 114$\pm$7 & 327$\pm$65 & 0.39$\pm$0.01 & 1.3$\pm$0.4\\
J1226+0415 &  64$\pm$3 & 360$\pm$40 & 0.13$\pm$0.01 & 3.6$\pm$1.0\\
J1311-0038 &  71$\pm$4 & 273$\pm$26 & 0.23$\pm$0.01 & 7.8$\pm$2.0\\
J1335+0801 & -14$\pm$0.5 & --- & 0 & --- \\
J1448-0110 &  -18$\pm$0.1& --- & 0 & --- \\
J1509+3731 &  12$\pm$1 & 400$\pm$27 & 0.05$\pm$0.03 & 0.9$\pm$0.3\\
J1608+3528 &  163$\pm$12 & 214$\pm$30 & 0.18$\pm$0.02 & 27.1$\pm$6.9\\
J1735+5703 & 64$\pm$4 & 460$\pm$ 47 & 0.09$\pm$0.01 & 2.3$\pm$0.6\\
J2302+0049 & 64$\pm$3 & 279$\pm$ 48 & 0.28$\pm$ 0.01 & 8.7$\pm$2.2\\
\hline
\end{tabular}
\end{center}
\tablenotetext{a}{Positive values denote net emission.}
\tablenotetext{b}{The triple-peaked \lya\ profile has two blue peaks and two associated flux minima.}
\label{table:lya}
\end{table*}

Following \citet{jaskot17}, we also measure the EW, the velocity weighted by absorption depth (\vchar), and the maximum velocity (\vmax) for the low-ionization lines \oi~$\lambda$1302, \sitwo~$\lambda$1190, $\lambda$1193, $\lambda$1260, $\lambda$1304, and \cii~$\lambda$1334 and the high-ionization lines \sithree~$\lambda$1206 and \siiv~$\lambda$1394 and $\lambda$1403. For consistency, we re-measure these parameters for the GPs in the \citet{henry15} sample. We calculate uncertainties using a Monte Carlo method. We generate 1000 possible spectra based on the per-pixel uncertainties and an additional 10\%\ continuum normalization uncertainty and re-measure all UV metal absorption and emission line parameters. We also include the G140L measurements of \sitwo~$\lambda$1260 EW and velocity from \citet{chisholm17} for 5 LCE GPs. The other GPs with COS observations do not have sufficient S/N in the continuum for reliable LIS line measurements. Finally, we exclude all lines that may be affected by Milky Way or geocoronal emission lines. 

We take covering fraction (\fcov) measurements for our sample from \citet{mckinney19} and for an additional 10 GPs, including 5 LCE GPs, from \citet{gazagnes18}. Covering fractions are derived from the residual intensity of the \sitwo\ lines at line center. Because gas clumps that cover the continuum source may absorb at other velocities, our measured \fcov\ is only a lower limit on the true gas covering fraction \citep[\eg][]{jones13, riverathorsen15}. Nevertheless, we expect that the measured \fcov\ should correlate with the true covering fractions in our sample of GPs. Of the GPs with \fcov$<0.5$, the five with LyC measurements are all confirmed LCEs \citep{gazagnes18}, which suggests that they genuinely have low covering fractions. Most of the remaining GPs have similar absorption line kinematics (\vchar, \vmax, FWHMs) at both low and high \fcov\ \citep[\eg][]{mckinney19}. Among GP galaxies, lower residual intensities in \sitwo\ therefore likely result from lower gas covering fractions rather than from absorption at a broader range of velocities. However, we caution that the \sitwo-derived \fcov\ may not be exactly equivalent to the true gas covering fraction.

\subsection{UV Morphologies}
\begin{figure*}
\epsscale{1}
\plotone{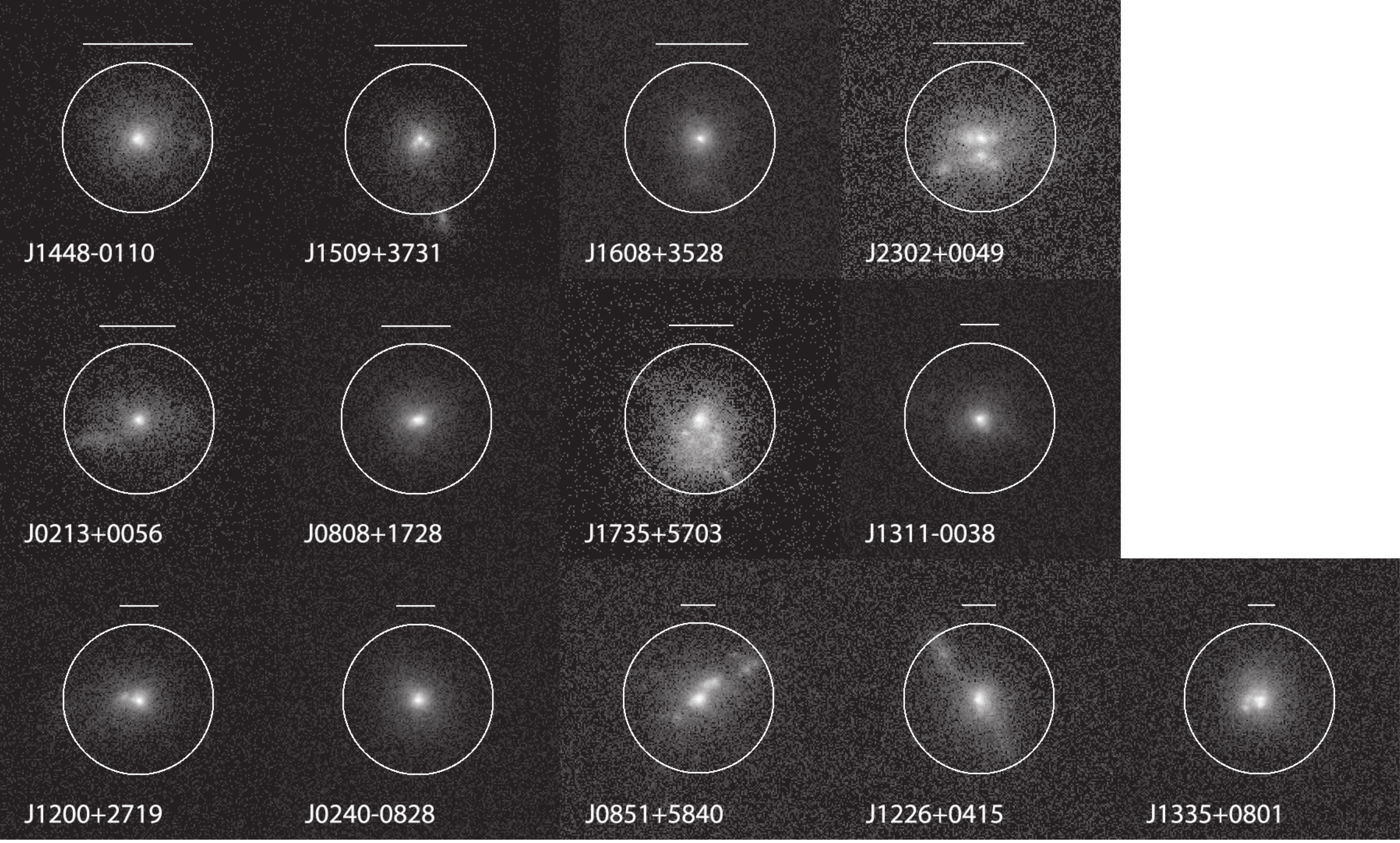}
\caption{COS NUV acquisition images of the sample, ordered by increasing redshift from left to right. The images are displayed using a logarithmic brightness scale. The scale bars in each image represent 1 kpc, and the circle is the 2.5\arcsec\ diameter aperture, centered on the image centroid.} 
\label{fig:cosmorph}
\end{figure*}

Our sample is at lower redshift than previous GP observations \citep[\eg][]{henry15, yang17}, yet the COS acquisition images still show compact UV morphologies (Fig.~\ref{fig:cosmorph}). Even in cases where multiple clumps are present, a single clump generally dominates the UV emission. As with the nearby GP analog NGC 2366, one young star-forming region may be primarily responsible for the GPs' nebular properties. Like higher redshift GPs \citep[\eg][]{cardamone09}, some GPs in this sample also show morphological disturbances, tidal tails, or possible companions (e.g., J1509+3731, J1735+5703, J1226+0415), which suggests that mergers may have triggered some of these starbursts.

In this low-redshift sample, both the stellar FUV and \lya\ emission trace compact regions. Following \citet{yang17a}, we generate exponential profiles of various effective radii, convolve these profiles with the instrumental point-source profile, and compare the resulting FWHMs with the measured widths in the cross-dispersion directions for the two-dimensional spectra. For our targets at $z\geq0.08$, the FUV continuum in the cross-dispersion direction is indistinguishable from a point-source profile. For the lower redshift GPs, the estimated FWHMs are 0.2-0.3 kpc in the continuum and 0.4-1.2 kpc for \lya; J1735+5703 is slightly more extended with a continuum FWHM$=$0.8 kpc. In both FUV and \lya\ emission, our sample appears more compact than the higher-redshift GPs in \citet{yang17a}. Since our sample is at $z\lesssim0.1$, COS may be able to spatially resolve star-forming knots that would be blended in observations of higher-redshift GPs. 

Although the \lya\ emission is more spatially extended than the continuum, it still represents a relatively compact spatial area. Similar to \citet{yang17a}, we generally find no major differences in the spatial origin of the blue and red sides of double-peaked \lya\ profiles. One exception is J1509+3731, whose red \lya\ peak appears spatially offset toward the weaker of the two FUV-emitting knots within the aperture. The $\lesssim$1 kpc extent of the \lya\ emission in this sample indicates that the observed \lya\ predominantly originates from the major star-forming regions, with a minimal contribution from scattered halo light. The \lya\ spectral properties will therefore reflect the optical depth and other physical conditions of the starburst region.

\section{\lya\ in High Ionization GPs}
\subsection{\lya\ Profiles}
\begin{figure*}
\epsscale{1.15}
\plotone{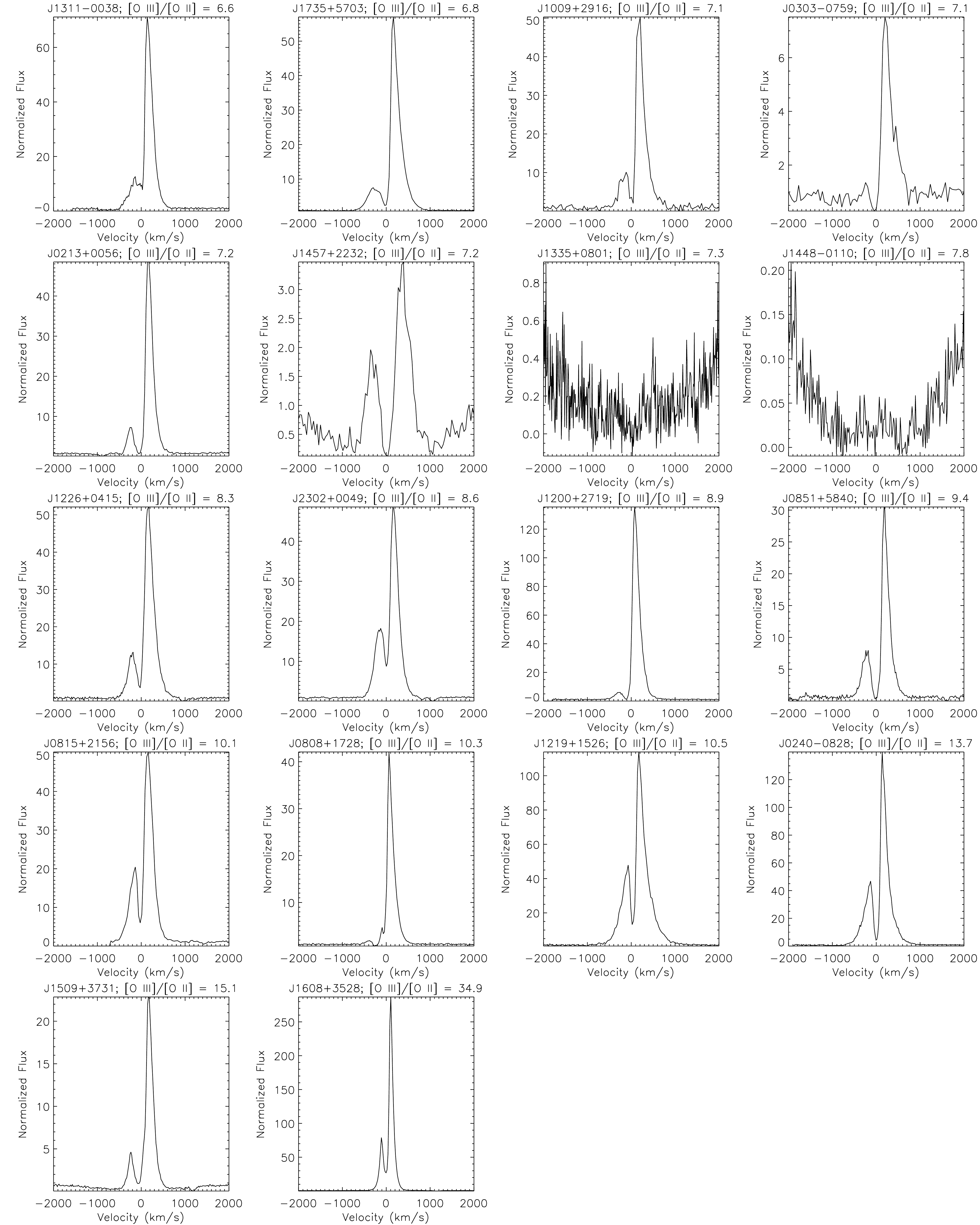}
\caption{The \lya\ profiles of GPs with high \oiii/\oii, normalized to the continuum level. We include 5 GPs from \protect\citet{jaskot14}, \protect\citet{henry15}, and \protect\citet{yang17}, which also have \oiii/\oii$>6.6$. The profiles are shown in order of increasing \oiii/\oii.} 
\label{fig:lya_gallery}
\end{figure*}

Figure~\ref{fig:lya_gallery} shows a gallery of the observed \lya\ profiles of the new sample, in order of increasing \oiii/\oii. As with previous GP samples \citep[\eg][]{henry15, yang17, verhamme17}, a high fraction of these extreme ionization GPs are LAEs, with EWs $>25$\AA. However, not all high \oiii/\oii\ GPs show \lya\ in emission. The deep absorption troughs of J1335+0801 and J1448-0110 imply a high line-of-sight column density and show that high \oiii/\oii\ does not guarantee a low optical depth in all directions. The GPs with \lya\ in emission generally exhibit double- (or in one case, triple-) peaked \lya\ profiles, with weaker blue peaks, consistent with net gas motion toward the observer. Many of the high ionization GPs also have significant residual flux between the emission peaks. In addition to having the highest \oiii/\oii\ ratio of the sample, J1608+3528's has the narrowest \lya\ profile, with \vpeaks$=214$ \kmps, and highest residual flux at line center. Although only 18\% of its \lya\ photons escape, those photons that do make it out of the galaxy do so without significant scattering, which suggests that they have traveled along low column density sight lines.

\begin{figure*}
\epsscale{1.1}
\plotone{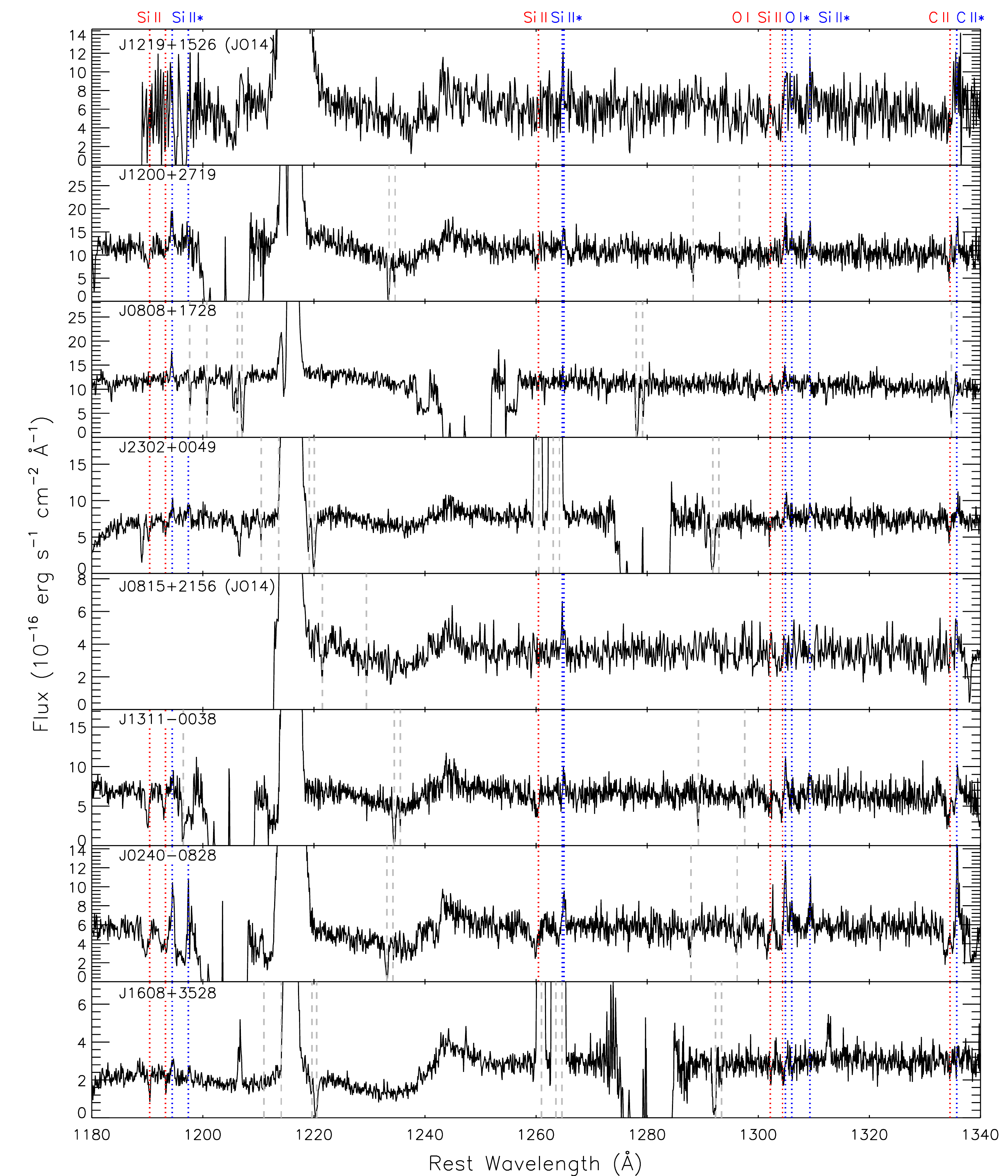}
\caption{ Rest-frame UV spectra of high \oiii/\oii\ GPs in order of decreasing \fesclya. The label ``(JO14)" identifies the four GPs from \protect\citet{jaskot14}. Red dotted lines show the positions of resonant LIS lines, and blue dotted lines show the corresponding transitions to the first fine structure level. Gray dashed lines show the positions of Milky Way or geocoronal features.} 
\label{fig:fullspec1}
\end{figure*}

\begin{figure*}
\epsscale{1.1}
\plotone{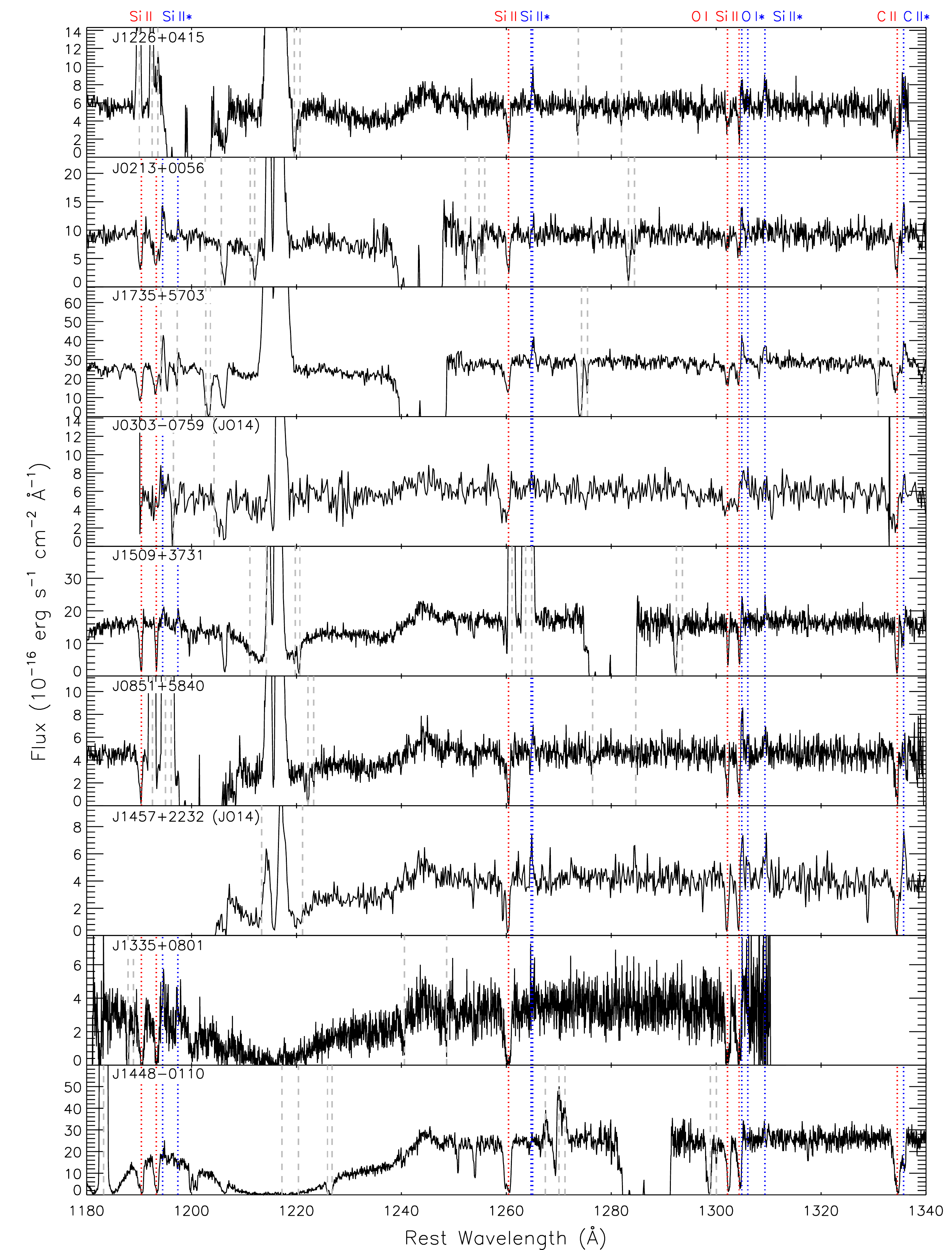}
\caption{ Same as Figure~\ref{fig:fullspec1}.} 
\label{fig:fullspec2}
\end{figure*}

As discussed by \citet{mckinney19}, many GPs in the sample simultaneously show strong \lya\ emission and underlying absorption troughs (Figures~\ref{fig:fullspec1} and \ref{fig:fullspec2}). Underlying \lya\ absorption is most noticeable in J1509+3731. Weak blue-shifted absorption appears in J1311-0038 and J0240-0828, and we detect broad, shallow \lya\ absorption in J1608+3528, J1226+0415, J0213+0056, and J0851+5840 \citep{mckinney19}. The presence of both \lya\ absorption and narrow \lya\ emission suggests that the neutral gas in the GPs is inhomogeneous, with both high column density \hi\ and more transparent regions within the COS aperture \citep{mckinney19}.

These \lya\ absorption troughs are present in a higher fraction of our sample compared with previous GP samples. Nine of the thirteen galaxies in our new sample show a \lya\ absorption component, whereas only eight of the 44 GPs in \citet{yang17} show the same trait. In galaxies with high inferred neutral gas covering fractions (e.g., J1509+3731, J1335+0801; \citealt{mckinney19}), \lya\ absorption will be deep and easily detectable. However, many GPs show low \fcov\ \citep{gazagnes18, mckinney19}, which will lead to shallow \lya\ absorption. In this case, the \lya\ absorption will only be detected when the continuum signal-to-noise is sufficiently high and when \lya\ emission does not overlap with the absorption region. For instance, GPs with broad \lya\ emission, such as J1219+1526, can mask \lya\ absorption within 1000 \kmps\ of the systemic velocity, where it would be deepest and most noticeable. 

In addition to higher signal-to-noise, our sample has systematically narrower optical emission lines compared with the GPs in \citet{yang17}, likely because our GPs have systematically lower luminosities (Figure~\ref{fig:sample}) and lower dynamical masses. The SDSS emission lines for our sample are usually fit by a single Gaussian with median FWHM of 78 \kmps, whereas the GPs in the \citet{yang17} sample have emission line fits with median FWHM$=89$\kmps\ and typically require a second Gaussian to fit the emission line wings. As a result of narrow intrinsic \lya\ line profiles and/or less resonant scattering, our sample tends to have narrow observed \lya\ profiles. The high signal-to-noise continua and narrow \lya\ profiles of our sample may enable us to detect \lya\ absorption in galaxies with \fcov$\lesssim0.5$, such as J0213+0056 and J1608+3528 \citep{mckinney19}. 

\subsection{Correlations with \lya\ Properties}
\label{sec:lyacor}
Strong, narrow, double-peaked \lya\ profiles appear common among the GPs, yet it is not clear which of the GPs' properties are most closely linked to these \lya\ properties. Because \lya\ resonantly scatters on neutral hydrogen, the causes of strong \lya\ can be complex. High ionizing fluxes or strong rates of collisional excitation can increase \lya\ production, but \lya\ photons must then traverse the ISM to escape. High \hi\ column densities will broaden and weaken \lya\ profiles as \lya\ photons scatter away from resonance and increase the chance of absorption by dust \citep[\eg][]{verhamme15}. Outflows can also facilitate \lya\ escape by shifting \hi\ gas and \lya\ photons out of resonance with each other \citep[\eg][]{kunth98, mashesse03}. 

To shed light on the factors that influence the GPs' \lya\ emission, we investigate the correlations between \lya, UV, and optical properties. We consider four main \lya\ properties (\lya\ EW, \vpeaks, \fesclya, and the minimum residual intensity between the \lya\ peaks relative to the continuum level, \fmin/\fcont). Using the nonparametric Spearman rank method, we calculate correlation strengths (\rcor) between these \lya\ properties and other properties of the GPs, which we describe in the Appendix. For these correlations, we include the GPs listed in \citet{yang17} for a total of 56 galaxies. GP J0808+1728 has two blue peaks and two associated minima, and we list correlation strengths calculated with each of these values separately. In Tables~\ref{table:cor_ew}-\ref{table:cor_fmin}, we list the resulting \lya\ correlations with $\left | \rho \right |  \geq0.5$ for measurements with at least 10 GPs. In each section below, we first describe the main observed correlations and then suggest a possible physical interpretation.

\subsubsection{The Origins of Strong \lya}
Each \lya\ property correlates strongly with other \lya\ properties, such that GPs with high \lya\ EWs also have high \fesclya\ (\rcor=0.88), higher residual fluxes at profile minimum (\rcor=0.82-0.83), and narrower velocity peak separations (\rcor=0.69-0.72).  In general, the correlations between \lya\ properties are typically stronger than the correlations between \lya\ and most other nebular or galaxy properties. Each of these \lya\ parameters also correlates with weaker low-ionization absorption line EWs \citep[\cf][]{shapley03, henry15}.

These correlations suggest that low optical depth is one of the primary drivers of the GPs' \lya\ properties. Lower optical depths, either from low covering fractions or low column densities, are a likely cause of reduced \lya\ scattering and narrow \vpeaks\ \citep[\eg][]{verhamme15, dijkstra16} and would also result in weak low-ionization absorption lines. Reduced scattering in the neutral ISM may contribute to higher \lya\ escape in the GPs, leading to the observed correlation between narrow \vpeaks\ and higher \lya\ EWs.

\begin{figure*}
\epsscale{0.6}
\plotone{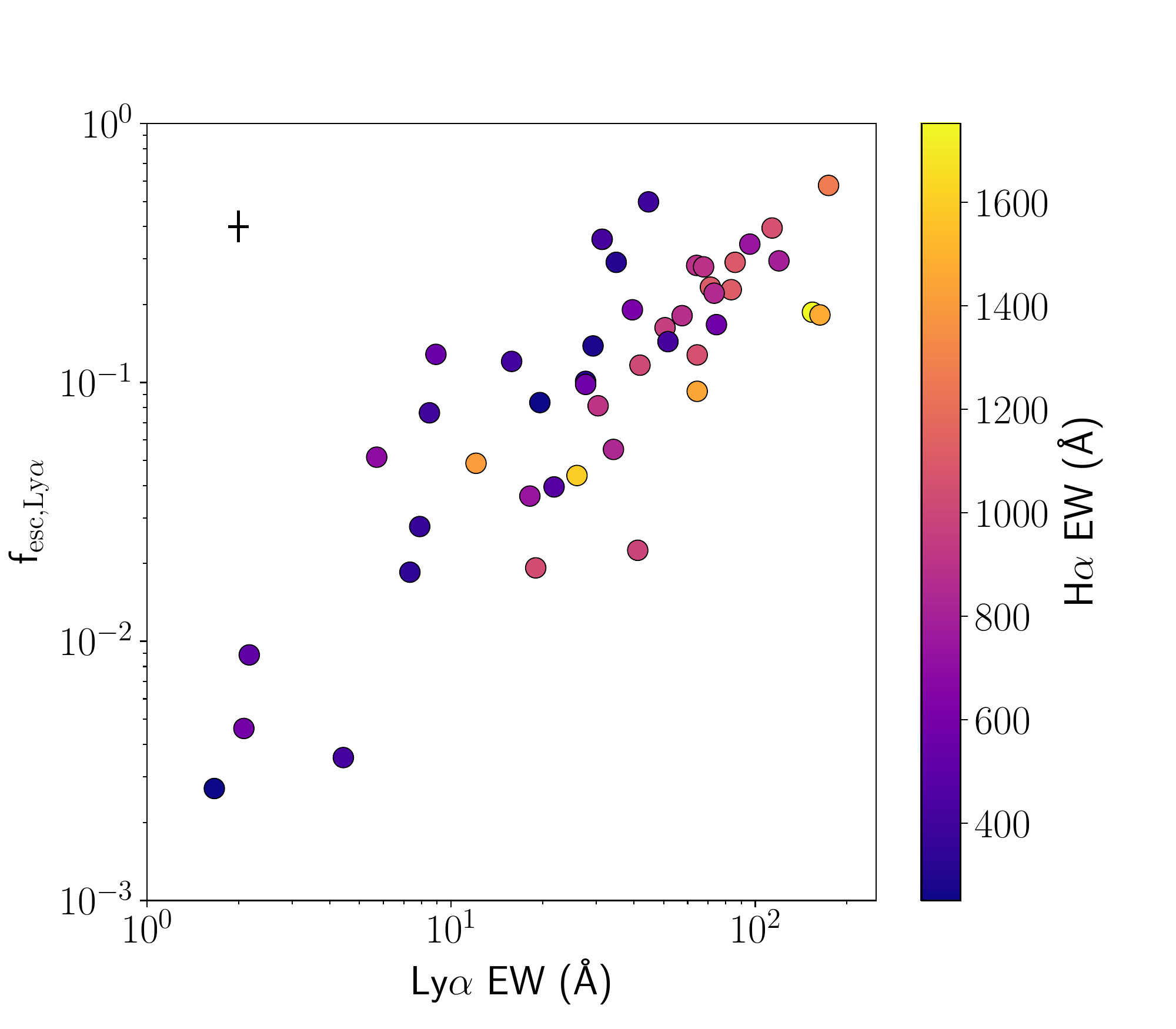}
\caption{\fesclya\ correlates with \lya\ EW. H$\alpha$ EW, a measure of the intrinsic \lya\ strength, is shown by the color scaling. Only GPs with net \lya\ emission are shown. Representative errors are indicated by the error bar in the upper left.} 
\label{fig:fesc_ew}
\end{figure*}

In addition to \lya\ escape, the production of \lya\ photons also affects the observed \lya\ EWs. The strong correlation between \lya\ EW and \fesclya\ is not surprising. {\bf However, even moderate \fesclya\ will result in strong \lya\ EWs, as long as the intrinsic production of \lya\ is strong.} Figure~\ref{fig:fesc_ew} shows the observed correlations between \lya\ EW, \fesclya, and H$\alpha$ EW for the GPs. While all GPs show high H$\alpha$ EWs, the highest H$\alpha$ EWs ($\gtrsim800$\AA) are more common at the high \lya\ EW end. Among the GPs with high \lya\ EWs, galaxies with lower \fesclya\ tend to show higher H$\alpha$ EWs. For instance, two GPs show \lya\ EWs $>100$\AA\ yet have \fesclya\ of $<20\%$. These same GPs show H$\alpha$ EWs $>$ 1400\AA. Since we find no systematic differences in the dust-corrected UV to optical continuum ratios among the strong \lya-emitting GPs, we infer that the high H$\alpha$ EWs imply elevated {\it intrinsic} \lya\ EWs and high \lya\ production. Because of their high \lya\ production, these GPs do not need extremely high \fesclya\ to attain high observed \lya\ EWs. 

The high detection rates of LyC emission among GPs may have a similar explanation. For instance, \citet{schaerer16} find high intrinsic production of LyC photons in the GPs. The high LyC production in this population of young starbursts may therefore ensure a strong LyC flux into the IGM, even for moderate \fesclyc. 

\subsubsection{\vpeaks\ and Optical Depth}
\label{subsec_vpeaks_optdepth}
\begin{figure*}
\epsscale{1}
\gridline{\fig{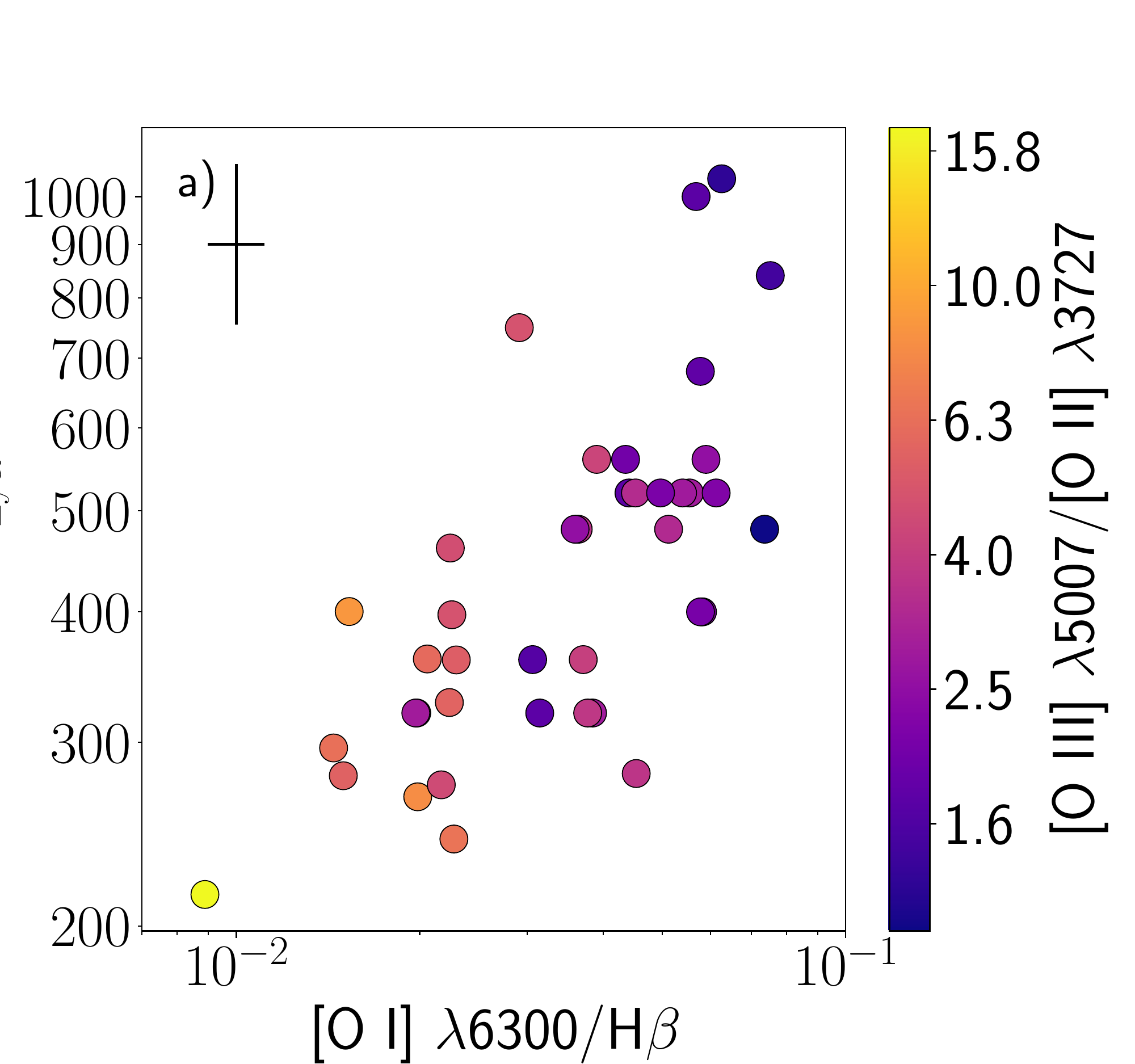}{0.36\textwidth}{}
	\fig{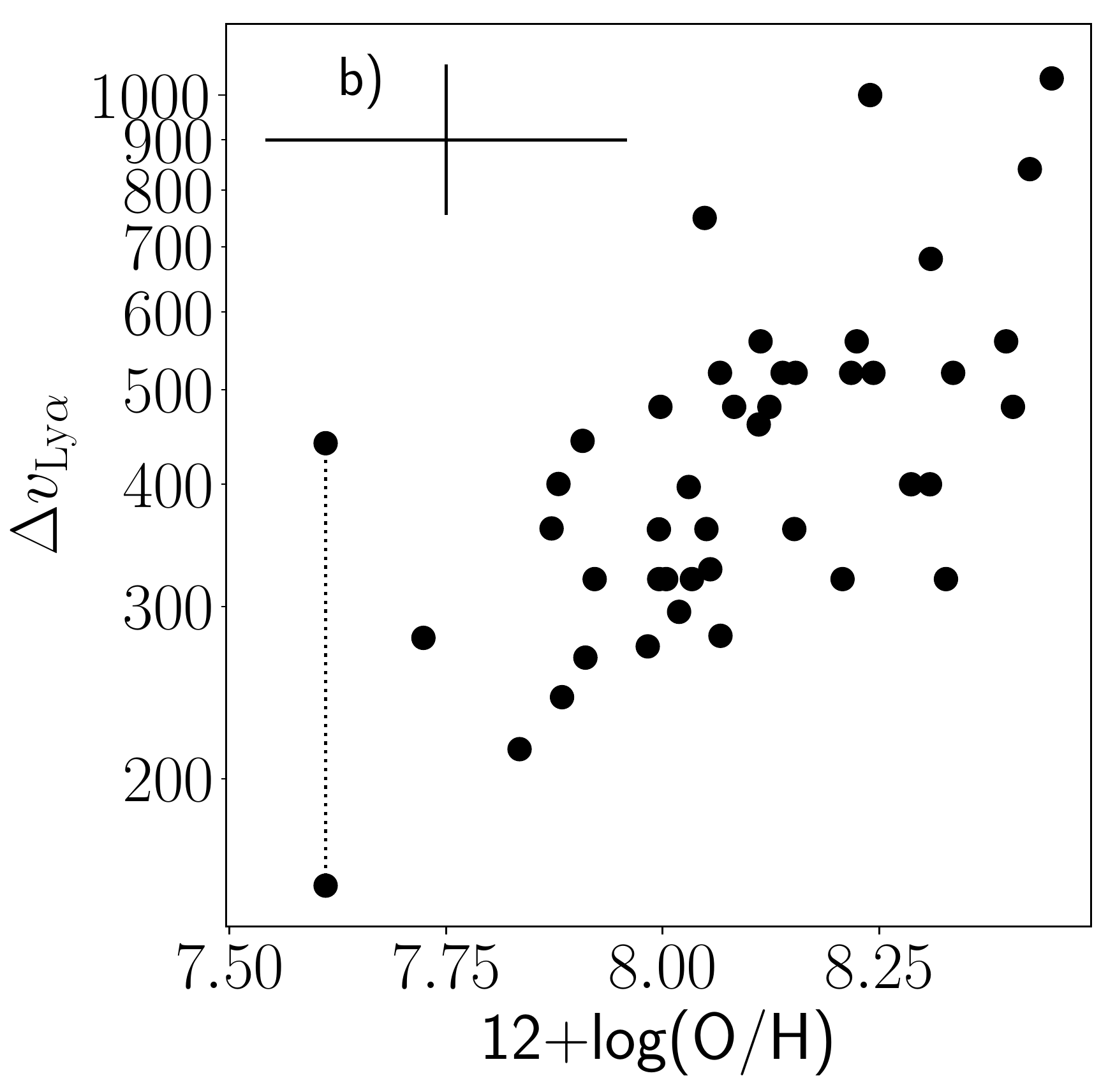}{0.3\textwidth}{}
	\fig{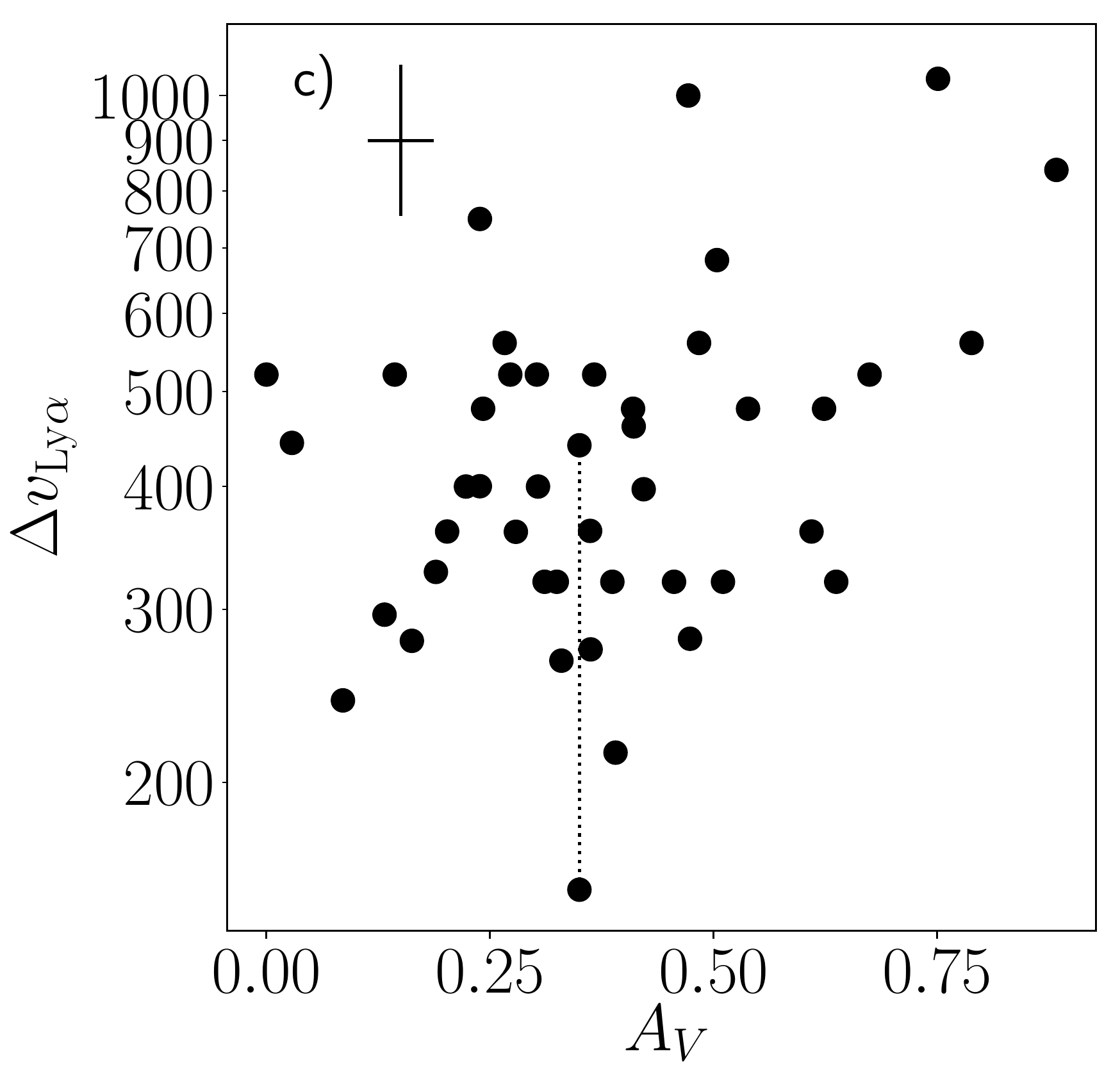}{0.3\textwidth}{}
	}
\caption{(a) Lower \vpeaks\ correlates with lower \oifb~$\lambda$6300/H$\beta$. Color shows \oiii/\oii. (b) Galaxies with lower gas-phase metallicities tend to have lower \vpeaks. (c) \vpeaks\ does not show any clear correlation with $A_V$. The dotted line connects the two values of \vpeaks\ for J0808+1728. Representative error bars are shown in the upper left of each panel.} 
\label{fig:vpeaks_cors}
\end{figure*}

Of all the \lya\ properties considered, \vpeaks\ is the property most closely associated with measures of the nebular ionization state. Optical nebular line ratios, such as \oifb~$\lambda$6300/H$\beta$ (\rcor=0.69), \oifb~$\lambda$6300/\oiii~$\lambda$5007 (\rcor=0.67), and \oiii~$\lambda$5007/\oii~$\lambda$3727 (\rcor=$-0.57$ to $-0.61$) show some of the strongest correlations with \vpeaks\ (Fig.~\ref{fig:vpeaks_cors}a; Table~\ref{table:cor_vpeaks}). Two of the strongest \vpeaks-line ratio correlations involve \oifb~$\lambda$6300, which suggests that \vpeaks\ is sensitive to the overall neutral gas content. 

Ratios such as \oifb/H$\beta$ and \oifb/\oiii\ are possible diagnostics of density-bounding \citep[\eg][]{iglesias02,stasinska15}, and because \oifb\ traces the neutral gas phase, ratios involving \oifb\ may be more reliable diagnostics than \oiii/\oii. However, as a weak line, \oifb\ may be difficult to measure at high redshift. As seen in Fig.~\ref{fig:vpeaks_cors}a, a sample selection based on \oiii/\oii\ tends to select galaxies with low ratios of neutral to ionized gas (i.e., low \oifb/H$\beta$) and low \vpeaks.

This association between weaker \oifb\ and narrower \lya\ profiles is likely related to the ionization state of the ISM. Although shocks can contribute to \oifb\ emission, \oifb\ also has a non-shock origin in neutral and partially ionized regions. It is also not clear why higher \oiii/\oii\ would correlate with weaker shock emission. In the outer edges of \hii\ regions, charge exchange reactions produce neutral O, which can then generate \oifb~$\lambda$6300 emission via collisional excitation by electrons \citep{iglesias02}. Consequently, \oifb/H$\beta$\ could serve as a diagnostic of density-bounded nebulae \citep{iglesias02, stasinska15}. However, \oifb\ can also originate from the photodestruction of OH in the hot neutral zones within photodissociation regions (PDRs; \citealt{storzer98}) and from collisional excitation in the diffuse ionized gas (DIG) of the ISM \citep[\eg][]{voges06}. Starburst galaxies have a higher ratio of \hii\ region to DIG luminosity \citep{hanish10}, and similarly, the most highly ionized GPs could have a weaker DIG contribution and lower \oifb/H$\beta$. In summary, GPs with high \oiii/\oii\ could have lower \oifb/H$\beta$ ratios due to more density-bounded \hii\ regions, weaker PDR emission, and/or a weaker DIG contribution, all of which could result from a more highly ionized ISM.

\subsubsection{\vpeaks\ and \lya\ Escape}
\label{sec:fesclya_vpeaks}
\begin{figure*}
\epsscale{1}
\plottwo{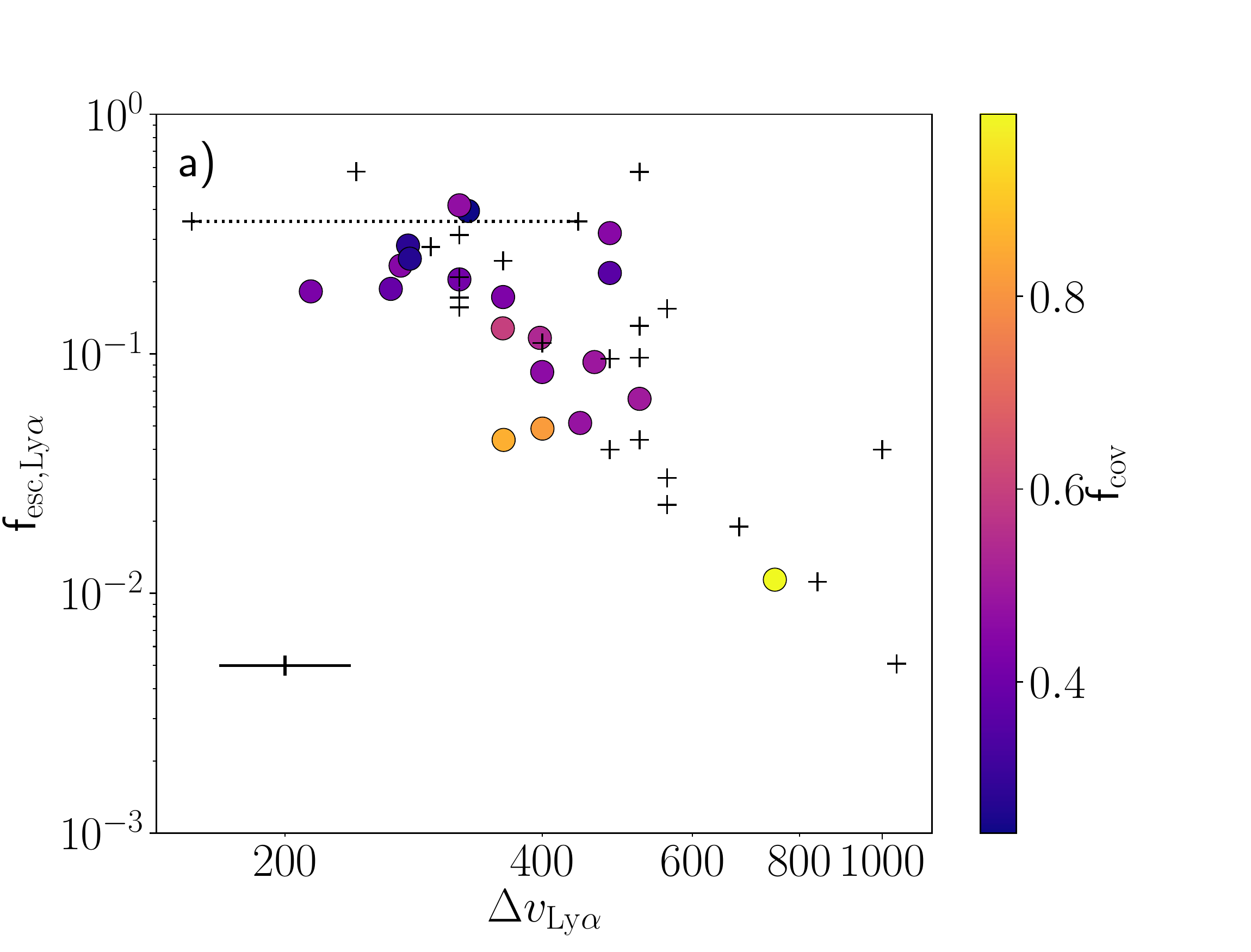}{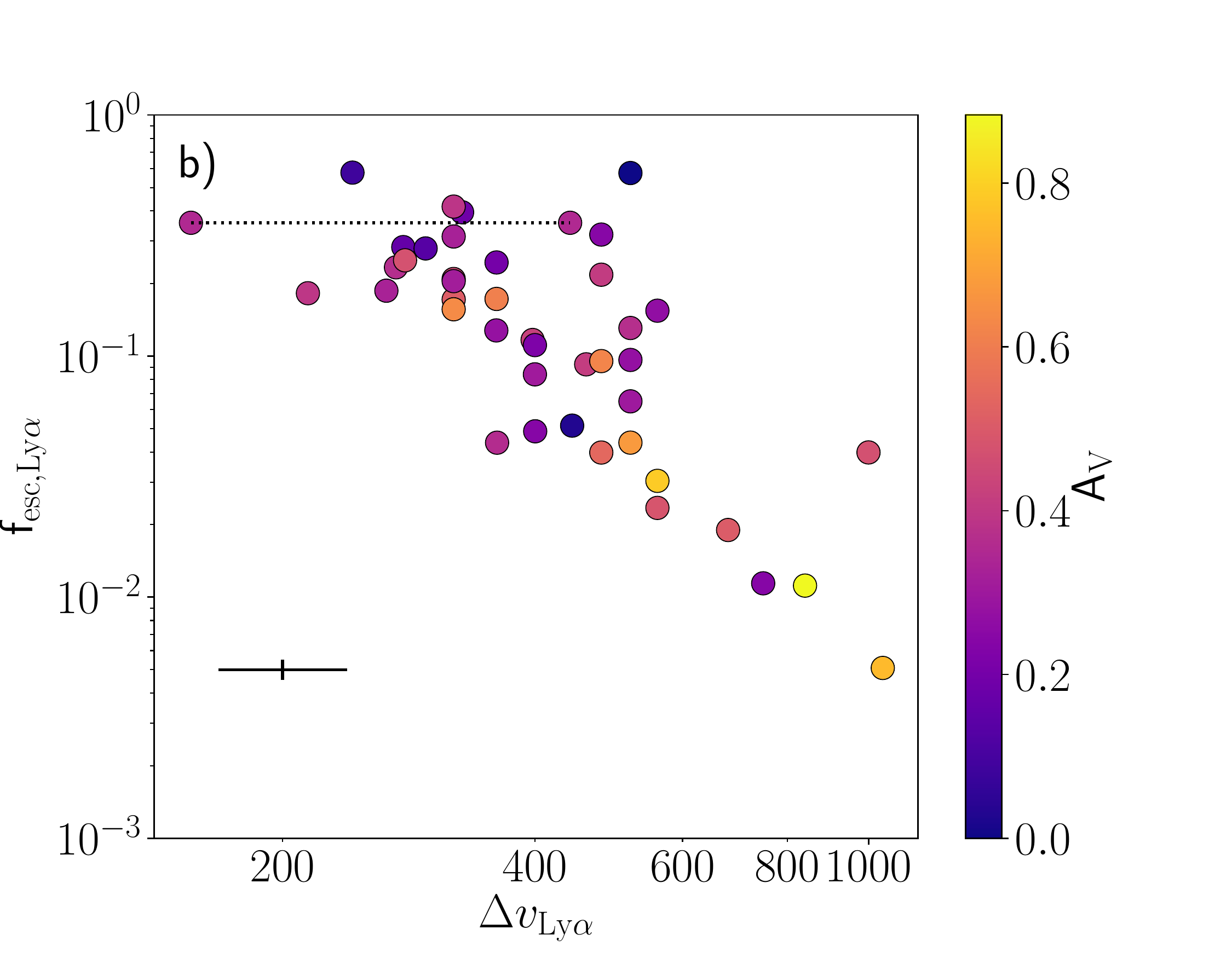}
\caption{\vpeaks\ and \fesclya\ anti-correlate, but with substantial scatter at low \vpeaks. The dotted line connects the two values of \vpeaks\ for J0808+1728. Representative error bars are shown in the lower left corner. (a) Color shows \fcov. At a given \vpeaks\ value, galaxies with lower \fcov\ tend to show higher \fesclya. Data points without \fcov\ estimates appear as small, black points. (b) Color shows \AV. Galaxies with lower \AV\ also tend to show higher \fesclya\ for a given value of \vpeaks.} 
\label{fig:fesc_vpeaks}
\end{figure*}

Both \vpeaks\ and \fesclya\ are potential diagnostics of LyC escape \citep[\eg][]{verhamme15, verhamme17, dijkstra16, izotov18b}, but the relationship between these parameters is not yet clear. As with other \lya\ properties, high \fesclya\ and narrow \vpeaks\ do correlate with each other, although their relationship has substantial scatter (Fig.~\ref{fig:fesc_vpeaks}; \rcor=-0.67 to -0.71; \cf\ \citealt{yang17}). 

We see a much tighter anti-correlation between \fesclya\ and \fcov\ (\rcor=-0.82; Table~\ref{table:cor_fesc}), as shown by \citet{riverathorsen15} and \citet{mckinney19}. In fact, the available \fcov\ measurements suggest that lower covering fractions can account for some of the spread in \fesclya\ at a given \vpeaks\  (Fig.~\ref{fig:fesc_vpeaks}a). \citet{yang17} propose that dust extinction may also influence the scatter. We find that low dust extinction correlates with higher \fesclya\ at a given \vpeaks\ in some cases (Fig.~\ref{fig:fesc_vpeaks}b). However, the trend is fairly weak \citep[\cf][]{giavalisco96, atek09, hayes13}. 

Although both \vpeaks\ and \fcov\ anti-correlate with \fesclya, they do not correlate as strongly with each other (\rcor=0.56; Table~\ref{table:cor_vpeaks}). As discussed above, \vpeaks\ is closely related to ionization state (e.g., Fig.~\ref{fig:vpeaks_cors}a). However, neither \fcov\ nor \fesclya\ show the same connection. The correlation coefficients for \fcov\ with \oifb/H$\beta$\ and \oiii/\oii\ are only \rcor=-0.10 and 0.22, and the correlations for \fesclya\ are similarly weak (\rcor=-0.36 and 0.32, respectively).

Studies of the GPs typically posit either a density-bounded or ``picket-fence" \hi\ geometry to explain their LyC escape \citep[\eg][]{izotov18a,chisholm18}. However, our observed \lya\ correlations imply a more complex geometry. While the ionization dependence of \vpeaks\ suggests a sensitivity to density-bounding (Section \ref{subsec_vpeaks_optdepth}), the connection between \fcov\ and \fesclya\ points to \lya\ escape through a patchy neutral ISM, as in the ``picket-fence" model \citep[\eg][]{heckman11} or clumpy gas model of \citet[\eg][]{gronke17}. 

{\bf Hence, the \vpeaks\ and \fcov\ parameters may play different roles in \lya\ escape, with \vpeaks\ tracing column density, while \fcov\ traces porosity.} The coexistence of \lya\ absorption with \lya\ emission and the low derived covering fractions in many GPs suggest that the neutral gas is not smoothly distributed \citep{mckinney19}. The relative fraction of low-column-density channels along the line of sight, as traced by \fcov, has a significant effect on the observed \fesclya\ and likely on \fesclyc\ as well \citep[\eg][]{gazagnes18, chisholm18, mckinney19}. A more porous medium can make a low line-of-sight \fcov\ more likely, while also decreasing \vpeaks, as the \lya\ photons have fewer dense gas clumps on which to scatter \citep{hansen06, dijkstra16}. However, \lya\ scatters even at low column densities ($\tau\sim$1 for \nhi$=10^{12}$ cm$^{-2}$; \citealt{verhamme15}). Consequently, the column density of the inter-clump medium may also leave an imprint on the \lya\ profile by affecting \vpeaks\ \citep[\cf][]{kakiichi19}, and in LCEs, \vpeaks\ may reveal the \hi\ column density of the channels through which LyC escapes. The degree of ionization could control the residual inter-clump column density, with more transparent pathways producing narrower \lya\ profiles. Because \lya\ is sensitive even to low \nhi, \lya\ profile shape may therefore give insight into the diffuse gas between dense clouds. 

Figure~\ref{fig:fesc_vpeaks} also suggests a lower envelope of values, where \fesclya\ sets a lower limit on \vpeaks\ or vice versa. In \lya\ radiative transfer models with a clumpy gas distribution, \lya\ photon scatterings increase with clump filling factor \citep{hansen06, dijkstra16}. The porosity of the medium could therefore set a minimum \vpeaks, with higher column densities between clumps increasing \vpeaks\ above that minimum value. If porosity drives \fesclya, we might then expect to see a lower limit on the \lya\ profile width at a given \fesclya. 

Although we do not observe any obvious trends between redshift and \lya, we also note that aperture size may increase the scatter between \fesclya\ and \vpeaks. As redshift increases, the COS aperture will subtend a larger physical area, covering a diameter of 1.4 kpc at $z=0.027$ vs.\ a diameter of 12.5 kpc at $z=0.356$, the range of the full GP sample. At higher redshift, the aperture may capture a higher fraction of the galaxies' scattered \lya\ halos. Since \lya\ escapes via scattering, \fesclya\ may only set an upper limit on \fesclyc\ \citep{verhamme17}, and the connection between \vpeaks\ and \fesclyc\ could be tighter than the relationship between \vpeaks\ and \fesclya\ would suggest.

\subsubsection{\vpeaks, Ionization Parameter, and Metallicity}
\label{sec:vpeaks_u_z}

Interestingly, we find that narrower \vpeaks\ is also associated with lower metallicities (\rcor=0.61-66; Fig.~\ref{fig:vpeaks_cors}b). The correlation between \vpeaks\ and metallicity is as strong as the relationship between \vpeaks\ and other potential optical depth diagnostics, such as \oiii/\oii\ or \fesclya\ (Table~\ref{table:cor_vpeaks}; Fig.~\ref{fig:vpeaks_cors}b). The metallicity trend likely does not represent a trend with galaxy mass or dust extinction, as neither galaxy luminosity nor \AV\ (Fig.~\ref{fig:vpeaks_cors}c; Table~\ref{table:cor_vpeaks}) correlates as strongly with \vpeaks. Metallicity is strongly associated with ionization state (e.g., \oiii/\oii, \neiii/\oii) but not with other properties related to \lya\ escape, such as \lya\ EW, \fesclya, or covering fraction. {\bf Thus, metallicity appears to be linked specifically to \lya\ optical depth.}

Metal-poor \hii\ regions have long been associated with higher ionization parameter \citep[\eg][]{mcgaugh91} for reasons that are not fully understood. The connection is often attributed to a combination of harder ionizing spectra and higher electron temperatures at low metallicity \citep[\eg][]{stasinska15}. The BPASS stellar population models \citep{eldridge17} show that ionizing photon production in young ($<10$ Myr) starbursts can increase by a factor of two between $Z=0.002$ and $Z=0.008$, the metallicity range of the full GP sample. Other factors, such as a metallicity-dependent stellar initial mass function \citep[\eg][]{bromm99, larson05, marks12, schneider18}, could raise the ionizing photon production rate higher than the model predictions. 

Higher ionizing luminosities at lower metallicity are therefore one possible explanation for the trend we observe between metallicity, \oiii/\oii, and \vpeaks. The higher ionizing luminosities could generate lower \hi\ column densities, thereby reducing \vpeaks. In this case, we would expect to observe an anti-correlation between \vpeaks\ and H$\alpha$\ strength. However, low \vpeaks\ is associated with {\it lower} H$\alpha$\ luminosities (\rcor=0.55-0.59) and shows no strong relationship with H$\alpha$ EW (\rcor=-0.33).

Instead, the trend between metallicity and \vpeaks\ points to a more profound effect. At low metallicity, stellar winds are far weaker \citep[\eg][]{vink01}, and the highest mass stars may not explode as supernovae \citep[\eg][]{ramachandran19}, delaying the onset of energy-driven feedback. Thus, the cluster environment remains denser at the earliest ages. In particular, this denser gas is likely clumpier, with clouds of natal molecular gas present, which take longer to disrupt even after supernova feedback begins. This high-density ionized gas, in closer proximity to the ionizing cluster, therefore has a much higher ionization parameter. In the case of suppressed superwinds at low metallicity \citep[\eg][]{jaskot17, oey17}, this effect is further enhanced. The presence of clumpy, dense gas generates the porous geometry evidenced above and provides a low column density inter-clump medium through which \lya\ photons can travel. This inter-clump medium has a lower optical depth than would be seen for the same total gas mass distributed more uniformly, such as in an expanding superbubble shell driven by cluster superwinds at higher metallicity. 

\begin{figure*}
\plotone{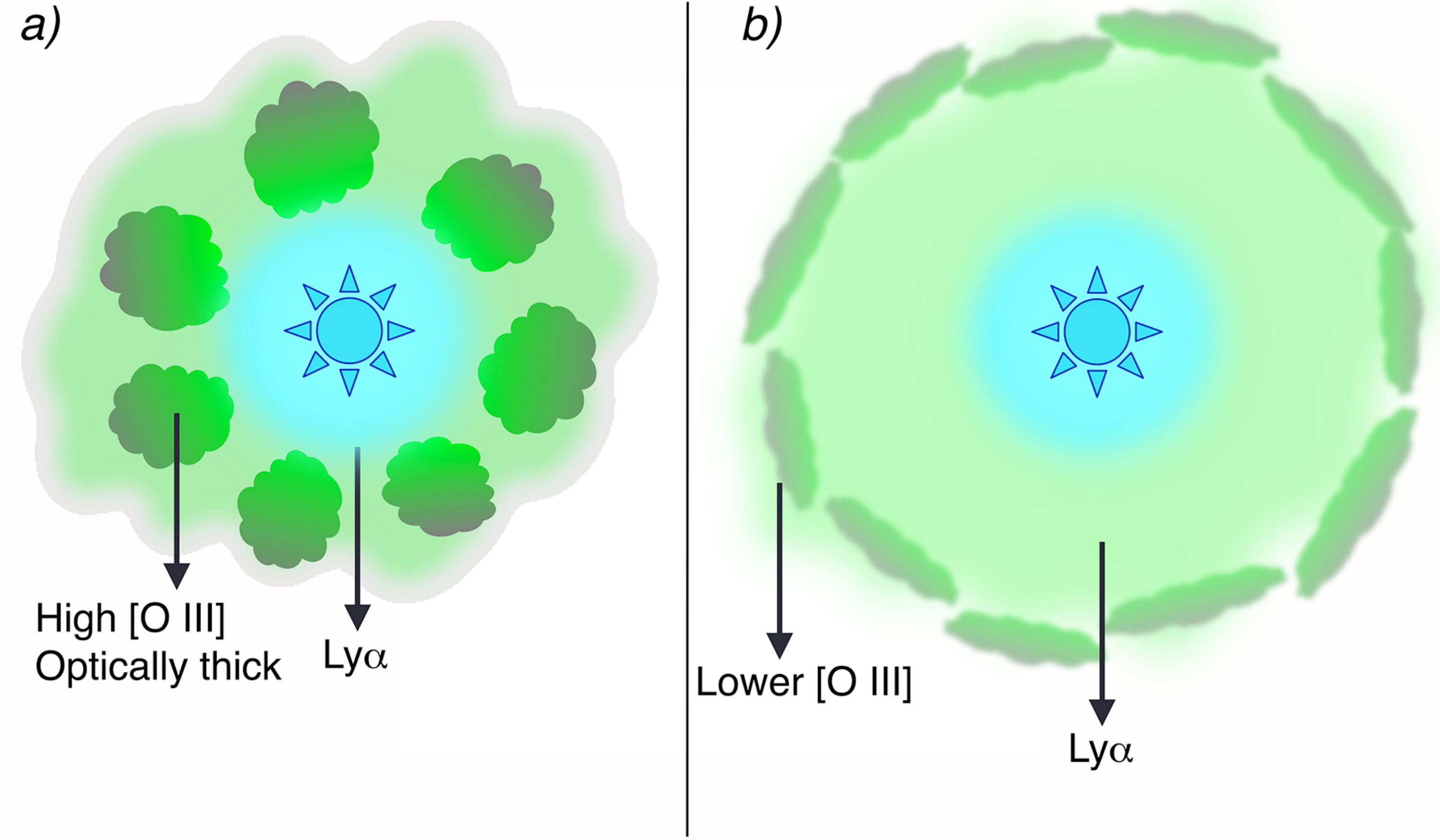}
\caption{An illustration of the proposed gas geometry at low vs. high metallicity. (a) At low metallicity, mechanical feedback is weaker, inducing catastrophic cooling. Thus, dense gas forms clumps and remains near the SSC. The high-density clumps have high ionization parameters and strong \oiii\ emission and generate the observed \lya\ and saturated low-ionization absorption (e.g., \sitwo, \cii, \oi). The \lya\ emission escapes from the low-density inter-clump medium. (b) At higher metallicity, expanding superbubbles drive gas farther from the SSC, resulting in lower ionization parameters. This gas is also more uniform, and \lya\ photons therefore traverse gas with a higher average column density. } 
\label{fig:superbubble}
\end{figure*}

Figure~\ref{fig:superbubble} illustrates our proposed model relationship between the ionizing super star cluster (SSC) and the gas geometry at low metallicity (panel $a$) and high metallicity (panel $b$). For metal-poor conditions, the weak stellar winds and lack of early supernovae \citep[\eg][]{heger03, sukhbold16} promote catastrophic cooling conditions where mechanical feedback is suppressed \citep[\eg][]{silich04, silich17}. Cooling generates a more clumped and compact geometry, promoting the formation of a picket-fence configuration (Section~\ref{sec:fesclya_vpeaks}). Gas in the high-density clumps is exposed to the ionizing radiation at close proximity, generating extreme ionization parameters and high \hii\ luminosities and surface brightness. At the same time, the gas in the low-density, inter-clump passageways has low \hi\ column density and is more optically thin to \lya, generating signatures of low optical depth in neutral gas, such as reduced \vpeaks\ and weaker \oifb\ and \oii\ emission. The radiative feedback dominating this environment is also conducive to clumping and the formation of optically thin channels \citep{krumholz_thompson12}.

On the other hand, for metal-rich conditions, strong stellar winds drive a superwind from the SSC, launching a superbubble that quickly drives gas much farther from the SSC. The ionization parameter is therefore lower. Larger quantities of ISM gas are swept up, with less clumping, and so the likelihood of encountering optically thin channels is also lower, increasing the mean observed \hi\ column. In principle, higher metallicity should enhance cooling, therefore implying that additional factors must be important to generate the correlation between column densities and metallicity. While weak mechanical feedback is one such factor, catastrophic cooling conditions occur for SSCs that are both extremely massive ($>10^5 \Msol$) and compact ($\lesssim 5$ pc; \citealt{silich17}). Our results may suggest that the formation of such systems is enhanced at low metallicity.

\subsubsection{The Flux at \lya\ Profile Minimum}
The double-peaked \lya\ profiles of GPs often show net flux at the profile minimum \citep[\eg][]{jaskot14, henry15, verhamme17}, where the \lya\ optical depth should be highest. Radiative transfer models differ in their interpretation of this residual flux, depending on whether the model geometry consists of a homogeneous shell or dense clumps. In the shell models, higher residual intensities occur in lower column density models and should show an association with lower \vpeaks\ \citep{orlitova18}. Clumpy models also predict that high residual intensities arise from low optical depths, which in these models come from low porosity \citep{gronke16b}. Observationally, \citet{orlitova18} noted that lower \vpeaks\ does correlate with higher relative \fmin.

\begin{figure*}
\epsscale{1}
\plottwo{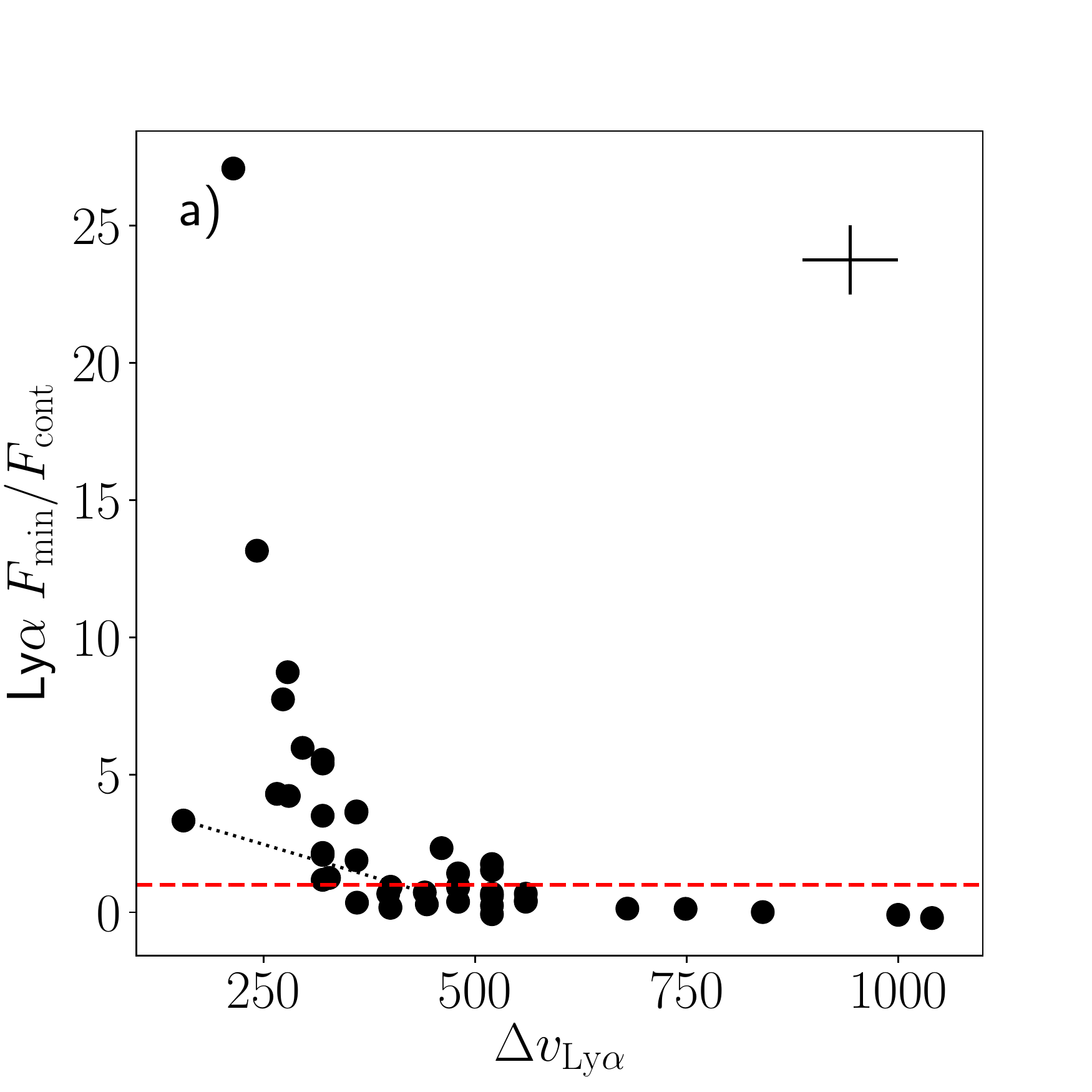}{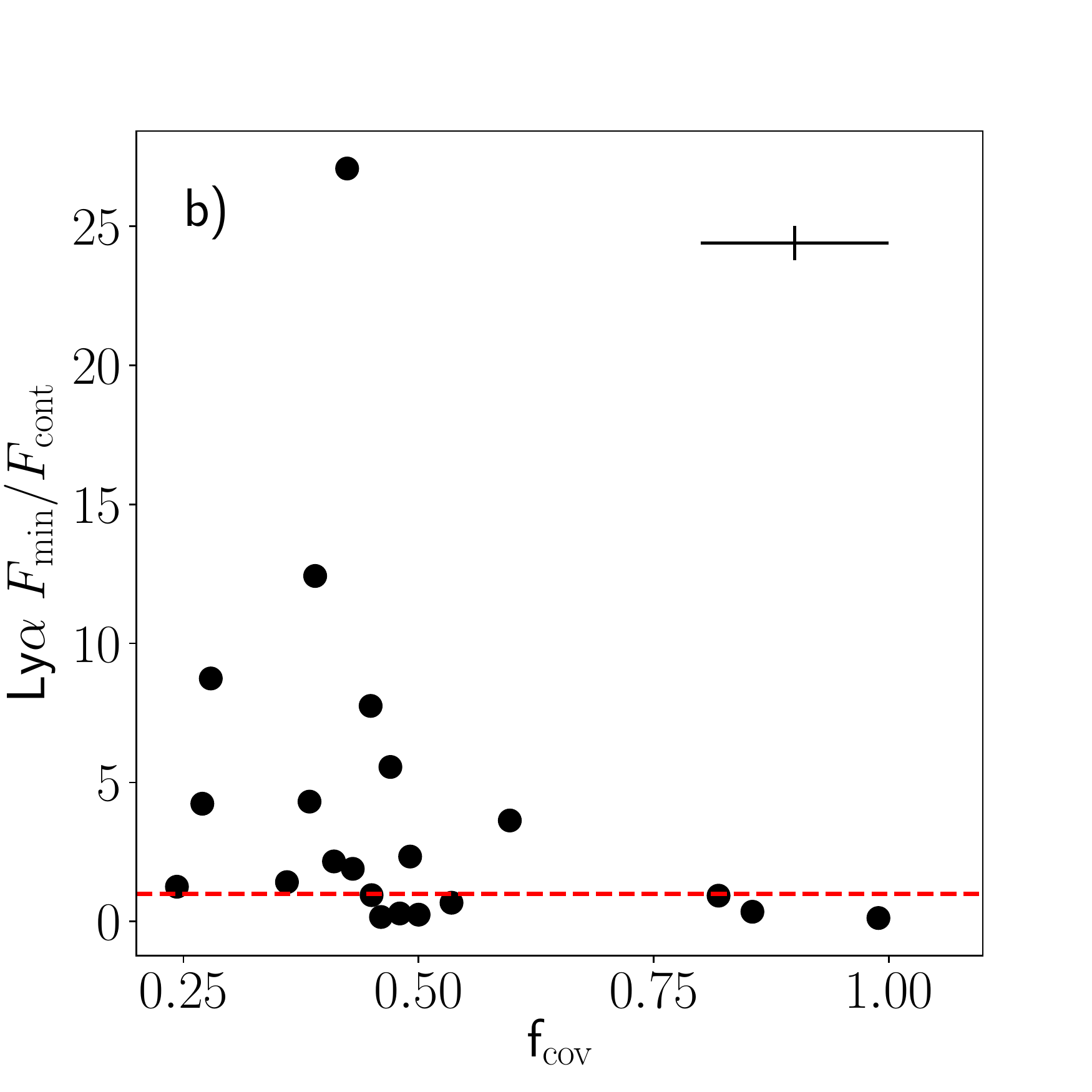}
\caption{ (a) The ratio of the flux at \lya\ profile minimum to the continuum level (\fmin/\fcont) vs. \vpeaks. Galaxies above the red dashed line have net flux above the continuum. The dotted line connects the measurements for each of J0808+1728's blue peaks and associated minima. Narrower \lya\ profiles show a greater residual flux between the two \lya\ peaks. (b) \fmin/\fcont\ vs. \fcov. Galaxies with high \fcov\ show low residual flux between the \lya\ peaks, but otherwise \fmin/\fcont\ shows no strong trend with \fcov. Representative error bars appear to the upper right of each panel.} 
\label{fig:fmin}
\end{figure*}

With a larger sample that extends to lower \vpeaks\ values, we corroborate this trend (Fig.~\ref{fig:fmin}a); lower \vpeaks\ and higher \fmin/\fcont\ correlate with \rcor\ $=0.82$. In contrast, the trend between \fmin/\fcont\ and \fcov\ is weaker (Fig~\ref{fig:fmin}b; \rcor = -0.56). GPs with high covering fractions universally have low \fmin\ values, but the scatter increases at low \fcov. Part of this scatter could be due to the fact that the \fcov\ measured along our line of sight is not necessarily representative of the average gas porosity. Clumpy models also predict that the residual flux is related to the fraction of the CGM captured by the aperture, but we see no trend with redshift (\rcor=$-0.13$ to $-0.16$). 

In summary, we find some support for the predicted association between high \fmin\ and low \hi\ column density. However, as we argue above, the GPs are likely neither pure shells nor a distribution of clumps with completely evacuated holes. More accurate \lya\ models may come from considering more complex gas distributions that include a range of column densities.

\subsubsection{The Role of \oiii/\oii}

The \oiii/\oii\ ratio is a possible diagnostic of density-bounding \citep[\eg][]{jaskot13,nakajima14}, and high \oiii/\oii\ may correlate with \lya\ and LyC emission \citep[\eg][]{izotov16b, yang17}. However, high \oiii/\oii\ alone does not guarantee high LyC escape \citep[\eg][]{jaskot14, stasinska15} and the correlation between \oiii/\oii\ and \fesclyc\ appears to show large scatter \citep{izotov18b}. 

\begin{figure*}
\epsscale{0.6}
\plotone{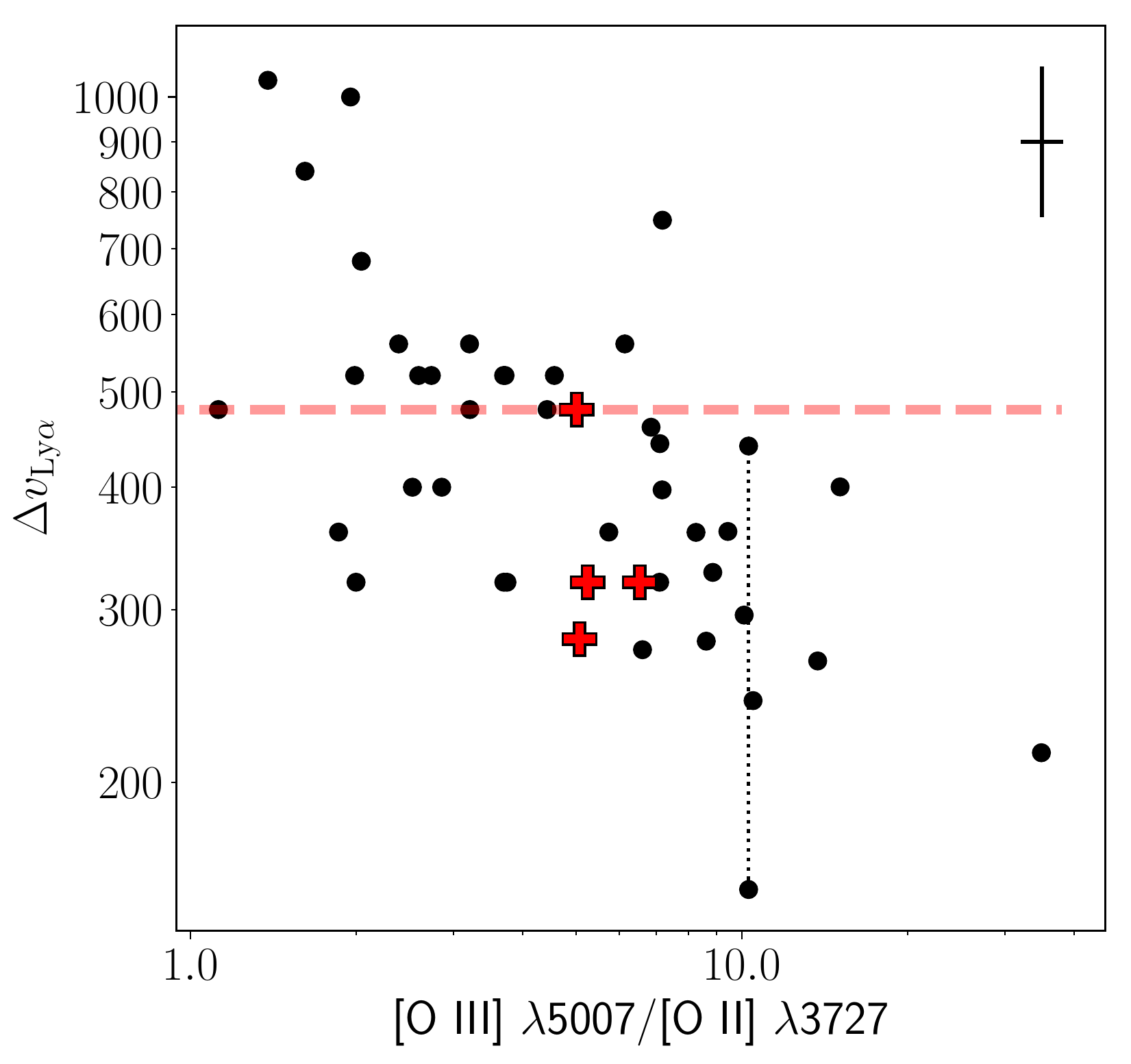}
\caption{ \vpeaks\ vs. \oiii/\oii. Red plus symbols show confirmed LyC emitters from \protect\citet{izotov16a, izotov16b}. Galaxies below the horizontal dashed line have \vpeaks\ lower than or comparable to confirmed LCEs with \fesclyc $>0.05$ and may be more common at higher \oiii/\oii. GPs with single-peaked \lya\ profiles or pure absorption are not plotted. The dotted line connects the two \vpeaks\ measurements for J0808+1728. A representative error bar is shown in the upper right.} 
\label{fig:vpeaks_o3o2}
\end{figure*}

Of the various \lya\ parameters we consider, we find that \oiii/\oii\ is most closely associated with \vpeaks\ (Fig.~\ref{fig:vpeaks_o3o2}). The average \vpeaks\ value decreases with higher \oiii/\oii, although the data are still sparse at the most extreme ionization end. High scatter is also apparent; low \vpeaks\ is more common at high \oiii/\oii, but highly ionized galaxies can also have \vpeaks$>500$\kmps\ (e.g., J1457+2232). 

As argued above (Section~\ref{sec:vpeaks_u_z}), galaxies with high ionization parameters may have more inhomogeneous gas distributions and a higher fraction of low-column-density passageways, some of which may provide direct channels for LyC escape. The lower column density of these paths could then lead to reduced \lya\ scattering and narrower \vpeaks. However, because the GPs' gas is inhomogeneous, galaxy orientation may determine whether or not any optically thin channels align with our line of sight.

\begin{figure*}
\epsscale{0.6}
\plotone{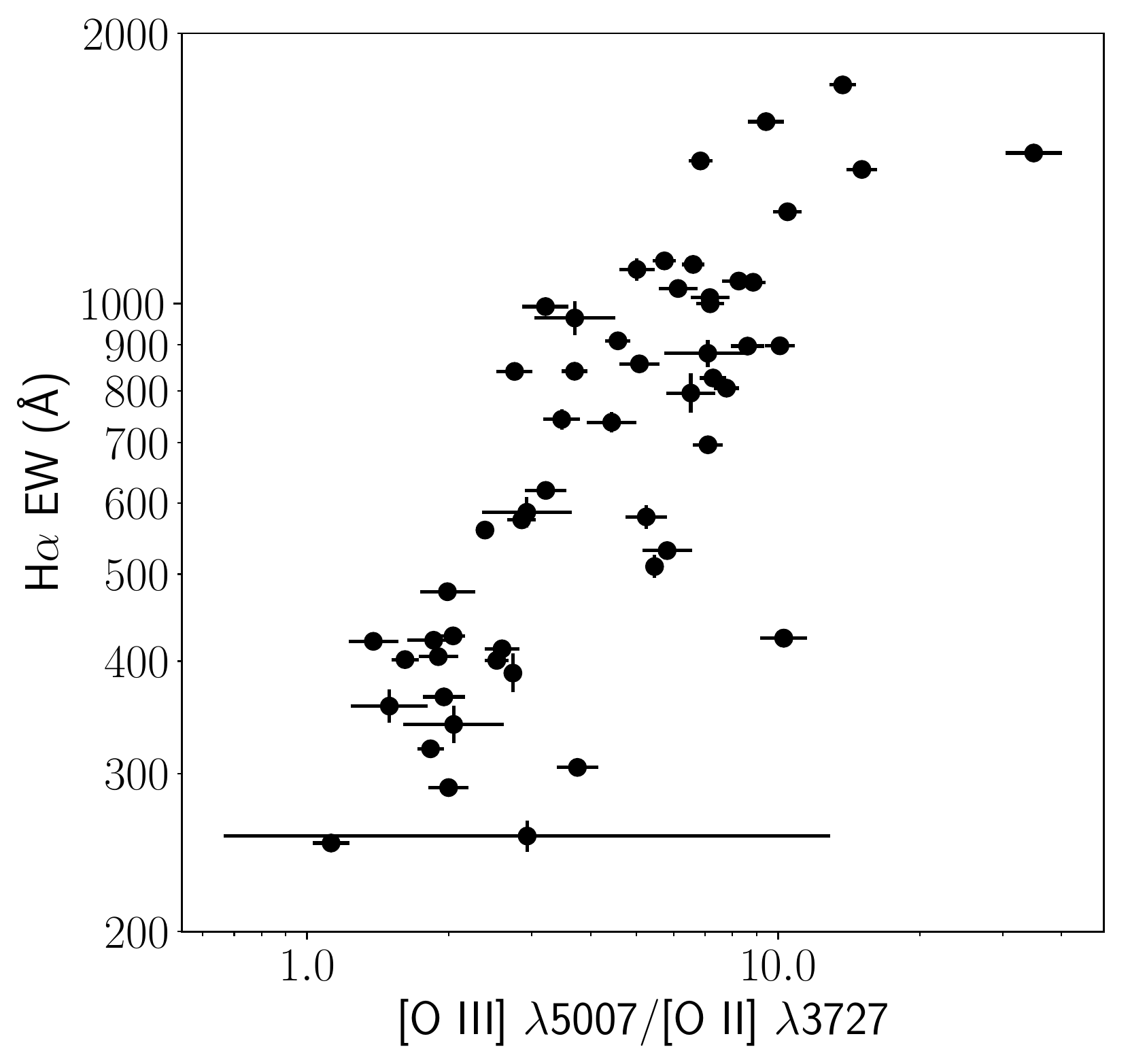}
\caption{ Rest-frame H$\alpha$ EW vs. \oiii/\oii. Galaxies with high \oiii/\oii\ ratios also show high H$\alpha$ EWs and consequently high intrinsic \lya\ EWs.} 
\label{fig:haew_o3o2}
\end{figure*}

Previous GP observations \citep{yang17} and stacked spectra of LAEs \citep{trainor16} indicate that high \lya\ EWs are also more prevalent at high ionization. \citet{yang17} find that \oiii/\oii\ correlates with \lya\ EW and \fesclya\ with \rcor\ $=0.52$ and $0.40$, respectively. We likewise find that the correlation with \lya\ EW is stronger than that with \fesclya; however, the addition of more GPs at the higher ionization end lowers the correlation coefficients to \rcor\ $=0.42$ and $0.31$. In contrast, high \oiii/\oii\ correlates quite strongly with high H$\alpha$ EWs (Fig.~\ref{fig:haew_o3o2}; \rcor$=0.78$).

We suggest that the association between strong \lya\ emission and high \oiii/\oii\ likely arises because of strong \lya\ production in GPs. The tight correlation between  \oiii/\oii\ and H$\alpha$ EW is consistent with a high inferred ionizing photon production rate among GP-like galaxies \citep{schaerer16, nakajima16}. In contrast, \lya\ escape is not tightly correlated with ionization. The high \lya\ EWs of highly ionized GPs may therefore result more from their strong intrinsic production than from unusually high escape fractions. This strong ionizing photon production is consistent with a link between high \oiii/\oii\ and catastrophic cooling (Section~\ref{sec:vpeaks_u_z}; Figure~\ref{fig:superbubble}), which occurs in the most massive young clusters \citep[\eg][]{silich17}.

High LyC production combined with the presence of low-density channels may explain the large number of known GP LCEs. GPs have a high fraction of galaxies with low covering fractions \citep{mckinney19}, yet these low covering fractions do not appear to be caused by enhanced mechanical feedback \citep[\eg][]{jaskot17, chisholm17}; indeed, there is evidence that the most extreme objects have suppressed superwinds \citep{jaskot17}. As we discuss in Section~\ref{sec:vpeaks_u_z}, intrinsically clumpy gas distributions associated with weak mechanical feedback could create low-column-density channels in these GPs. Feedback from a previous burst of star formation could also play a role by carving out the ISM before the current star formation episode \citep[\eg][]{micheva18}.

\section{Low-ionization Spectral Lines}
\label{sec:lis}
\begin{figure*}
\epsscale{1}
\plotone{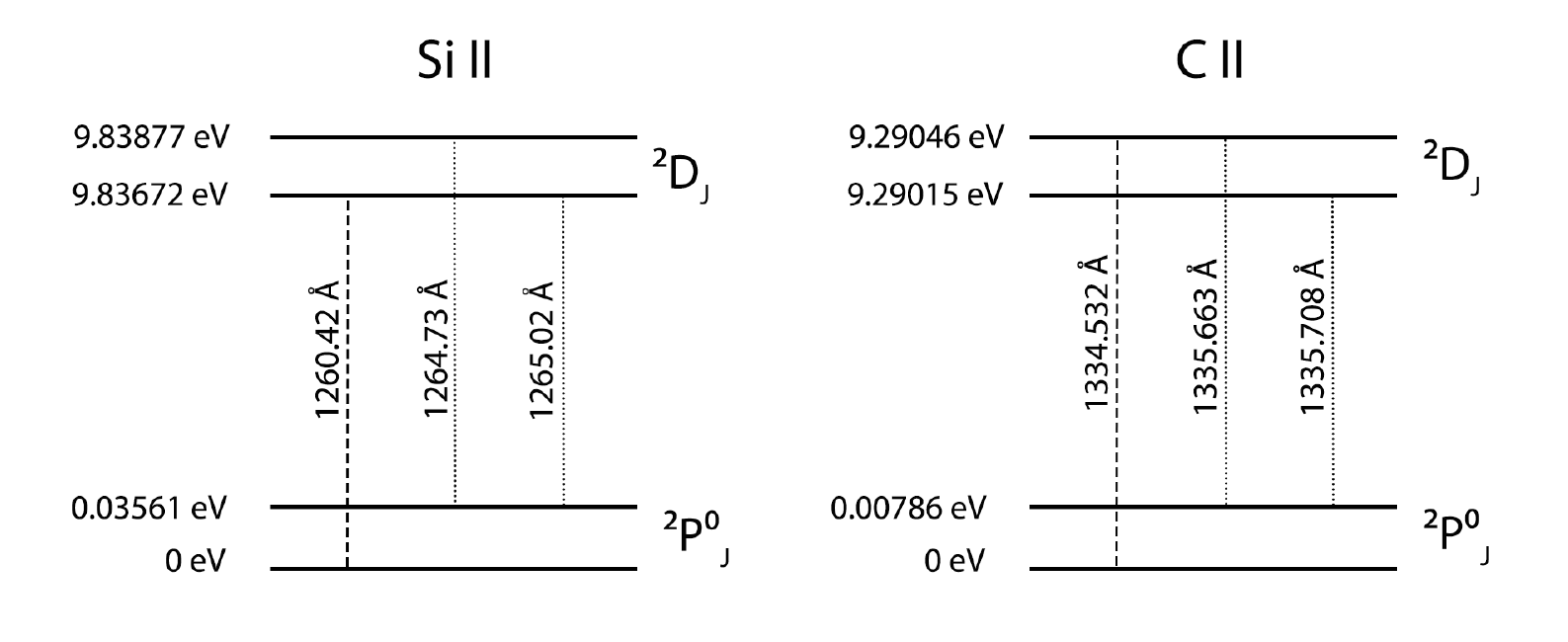}
\caption{Energy level diagrams for the \sitwo~$\lambda$1260 and \cii~$\lambda$1334 resonant transitions (dashed lines) and associated non-resonant \sitwo* and \cii* transitions (dotted lines).} 
\label{fig:energylevels}
\end{figure*}

The GPs' spectra show several low-ionization resonant absorption lines (\sitwo~$\lambda$1190, $\lambda$1193, $\lambda$1260, $\lambda$1304, \oi~$\lambda$1302, and \cii~$\lambda$1334), as well as their corresponding non-resonant transitions to the first fine-structure level above the ground state, denoted with an asterisk. Figure~\ref{fig:energylevels} shows energy level diagrams for the \sitwo~$\lambda$1260 and \cii~$\lambda$1334 transitions. While the absorption lines trace neutral or low-ionization gas along the line of sight, non-resonant emission can be emitted into the line of sight from other directions. The combination of low-ionization absorption and emission lines is therefore sensitive to the geometry of low-ionization gas \citep[\eg][]{prochaska11, jaskot14, scarlata15}. The ratio of non-resonant to resonant emission is set by the relative transition probabilities, but resonant emission from other directions can also fill in the absorption line profiles. 

\subsection{Low-Ionization Lines and \lya}
\begin{figure*}
\epsscale{1.1}
\plotone{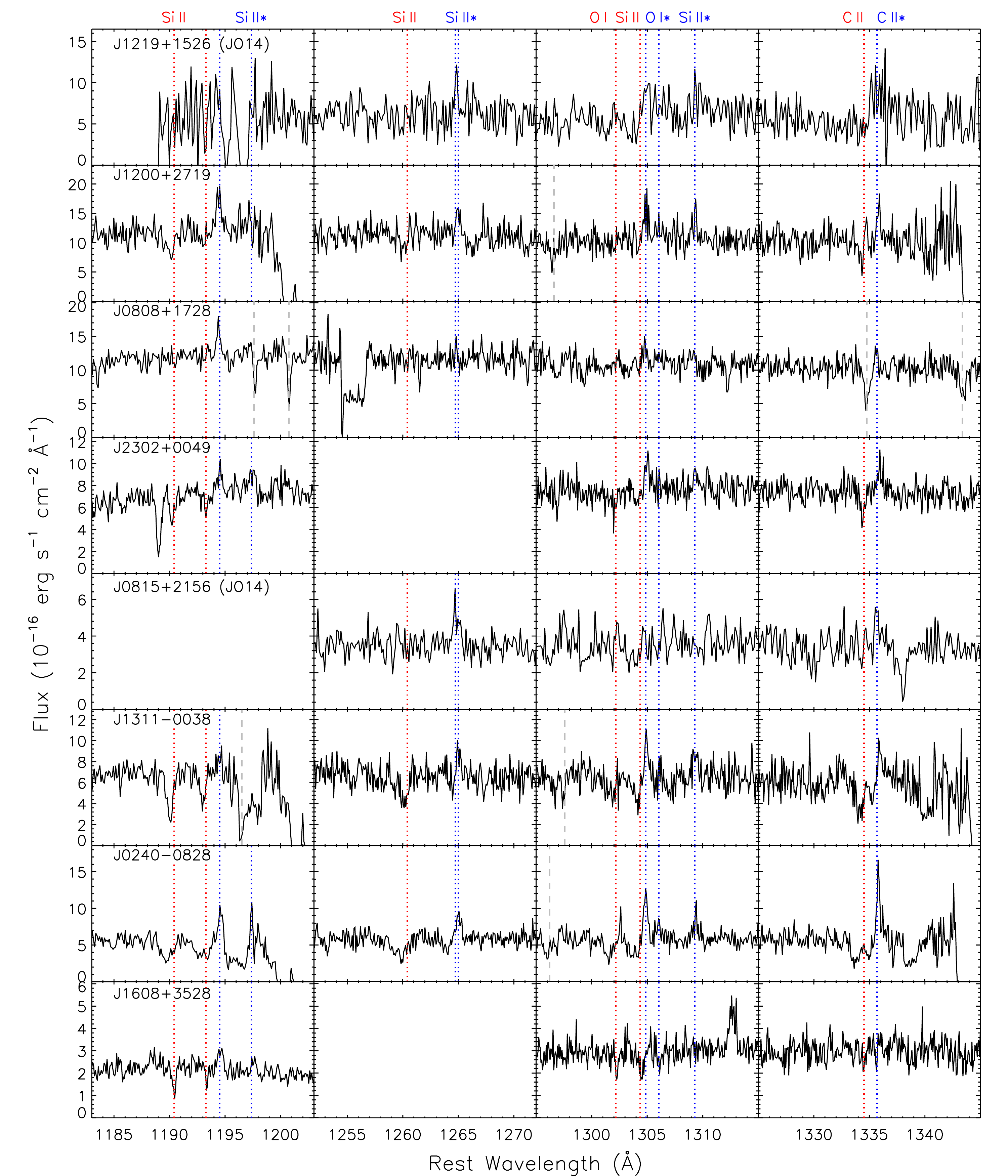}
\caption{LIS absorption and emission lines in high \oiii/\oii\ GPs in order of decreasing \fesclya. Labels are the same as in Figure~\ref{fig:fullspec1}.}
\label{fig:LIS1}
\end{figure*}

\begin{figure*}
\epsscale{1.1}
\plotone{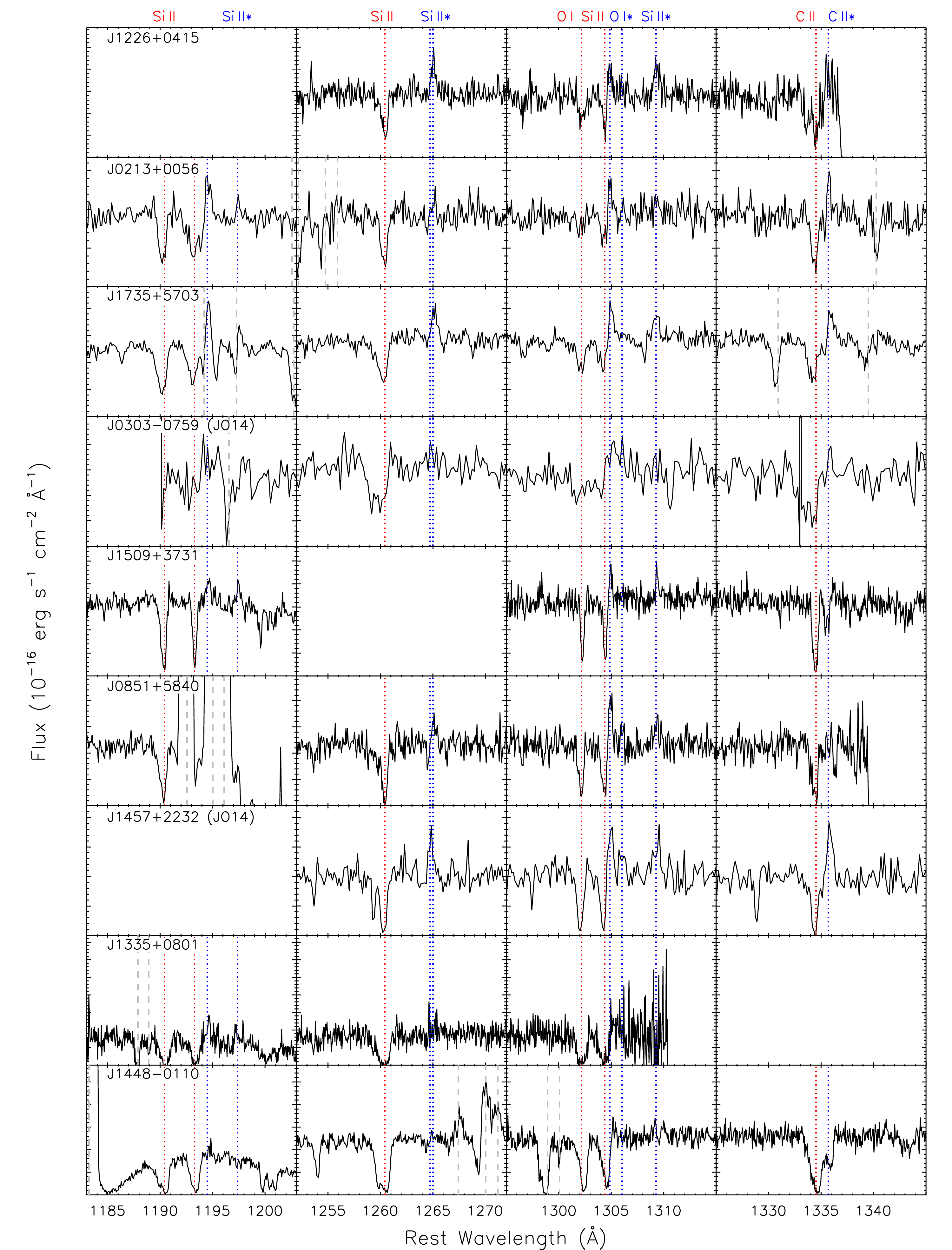}
\caption{A continuation of Figure~\ref{fig:LIS1}.}
\label{fig:LIS2}
\end{figure*}

We detect LIS absorption and emission lines with varying strengths in our sample of high-ionization GPs (Tables~\ref{table:lis_abs} and \ref{table:lis_emis}; Figs.~\ref{fig:LIS1} and \ref{fig:LIS2}), and these lines appear related to the \lya\ spectral profiles. The two GPs without \lya\ emission (J1335+0801 and J1448-0110) show deep LIS absorption with weak or non-existent non-resonant LIS emission. GPs with both strong LIS absorption and strong LIS fluorescent emission (J1457+2232, J0851+5840, and J1509+3731) also show \lya\ absorption troughs, but with double-peaked \lya\ emission within the troughs (Figures~\ref{fig:fullspec1} and \ref{fig:fullspec2}; see Figure 4 of \citealt{mckinney19} for fits to the \lya\ absorption). Finally, a majority of the sample shows weak LIS absorption and weak LIS emission, with some objects (J0808+1728, J0815+2156, and J1608+3528) showing barely any LIS features.
 
\begin{table*}
\begin{center}
\caption{LIS Resonant Line EWs of High Ionization GPs\tablenotemark{a}}
\begin{tabular}{lcccccc}
\hline
\hline
Galaxy & \sitwo~$\lambda$1190 & \sitwo~$\lambda$ 1193 & \sitwo~$\lambda$1260 &  \sitwo~$\lambda$1304 & \oi~$\lambda$1302 & \cii~$\lambda$1334 \\
\hline
J0213+0056 & $-0.44\pm0.10$ & $-0.66\pm0.20$ & $-0.49\pm0.14$ & $-0.21\pm0.12$ & $-0.15\pm0.13$ & $-0.52\pm0.13$ \\
J0240-0828 &  $-0.63\pm0.32$ & $-0.56\pm0.29$ & $-0.64\pm0.36$ & $-0.32\pm0.20$ & $-0.38\pm0.19$ & $-0.50\pm0.29$ \\
J0808+1728 & $>-0.06$\tablenotemark{b} & $>-0.05$\tablenotemark{b} & $>-0.04$\tablenotemark{b} & $>-0.18$\tablenotemark{b} & $>-0.30$\tablenotemark{b} & ---\tablenotemark{c} \\
J0851+5840 & $-0.70\pm0.14$ & ---\tablenotemark{c} & $-0.85\pm0.18$ & $-0.42\pm0.09$ &  $-0.44\pm0.10$ & $-0.58\pm0.14$ \\
J1200+2719 & $-0.37\pm0.19$ & $>-0.14$\tablenotemark{b} & $-0.16\pm0.11$ & $>-0.25$\tablenotemark{b} & $>-0.19$\tablenotemark{b}  & $-0.16\pm0.10$ \\
J1226+0415 & ---\tablenotemark{c} & ---\tablenotemark{c} & $-0.50\pm0.13$ & $-0.22\pm0.09$ & $-0.25\pm0.13$  & $-0.60\pm0.21$ \\
J1311-0038 & $-0.41\pm0.11$ & $-0.20\pm0.10$ & $-0.33\pm0.16$ & $-0.24\pm0.14$ & $-0.26\pm0.21$  & $-0.39\pm0.04$ \\
J1335+0801 & $-0.95\pm0.22$ & $-0.69\pm0.22$ & $-1.21\pm0.26$ & $-0.81\pm0.28$ & $-0.93\pm0.32$ & ---\tablenotemark{c} \\
J1448-0110 &  $-1.12\pm0.17$ & $-0.88\pm0.09$ & $-1.36\pm0.13$ & $-0.68\pm0.10$ & $-0.91\pm0.15$  & $-1.27\pm0.11$ \\
J1509+3731 & $-0.51\pm0.07$ & $-0.34\pm0.05$ & ---\tablenotemark{c} & $-0.33\pm0.07$ & $-0.31\pm0.07$  & $-0.54\pm0.05$ \\
J1608+3528 & $-0.17\pm0.09$ & $>-0.04$\tablenotemark{b} & ---\tablenotemark{c} & $-0.19\pm0.12$  & $-0.11\pm0.08$ & $>-0.06$\tablenotemark{b} \\
J1735+5703 & $-0.67\pm0.17$ & $-0.45\pm0.12$ & $-0.67\pm0.23$ & $-0.31\pm0.14$  & $-0.47\pm0.35 $ & $-0.55\pm0.32$ \\
J2302+0049 & $-0.15\pm0.09$ & $-0.08\pm0.07$ & ---\tablenotemark{c} & $>-0.12$\tablenotemark{b} & $>-0.33$\tablenotemark{b} &$-0.14\pm0.10$ \\
J0303-0759\tablenotemark{d} & ---\tablenotemark{c} & ---\tablenotemark{c} & $-0.80\pm0.42$ & $-0.46\pm0.33$ & $-0.77\pm0.36$  & $-0.85\pm0.47$ \\
J1219+1526\tablenotemark{d} & ---\tablenotemark{c} & ---\tablenotemark{c} & $>-0.36$\tablenotemark{b} & $>-0.54$\tablenotemark{b} & $>-0.58$\tablenotemark{b} & $>-0.57$\tablenotemark{b} \\
J0815+2156\tablenotemark{e} & ---\tablenotemark{c} & ---\tablenotemark{c} & $>-0.40$\tablenotemark{b} & $>-0.35$\tablenotemark{b} & $>-0.56$\tablenotemark{b} & $>-0.70$\tablenotemark{b} \\
J1457+2232\tablenotemark{e} & ---\tablenotemark{c}  & ---\tablenotemark{c} & $-0.74\pm0.16$ & $-0.54\pm0.18$ & $-0.63\pm0.19$  & $-0.85\pm0.26$ \\

\hline
\end{tabular}
\end{center}
\tablenotetext{a}{All EWs are rest-frame values in \AA. Positive values of EW denote net emission.}
\tablenotetext{b}{$1\sigma$ limit.}
\tablenotetext{c}{Line is not covered by the spectrum or is obscured by a Milky Way or geocoronal feature.}
\tablenotetext{d}{From Program GO-12928 (P.I. Henry).}
\tablenotetext{e}{From Program GO-13293 (P.I. Jaskot).}
\label{table:lis_abs}
\end{table*}

\begin{table*}
\begin{center}
\caption{LIS Non-Resonant Line EWs of High Ionization GPs\tablenotemark{a}}
\begin{tabular}{lcccccc}
\hline
\hline
Galaxy & \sitwo* $\lambda$1195 & \sitwo* $\lambda$1197 & \sitwo* $\lambda$1265 & \sitwo* $\lambda$1309 & \oi* $\lambda$1305 & \cii* $\lambda$1335 \\
\hline
J0213+0056 & $0.33\pm0.18$ & $<0.13$\tablenotemark{b} & $0.09\pm0.09$ & $<0.11$\tablenotemark{b}  & $0.17\pm0.13$ & $0.21\pm0.13$\\
J0240-0828 & $0.32\pm0.15$ & $0.15\pm0.13$ & $<0.25$\tablenotemark{b} & $0.17\pm0.14$ & $0.45\pm0.15$ & $0.39\pm0.27$\\
J0808+1728 & $0.13\pm0.12$ & ---\tablenotemark{c} & $<0.05$\tablenotemark{b} & $0.08\pm0.08$ & $0.10\pm0.09$ & $0.07\pm0.06$\\
J0851+5840 & ---\tablenotemark{c} & ---\tablenotemark{c}& $<0.16$\tablenotemark{b}  & $0.15\pm0.11$ & $0.27\pm0.12$  & $<0.06$\tablenotemark{b} \\
J1200+2719 & $0.24\pm0.15$ & $<0.07$\tablenotemark{b} & $0.16\pm0.12$ & $0.12\pm0.11$  &$0.19\pm0.14$ & $0.24\pm0.15$\\
J1226+0415 & ---\tablenotemark{c}  & ---\tablenotemark{c} & $0.26\pm0.14$ & $0.32\pm0.18$ & $0.20\pm0.11$  & $0.30\pm0.15$\\
J1311-0038 & $0.15\pm0.12$ & ---\tablenotemark{c} & $0.13\pm0.09$  & $0.18\pm0.14$ & $0.22\pm0.12$ & $<0.22$\tablenotemark{b}\\
J1335+0801 & $0.60\pm0.23$ & $0.63\pm0.18$ & $0.09\pm0.07$ & $<0.24$\tablenotemark{b} & $<0.09$\tablenotemark{b} & ---\tablenotemark{c}\\
J1448-0110  & $<0.06$\tablenotemark{b} & $<0.02$\tablenotemark{b} & $<0.04$\tablenotemark{b} & $<0.06$\tablenotemark{b}  & $<0.05$\tablenotemark{b} & $-0.30\pm0.10$\\
J1509+3731 & $0.29\pm0.19$ & $0.17\pm0.13$ & ---\tablenotemark{c} & $0.14\pm0.08$  & $0.11\pm0.07$ & $<0.14$\tablenotemark{b}\\
J1608+3528 & $0.30\pm0.18$ & $0.14\pm0.11$ & ---\tablenotemark{c} & $<0.05$\tablenotemark{b} &$<0.05$\tablenotemark{b} & $0.07\pm0.06$\\
J1735+5703 & ---\tablenotemark{c} & ---\tablenotemark{c} & $0.21\pm0.15$  & $0.19\pm0.18$  & $0.21\pm0.16$ & $<0.21$\tablenotemark{b}\\
J2302+0049 & $0.18\pm0.13$ & $0.16\pm0.12$ & ---\tablenotemark{c} & $0.11\pm0.09$  & $0.14\pm0.12$ & $0.16\pm0.11$\\
J0303-0759\tablenotemark{d} & ---\tablenotemark{c} & ---\tablenotemark{c}  & $<0.23$\tablenotemark{b} & $<0.18$\tablenotemark{b}  & $0.45\pm0.31 $ & $<0.21$\tablenotemark{b}\\
J1219+1526\tablenotemark{d}  & ---\tablenotemark{c}  & ---\tablenotemark{c}  & $0.26\pm0.26$ & $<0.29$\tablenotemark{b} & $0.34\pm0.27$ & $0.45\pm0.40$\\
J0815+2156\tablenotemark{e} & ---\tablenotemark{c} & ---\tablenotemark{c}  & $0.33\pm0.22$  & $<0.12$\tablenotemark{b} & $<0.13$\tablenotemark{b} & $0.34\pm0.26$\\
J1457+2232\tablenotemark{e} & ---\tablenotemark{c} & ---\tablenotemark{c} & $0.31\pm0.19$  & $0.43\pm0.28$  & $0.35\pm0.22$  & $0.50\pm0.29$\\

\hline
\end{tabular}
\end{center}
\tablenotetext{a}{EWs are rest-frame values in \AA\ and include any detected absorption and emission in the non-resonant line. Positive values of EW denote net emission.}
\tablenotetext{b}{$1\sigma$ limit.}
\tablenotetext{c}{Line is not covered by the spectrum or is obscured by a Milky Way or geocoronal feature.}
\tablenotetext{d}{From Program GO-12928 (P.I. Henry).}
\tablenotetext{e}{From Program GO-13293 (P.I. Jaskot).}
\label{table:lis_emis}
\end{table*}

The deep LIS absorption to zero or near-zero flux in several objects (J0851+5840, J1335+0801, J1448-0110, and J1509+3731) indicates high column densities of low-ionization and/or neutral gas. The \hi\ column densities derived from the \lya\ absorption troughs in these same galaxies are log(\nhi/cm$^{-2}$)$=20.35-21.49$ \citep{mckinney19}. One interesting exception is J1608+3528, which has \lya\ absorption corresponding to an \hi\ column density of log(\nhi)$=21.39$ \citep{mckinney19}, yet which also has weak low-ionization absorption lines and strong associated \lya\ emission. Although J1608+3528's metallicity is low (12+log(O/H)=7.83), its oxygen abundance should only weaken its low-ionization absorption lines by 32\%\ relative to GPs with 12+log(O/H)=8.0. Even after accounting for this reduction in absorption, J1608+3528's \sitwo~$\lambda$1190 absorption line EW would still be anomalously weak by $\sim0.7-0.8$ \AA\ compared to the $\sim$1 \AA\ EWs of J1335+0801 and J1448-0110, which have similar \nhi. The weakness of J1608+3528's LIS absorption lines is likely due to a lower \hi\ covering fraction compared to J1335+0801 and J1448-0110 \citep{mckinney19}. Consistent with this interpretation, a Voigt profile fit to J1608+3528's \lya\ absorption does not reach zero flux \citep{mckinney19}. Furthermore, since natural broadening, rather than kinematic broadening, dominates the \lya\ profile at high column densities, this low \lya\ absorption depth implies a low total line-of-sight covering fraction across all velocity bins.

\begin{figure*}
\epsscale{1.18}
\plotone{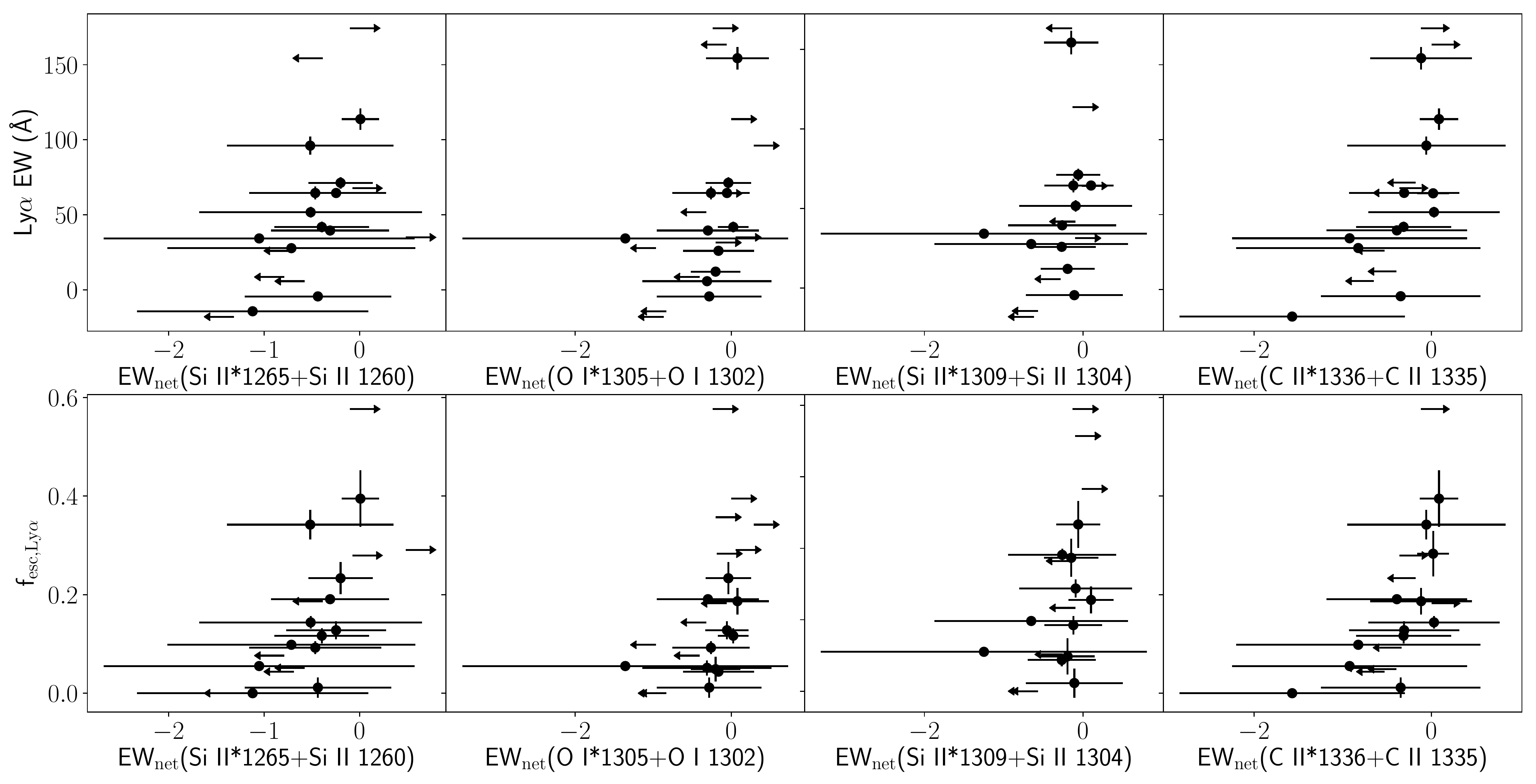}
\caption{ Correlations between \lya\ emission and \ewnet, the sum of resonant absorption and non-resonant emission EWs with the same upper energy level. Top panels show correlations with \lya\ EW, and bottom panels show correlations with \fesclya. We show limits for galaxies where only absorption or emission is measurable.} 
\label{fig:deltaEWs}
\end{figure*}

While the LIS resonant absorption lines show some association with \lya\ absorption, the combination of resonant absorption and non-resonant emission lines appears associated with \lya\ emission. As the strength of non-resonant emission (\sitwo*~$\lambda$1265, $\lambda$1309; \oi*~$\lambda$1305; \cii*~$\lambda$1335) increases relative to their corresponding resonant absorption lines, \lya\ emission strength increases. Studies of stacked $z\sim3$ LBG spectra suggest that \lya\ EW may correlate with the EW of LIS non-resonant emission (e.g., \sitwo*; \citealt{steidel18}). We do not observe this correlation in our sample. Instead, we find that \lya\ EWs and \fesclya\ correlate with \ewnet, the sum of resonant absorption and non-resonant emission EWs with the same upper energy level (Fig.~\ref{fig:deltaEWs}). Negative \ewnet\ indicates that resonant absorption dominates over non-resonant emission. 

The non-resonant emission EWs alone cannot predict whether an individual galaxy is a strong LAE. For example, strong non-resonant emission is sometimes present along with strong resonant absorption (e.g., J1509+3731 or J1457+2232). In these cases, \lya\ emission appears within an absorption trough as discussed above. The LIS absorption implies a high \nhi\ and high covering fraction along the line of sight, while the LIS emission indicates that photons may also scatter into the line of sight from other directions \citep[\eg][]{jaskot14}; \lya\ emission in such objects may therefore be scattered into the line of sight from these other neutral regions. For instance, the \lya\ emission component in J1457+2232 shows high \vpeaks\ indicative of scattering. 

In the most optically thin objects, we do not necessarily expect to see strong non-resonant emission. The absence of neutral gas along the line of sight will result in weak emission. If neutral gas off the line of sight is also optically thin or not captured by the aperture, the non-resonant emission lines will not be detected. Indeed, some strong LAEs in our sample show relatively weak or even non-existent emission from these transitions. J1608+3528 and J0808+1728 both have weak absorption lines as well as comparatively weak non-resonant emission lines. Intriguingly, both objects have narrow \lya\ components (\vpeaks$=156-214$\kmps), consistent with weak \lya\ scattering and low \hi\ column densities. While J0808+1728's extremely low metallicity could also explain its non-detected low-ionization lines, J1608+3528's metallicity is comparable to other galaxies in the sample. In these two galaxies, the \lya\ emission we observe may reach our line of sight via low-density pathways with minimal scattering. Gas off the line of sight may also be optically thin, contributing little LIS emission. 

\subsection{Non-Resonant Line Kinematics}
If an outflow is present, the velocities of the non-resonant emission lines can provide further clues to their spatial origin \citep[\cf][]{scarlata15,carr18}. For instance, net foreground emission will appear with a blue-shifted velocity, while gas oriented perpendicular to the line of sight will appear at zero velocity. In our sample, the detected non-resonant emission lines are generally centered near $v=0$, which is consistent with a spherically symmetric large-scale gas distribution, \citep{prochaska11, scarlata15}, perpendicularly-oriented biconical outflow \citep{carr18}, or static gas. High-velocity, line-of-sight outflow motions do appear to be absent for many objects in the sample (e.g., J0213+0056, J1226+0415, J1608+3528; \citealt{jaskot17}), as expected in the case of suppressed superwinds or if the gas is only outflowing in the perpendicular direction. We see no obvious sign of bipolar outflows aligned along or near to the line of sight; in this case, most emitting gas would be either approaching or receding, which should result in a dip in flux at $v=0$ in the \sitwo* emission \citep{carr18}.

\subsection{Non-Resonant Absorption}
Interestingly, we also detect non-resonant lines in {\it absorption} in several GPs. \cii* absorption is apparent in J0240-0828, J1311-0038, J1448-0110, J1509+3731, and J1735+5703, and \sitwo* $\lambda$1265 absorption is present in J0240-0828 and J0851+5840. Some of these GPs show both absorption and emission in \cii* $\lambda$1335 and \sitwo* 1265. Such absorption may be present in more objects, possibly the entire sample, but could be masked by the non-resonant emission component. In general, we detect non-resonant absorption in the objects with deep LIS absorption and weak non-resonant emission (J1448-0110, J1509+3731, J0851+5840) or faster-than-average outflows, such that the absorption appears distinct from the non-resonant emission (J0240-0828, J1311-0038, J1735+5703; \citealt{jaskot17}). The non-resonant lines in these latter galaxies exhibit a P-Cygni profile (Figs.~\ref{fig:LIS1} and \ref{fig:LIS2}), which indicates that the absorbing material consists of outflowing gas surrounding the UV source. 

With energies of only 0.0079 eV and 0.0356 eV above the ground state (Fig.~\ref{fig:energylevels}), collisions can easily excite the first fine structure levels of \cii\ and \sitwo, even at low temperatures ($\sim100$ K). This collisional excitation is responsible for the \ciif~$\lambda$157.7 $\mu$m fine-structure line commonly observed in galaxy spectra \citep[\eg][]{crawford85,wright91,boselli02}, and the presence of \cii* and \sitwo* absorption illustrates that a non-negligible fraction of atoms are in these excited fine-structure states. In addition to our sample, \cii* absorption is observed in I Zw 18 \citep{lebouteiller13}, damped \lya\ systems \citep[\eg][]{wolfe03}, and the Milky Way (e.g., the absorption features at $\sim$1275\AA\ and $\sim1284$\AA\ in the spectra of J1735+5703 and J0213+0056 in Figure~\ref{fig:fullspec2}). 

Absorption in \cii* and \sitwo* provides another means of populating the upper levels for the \cii* and \sitwo* transitions. In particular, since not all electrons are in the ground state, \cii~$\lambda$1334 absorption may be weaker. However, the \cii\ line could still appear stronger and broader in some spectra, if it is blended with the \cii* absorption line \citep{riverathorsen17a}. In addition, \cii* emission will appear stronger, provided it is offset in velocity from the \cii* absorption. Absorption from the ground state can only populate the 9.29046 eV upper level, whereas \cii* absorption can also populate the 9.29015 eV upper state, which has a higher radiative transition probability. The effects of \cii* and \sitwo* absorption should be included in future models of these UV transitions. 

\subsection{Stacked Spectrum}
\begin{figure*}
\centering
\gridline{\fig{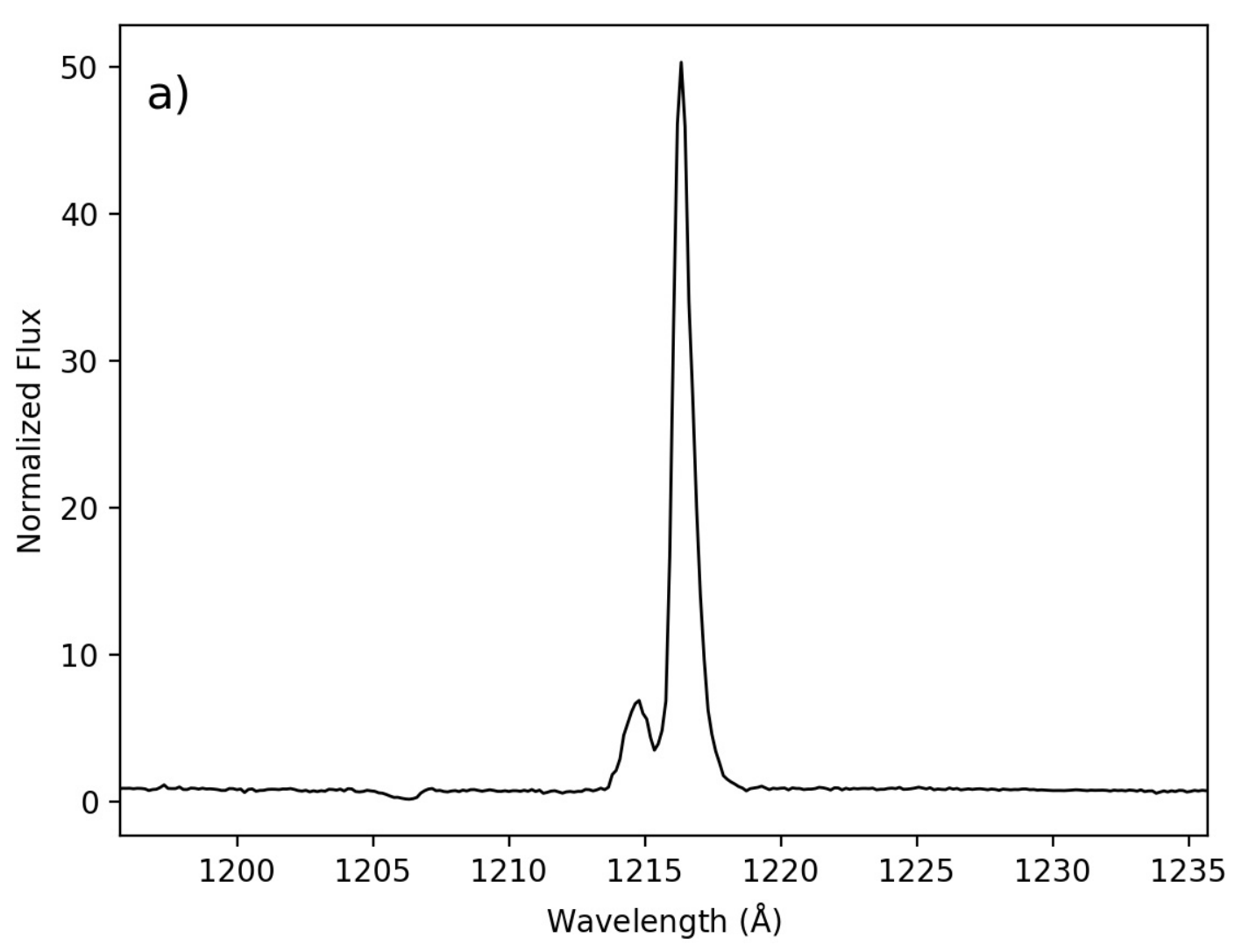}{0.4\textwidth}{}}
\vspace{-1cm}
\gridline{\fig{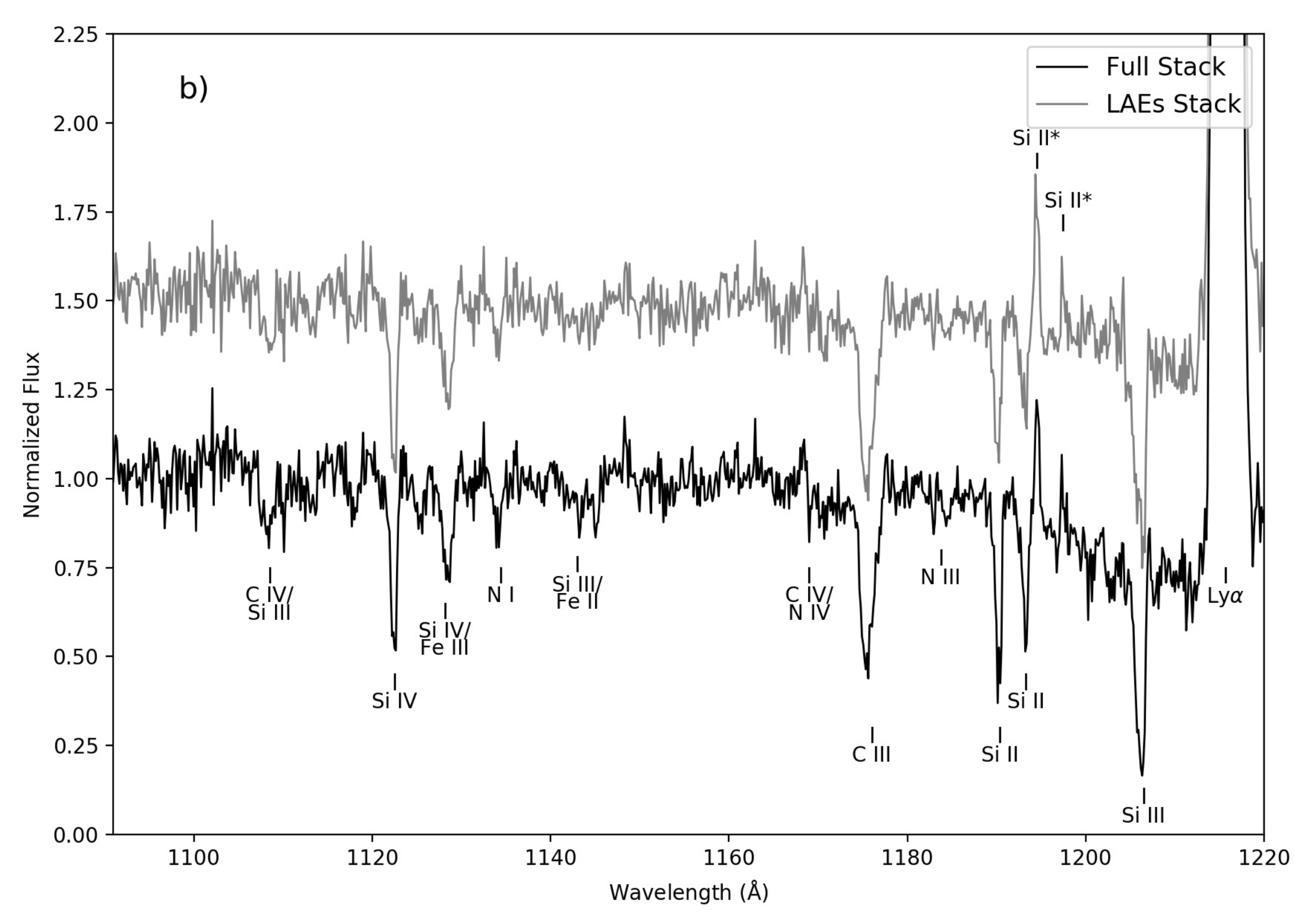}{0.65\textwidth}{}}
\vspace{-1 cm}
\gridline{\fig{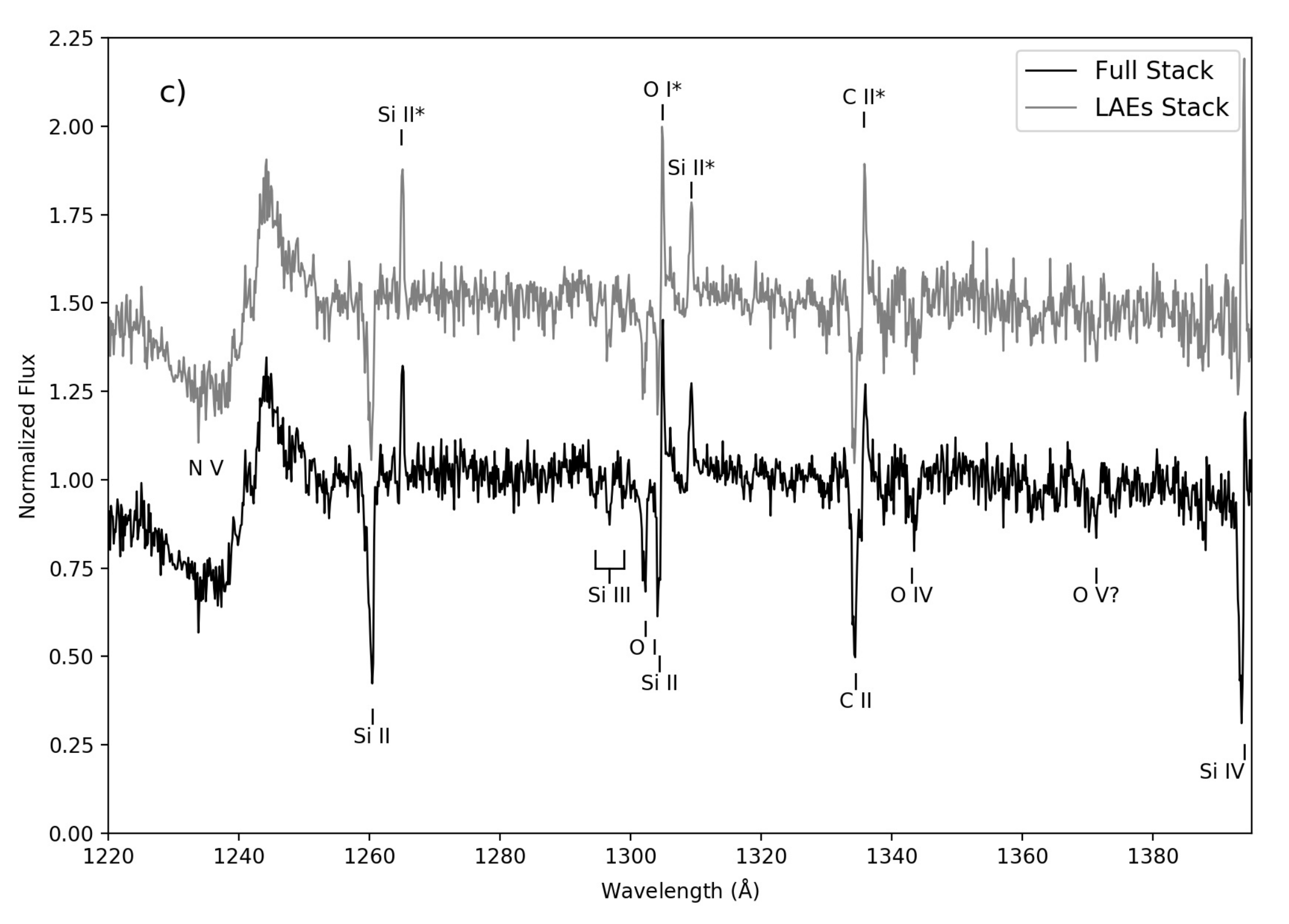}{0.65\textwidth}{}}

\vspace{-0.5 cm}
\caption{(a). Stacked UV spectrum of the \lya\ region. (b). Stacked UV spectrum ($\lambda<1220$) of the 13 high-ionization GPs (black) and of the 10 with \lya\ EW $>25$\AA\ (gray), offset by 0.5 for clarity. (c) Stacked UV spectrum ($\lambda>1220$) of the 13 high-ionization GPs (black) and of the 10 with \lya\ EW $>25$\AA\ (gray), offset by 0.5 for clarity. } 
\label{fig:stack}
\end{figure*}

For comparison with high-redshift studies, we create a composite UV spectrum of the 13 galaxies in our sample (Figure~\ref{fig:stack}). Before stacking, we mask out all geocoronal or Milky Way lines, shift each spectrum to rest-frame wavelengths, and bin the spectra to the lowest resolution of the sample (34 \kmps). We then normalize each spectrum by a linear fit to the UV continuum, excluding all regions with spectral lines. The resulting composite \lya\ profile is double-peaked, with positive flux at the systemic velocity, and weak, broad, blue-shifted absorption, consistent with most of the observed \lya\ profiles. 

The stacked spectrum shows high-ionization stellar features indicative of a young stellar population, such as \siiv~$\lambda$1128, \ciii~$\lambda$1176, and \nv~$\lambda$1240. Whereas the interstellar lines in the sample show considerable variation in depth, the depth of the \siiv~$\lambda$1128 and \ciii~$\lambda$1176 lines varies by less than 17\%\ from galaxy to galaxy, which suggests the sample shares a similar stellar population. In particular, the detectable bump from \civ~$\lambda$1169 near \ciii~$\lambda$1176 implies ages of 2-6 Myr \citep{eldridge17}, consistent with the strong \nv~P-Cygni profile. The spectrum also tentatively shows \ov~$\lambda$1371, a line produced by Very Massive Stars with $M>100$\Msol\ and another sign of young age \citep[\eg][]{crowther16,smith16}.

As discussed above, individual GP spectra show significant variation in the strength of LIS line absorption and emission, which will be hidden in a composite spectrum. The stacked spectrum shows both LIS resonant absorption and clear, but weaker emission from non-resonant transitions to the first fine-structure level. Although strong, the LIS absorption does not reach zero intensity. Such composite spectra therefore hide the existence of both strong absorbers and the most optically thin galaxies with weak absorption and emission. Since many high-redshift studies stack only LAEs, we also consider a composite spectrum of the 10 GPs listed in Table~\ref{table:lya} with \lya\ EW$>25$\AA\ (Figure~\ref{fig:stack}). In this composite, the LIS absorption lines are slightly weaker and LIS emission lines are slightly stronger; otherwise, the two composite spectra are comparable. 

Notably, the pattern of LIS absorption and emission line strengths in the composite spectra do not match any of the individual spectra. Several of the GPs contributing to the the LIS emission have weak LIS absorption and vice versa. Because these lines are so sensitive to gas geometry, model predictions of LIS line strengths should be treated with caution when applied to composite spectra.

\section{Discussion}
\subsection{Tracers of Neutral Gas Geometry}
Together, the \lya\ and LIS metal lines paint a coherent picture of neutral gas geometry, revealing the presence of both high column density clouds and lower column density paths. High column density clouds along the line of sight produce deep \lya\ absorption and saturated LIS absorption. Emission scattered from \lya\ haloes or transmitted through low column density gas may fill in the \lya\ absorption troughs; if the \lya\ photons have traveled through low \nhi\ gas, the resulting emission profile should be narrow, even though it may exist within an absorption trough. Similarly, if the line of sight intersects a high enough fraction of optically thin channels, the LIS lines will be saturated but will not reach zero flux \citep[\eg][]{heckman11}. This clumpy gas geometry appears common in low-redshift dwarf starbursts \citep{riverathorsen15} and LCEs \citep{gazagnes18}, with the LyC escaping along low column density channels within a ``picket fence" distribution of denser gas. Finally, the fluorescent non-resonant LIS emission lines can provide insight into the neutral gas distribution off the line of sight \citep[\eg][]{jaskot14}. Dense clouds that do not block the UV-emitting source in the line of sight will show up as strong LIS emission, rather than absorption. Hence, global LyC emitters, with optically thin pathways in many different directions should show weak fluorescent LIS emission as well as weak LIS absorption. 

Larger effective apertures and fainter continua may hinder the use of these diagnostics at higher redshifts. The \lya\ absorption prevalent in our sample may be masked in similar observations at higher redshifts. If the spectroscopic aperture subtends a wider physical area, extended \lya\ haloes may contribute additional \lya\ emission that makes absorption components more difficult to detect. In addition, high signal-to-noise in the continuum may be required to detect weak absorption components. At higher redshifts, faint galaxy continua often necessitate the use of composite spectra to analyze LIS lines \citep[\eg][]{shapley03,trainor16}. Such studies will show both LIS absorption and non-resonant emission and will be unable to trace the geometry in individual objects. The higher spatial resolution and magnification of lensed observations will be ideal for joint analyses of \lya\ and LIS lines at high redshift.

\subsection{Implications for LyC Searches}
Our analysis shows that many \lya\ properties correlate strongly with one another, so that any one of them could potentially work as a LyC diagnostic. However, each \lya\ characteristic traces subtly different galaxy properties. Because \lya\ EW traces both \fesclya\ and intrinsic \lya\ production, it may be a more appropriate proxy for the {\it net} escaping LyC flux, the product of ionizing photon production rate and \fesclyc, rather than \fesclyc\ alone. Indeed, \citep{steidel18} find a close correlation between the ratio of escaping LyC to UV flux and \lya\ EW in $z\sim3$ galaxies. Characterizing the reionization of the universe requires understanding this total escaping LyC flux, the product of both production and escape \citep[\eg][]{inoue06,robertson15,finkelstein15}, for which \lya\ EW may serve as an indirect indicator.

In contrast, the \fesclya\ and \vpeaks\ parameters may give insight into \fesclyc. Given its close anti-correlation with neutral gas covering fraction \citep{mckinney19}, \fesclya\ may be the property most closely related to \fesclyc. However, \fesclya\ is also increased by scattered \lya\ and is likely sensitive to the fraction of the \lya\ halo captured in the spectroscopic aperture. Larger samples of LCEs, at a range of redshifts, will be necessary to quantify this possible bias. 

Lastly, \vpeaks\ shows the closest association with ionization state and may be sensitive to a density-bounded geometry. This parameter apparently indicates the relative transparency of any low column density paths present. Its association with low-metallicity conditions and high ionization parameter suggests a clumped and patchy ISM model that is closely linked to the frequent success of high \oiii/\oii\ as a predictor of LyC escape (Figure~\ref{fig:superbubble}). Observationally, \vpeaks\ shows a close association with LyC escape \citep{izotov18b}, but larger samples will be necessary to reveal whether it is a more reliable tracer than \fesclya. 

As discussed above, LIS metal lines can trace optical depth on and off the line of sight. However, as with \lya, LIS metal absorption lines can be affected by infilling from an extended emitting halo \citep{prochaska11,scarlata15}. In addition, the presence of non-resonant absorption lines in our sample shows that not all electrons are in the ground state in LIS atoms. This effect may lead to underestimates of the column densities derived from these transitions. Weak fluorescent emission can indicate a low optical depth of gas off the line of sight. However, this emission can be masked by strong non-resonant absorption if the line of sight is sufficiently optically thick. Both LIS absorption and emission will also be affected by metallicity. Despite these caveats, our sample shows that the spectral patterns of the LIS lines are consistent with the conclusions drawn from the \lya\ emission.

For LyC searches at the highest redshifts ($z>6$), optical emission line diagnostics are ideal. At these redshifts, an increasingly neutral IGM may affect the availability of \lya\ diagnostics \citep[\eg][]{stark11,schenker14}, while UV absorption lines will require deep observations for unlensed targets. Direct LyC observations of the GPs \citep{izotov16b,izotov18b} and the \lya\ analyses in this work demonstrate that \oiii/\oii\ is a promising, but insufficient diagnostic of LyC escape. Samples selected by high \oiii/\oii\ do show a high fraction of LCE candidates. Our work suggests that the low metallicities and young ages of these galaxies may be associated with clumpy gas distributions that promote \lya\ and LyC escape. In our sample, 10 out of 13 galaxies have \vpeaks\ as narrow as confirmed LCEs with \fesclyc$>0.05$, while deep \lya\ absorption in two GPs suggest no line-of-sight LyC escape. The \oifb/H$\beta$ diagnostic is even more closely associated with \vpeaks\ but will be more difficult to observe. 

Much of the variation in optical depth among high \oiii/\oii\ galaxies could be due to chance orientation. Since both high and low column density gas exists in these galaxies, chance alignment could significantly affect whether we observe LyC or \lya\ escape. Observations of ionized gas \citep[\eg][]{zastrow13,keenan17} and simulations of LyC escape \citep[\eg][]{paardekooper15,cen15} both suggest that LyC escapes over narrow opening angles and depends strongly on galaxy orientation. Larger samples will be necessary to quantify the precise relationship between \fesclyc\ and \oiii/\oii. By measuring \fesclyc\ in 66 additional low-redshift galaxies, the upcoming {\it HST} Low-Redshift Lyman Continuum Survey (P.I. Jaskot) will shed light on the connection and observational scatter between LyC, \lya\ parameters, and nebular line ratios.

\section{Summary}
We have obtained {\it HST} COS spectra of 13 of the most highly ionized GPs, with \oiii/\oii\ $=6.6-34.9$. The sample appears compact in both optical and UV morphologies, and a single compact region generally dominates the UV emission with FWHM as low as 0.2-0.3 kpc. The UV emission is more compact than higher-redshift GP samples \citep[\eg][]{yang17a}, likely because the observations resolve star-forming knots that would be blended at higher redshift. As with previous GP samples, most of the targeted GPs are LAEs. We combine the sample with previous COS observations of GPs from the compilation of \citet{yang17} and investigate correlations between \lya\ spectral features, galaxy properties, and low-ionization absorption and emission lines. We summarize our main conclusions below.
\begin{enumerate}

\item Most ($\sim80$\%) GPs with high \oiii/\oii\ ($>6$) have double-peaked \lya\ emission profiles. However, 9 out of 13 galaxies in our new sample also show a \lya\ absorption component, which highlights the inhomogeneous nature of the gas in these galaxies. 

\item \lya\ spectral features correlate strongly with each other. Galaxies with high \lya\ EWs also tend to have high \fesclya, low \vpeaks, and high residual fluxes at the \lya\ profile minimum. Each of these \lya\ spectral properties also correlates with weaker LIS absorption line EWs, which demonstrates that they are likely related to low optical depth.

\item Even GPs with moderate \fesclya\ ($\lesssim20$\%) can exhibit extremely high \lya\ EWs ($>100$\AA). These GPs also show extremely high H$\alpha$ EWs, implying high intrinsic \lya\ production. GPs selected for high \oiii/\oii\ ratios tend to have high H$\alpha$ EWs and presumably, strong \lya\ production, but high \oiii/\oii\ does not correlate with high \fesclya. {\bf Hence, high intrinsic \lya\ EWs may explain the prevalence of detected \lya\ emission observed in GP samples.} Likewise, high LyC production in the GPs could result in a high fraction of detectable LCEs, even with moderate \fesclyc. Because of its sensitivity to both \lya\ production and escape, the \lya\ EW is a potential indirect tracer of the net escaping LyC flux, although it will also depend on the size of the spectroscopic aperture.

\item The \fesclya\ and \vpeaks\ parameters may be sensitive to different aspects of \lya\ radiative transfer. Of the \lya\ spectral properties we consider, \vpeaks\ shows the strongest trend with \oifb~$\lambda$6300/H$\beta$, \oiii/\oii, and other measures of ionization. In contrast,  \fesclya\ does not show a relationship with \oiii/\oii\ but does correlate strongly with low-ionization gas covering fraction. This is consistent with a scenario where {\bf low \nhi\ controls \vpeaks, whereas \fesclya\ is sensitive to the gas porosity along the line of sight.} Thus, we suggest that the GPs' derived covering fractions trace the fraction of low-column-density channels along the line of sight, while ionization plays a role in determining the transparency of those channels. Lower column-density channels could then result in narrower \vpeaks. Observationally, covering fraction can account for some of the spread in \fesclya\ at a given \vpeaks, and both covering fraction and column density may therefore influence \fesclya.

\item Lower metallicities correlate with both higher \oiii/\oii\ and narrower \vpeaks. Since neither dust nor luminosity can explain the trend, {\bf this correlation implies a fundamental link between low metallicity, high ionization parameter, and low \lya\ optical depth.} We suggest that weak mechanical feedback at low metallicity fails to disperse dense gas and thus promotes catastrophic cooling, inhibiting superwinds. This process promotes strong clumping, thereby generating the patchy gas geometry indicated by the low-ionization absorption lines. The dense clouds close to the ionizing SSC explain the extreme ionization parameters, compactness and high nebular luminosities, while the inter-cloud medium has low \nhi\ and is more optically thin to \lya. This clumpy geometry contrasts with the more isotropically uniform gas distributions and larger swept up gas masses generated by superwinds at higher metallicity.

\item The residual flux at \lya\ profile minimum anti-correlates strongly with \vpeaks, which may suggest a connection with low \nhi. 

\item The patterns of LIS resonant absorption and fluorescent emission in the GPs' spectra are connected with the shape of their \lya\ profiles. GPs with dominant \lya\ absorption troughs show deep LIS absorption. If \lya\ emission appears within the absorption troughs, the GPs also show strong LIS fluorescent emission. We find that the strength of fluorescent emission relative to absorption correlates with \lya\ EWs and \fesclya. However, GPs with low \vpeaks\ show weak or non-detected LIS emission and absorption, as expected if they contain a higher global fraction of low \nhi\ sight-lines.

\item Several GPs show LIS non-resonant transitions such as \cii*~$\lambda$1335 and \sitwo*~$\lambda$1265 in absorption, rather than in emission. This absorption indicates a significant population of electrons in the first fine-structure level above the ground state, which should be taken into account in modeling these transitions. P-Cygni profiles in several objects demonstrate that the absorbing gas originates in an outflow surrounding the central UV source. 

\item A stacked spectrum of the 13 high-ionization GPs in our sample shows features consistent with a young ($<6$ Myr) stellar population, including a possible detection of the \ov~$\lambda$1371 absorption line from Very Massive Stars. The stacked spectrum also shows both LIS resonant absorption and LIS fluorescent emission. Since these LIS line features depend strongly on gas geometry, their relative strengths in the stacked spectrum do not correspond to any of the individual GPs. Stacked spectra will hide the true variation in these features in individual galaxies and should be treated with caution. 

\end{enumerate}

Galaxy samples selected by high \oiii/\oii\ contain a high fraction of LAEs and a large number of candidate LCEs with low \vpeaks. However, by itself, high \oiii/\oii\ is insufficient as a diagnostic of LyC escape. In our sample of extreme GPs, we find galaxies with strong \lya\ absorption. Even among the \lya\ emitters, the frequent presence of underlying \lya\ absorption troughs shows that not all lines of sight in the GPs are optically thin. Because of this inhomogeneity, galaxy orientation will determine whether a given GP leaks LyC radiation along our line of sight. However, high \oiii/\oii\ may signal a compact, clumpy gas geometry with more optically thin inter-clump pathways. This geometry enhances the likelihood of global \lya\ and LyC escape, as well as toward our line of sight. The GPs' LIS and \lya\ features suggest that LyC escape occurs neither through completely evacuated holes within a picket fence nor through a homogeneous density-bounded medium. To understand the relationship between \lya\ and LyC escape, radiative transfer models may need to consider more complex gas geometries that include a range of \hi\ column densities.

\vspace{12pt}
 \acknowledgments{We are grateful to the anonymous referee for insightful discussions and comments that improved the clarity of the paper. We thank Ryan Trainor for assistance with the spectral stacking procedure and Tim Heckman for useful discussions. AEJ acknowledges support by NASA through Hubble Fellowship grant HST-HF2-51392. AEJ, JM, and MSO acknowledge support from NASA through grant HST-GO-14080 from STScI. STScI is operated by AURA under NASA contract NAS-5-26555. TD acknowledges support from the Massachusetts Space Grant Consortium. Funding for SDSS-III has been provided by the Alfred P. Sloan Foundation, the Participating Institutions, the National Science Foundation, and the U.S. Department of Energy Office of Science. The SDSS-III web site is http://www.sdss3.org/.

SDSS-III is managed by the Astrophysical Research Consortium for the Participating Institutions of the SDSS-III Collaboration including the University of Arizona, the Brazilian Participation Group, Brookhaven National Laboratory, Carnegie Mellon University, University of Florida, the French Participation Group, the German Participation Group, Harvard University, the Instituto de Astrofisica de Canarias, the Michigan State/Notre Dame/JINA Participation Group, Johns Hopkins University, Lawrence Berkeley National Laboratory, Max Planck Institute for Astrophysics, Max Planck Institute for Extraterrestrial Physics, New Mexico State University, New York University, Ohio State University, Pennsylvania State University, University of Portsmouth, Princeton University, the Spanish Participation Group, University of Tokyo, University of Utah, Vanderbilt University, University of Virginia, University of Washington, and Yale University.}

 \appendix
As described in Section~\ref{sec:lyacor}, we calculate the Spearman rank correlations between \lya\ parameters and a variety of UV and optical properties, which we describe below. Correlations with a coefficient of $\left | \rho \right | >0.5$ are listed in Tables~\ref{table:cor_ew}-\ref{table:cor_fmin}. If a parameter does not appear in the table, the correlation coefficient was $<0.5$.

From the COS UV spectra, we consider \lya\ and metal line properties. In addition to \lya\ EW, \vpeaks, \fesclya, and \fmin/\fcont, we include other \lya\ properties, namely the velocities of the blue (\vblue) and red (\vred) peaks, the velocity of the profile minimum, and the ratio of the total blue peak flux to the total red peak flux. We also examine correlations with low- and high-ionization UV metal line properties, specifically, EW, \vchar, \vmax, \ewnet, and \fcov. We derive the \fcov\ measurements from an average of the residual intensity of all detected \sitwo\ lines, as described in \citet{mckinney19}. Finally, we also calculate correlations with UV morphological parameters, such as the half-light radius ($r_{50}$) and the FWHM of the FUV continuum and \lya\ from the two-dimensional spectra. 

We also consider properties derived from the SDSS optical spectra. All SDSS line measurements and dust corrections are described in Section~\ref{sec:sdss}. We include galaxy redshift from the SDSS catalog \citep{ahn14} and the $A_V$ and emission line velocity widths derived from our line measurements (Section~\ref{sec:sdss}). In addition, we examine correlations from emission line EWs, line ratios, and derived nebular properties.  For correlations with optical EWs, we use the H$\alpha$, \oiii~$\lambda$5007, and \heii~$\lambda$4686 emission lines, representative of various ionization states. The line ratios we consider are \oii~$\lambda$3727/H$\beta$, \neiii~$\lambda$3869/\oii$\lambda$3727, \heii~$\lambda$4686/H$\beta$, \oiii~$\lambda$5007/\oii~$\lambda$3727, \oiii\ $\lambda$5007/H$\beta$, \oifb~$\lambda$6300/\oii~$\lambda$3727, \oifb\ $\lambda$6300/H$\beta$, \oifb\ $\lambda$6300/\oiii\ $\lambda$5007, and \sii~$\lambda$6716/H$\alpha$, all of which are corrected for Milky Way and internal extinction. We also include the \hei~$\lambda$7065/$\lambda$6678 and \hei~$\lambda$3889/$\lambda$6678 ratios proposed by \citet{izotov17b} as a LyC escape diagnostic. In our correlation calculations, we include three nebular parameters: electron temperature ($T_e$), electron density, and oxygen abundance. We derive the electron temperature and density from the \oiii~$\lambda$4363/\oiii~$\lambda$5007 and \sii~$\lambda$6716/\sii~$\lambda$6730 line ratios, respectively, using the {\tt PyNeb} package \citep{luridiana15}. We calculate oxygen abundances with {\tt PyNeb} using the direct method and ionization correction factors from \citet{perezmontero17}. For GPs without an \oiii~$\lambda$4363 detection, we adopt an electron temperature of 13451 K, the mean of the sample. Similarly, for purposes of deriving the ionization correction factor, we adopt the average \heii~$\lambda$4686/H$\beta$ ratio of the sample if \heii\ is not detected. Lastly, to examine correlations with luminosity, we include the 6000\AA\ continuum luminosity ($L_{6000}$) and H$\alpha$ luminosity ($L$(H$\alpha$)) in our calculations. 

\begin{table*}
\begin{center}
\caption{\lya\ EW Correlations\tablenotemark{a}}
\begin{tabular}{lcc}
\hline
\hline
Parameter & \rcor& Number of GPs  \\
\hline
\fesclya & 0.88 & 54\\
\fmin/\fcont & 0.83; 0.84 & 52\\
\ewnet(\cii*1335+\cii1334) & 0.80 & 12\\
\vpeaks & -0.69; -0.72 & 43 \\
\oi~$\lambda$1302 EW & 0.68 & 17 \\
\fcov & -0.68 & 23 \\
\ewnet(\oi*1305+\oi1302) & 0.67 & 11\\
\sitwo$\lambda$1304 EW & 0.65 & 16 \\
\ewnet(\sitwo*1265+\sitwo1260) & 0.62 & 12\\
\sitwo$\lambda$1190 EW & 0.62 & 14 \\
\cii~$\lambda$1334 EW & 0.61 & 17 \\
\sitwo~$\lambda$1260 EW & 0.60 & 22\\
\sitwo~$\lambda$1193 EW & 0.55 & 10\\
\heii~$\lambda$4686 EW & 0.52 & 37\\
H$\alpha$ EW & 0.51 & 54\\
\siiv~$\lambda$1403 EW & 0.50 & 11\\

\hline
\end{tabular}
\end{center}
\tablenotetext{a}{The table shows all correlations with $\left | \rho \right | \geq0.5$, where measurements exist for at least 10 GPs. When two \rcor\ values are listed, the correlations have been calculated separately for each of J0808+1728's \vpeaks\ and \fmin/\fcont\ values.}
\label{table:cor_ew}
\end{table*}

\begin{table*}
\begin{center}
\caption{\vpeaks\ Correlations\tablenotemark{a}}
\begin{tabular}{lcc}
\hline
\hline
Parameter & \rcor& Number of GPs  \\
\hline
\ewnet(\oi*1305+\oi1302) & -0.85 & 11\\
\vblue & -0.83; -0.82 & 43\\
\fmin/\fcont & -0.82; -0.82 & 43\\
\ewnet(\cii*1335+\cii1334) & -0.77 & 10\\
\cii~$\lambda$1334 EW & -0.73 & 15\\
\fesclya & -0.71; -0.67 & 42 \\
\oifb~$\lambda$6300/H$\beta$ & 0.69 & 40 \\
\oi~$\lambda$1302 EW & -0.69 & 15\\
\lya\ EW & -0.69; -0.72 & 43\\
\oii~$\lambda$3727/H$\beta$ & 0.68; 0.63 & 43\\
\oifb~$\lambda$6300/\oiii~$\lambda$5007 & 0.67 & 40 \\
12+log(O/H) & 0.67; 0.61 & 43\\
\sitwo~$\lambda$1304 \vmax & -0.66 & 11\\
\sitwo* $\lambda$1309 EW & 0.64; 0.42 & 14 \\
\ewnet(\sitwo*1265+\sitwo1260) & -0.64 & 10\\
$T_e$ & -0.61; -0.56 & 43\\
\oiii~$\lambda$5007/\oii~$\lambda$3727 & -0.61; -0.57 & 43\\
\neiii~$\lambda$3869/\oii~$\lambda$3727 & -0.61; -0.58 & 43\\
\sitwo~$\lambda$1304 EW & -0.60 & 14\\
\sii~$\lambda$6716/H$\alpha$ & 0.60; 0.55 & 42 \\
$L$(H$\alpha$) & 0.59; 0.55 & 42 \\
$L_{6000}$ & 0.57; 0.53 & 43\\
\vred & 0.57; 0.52 & 43\\
\heii~$\lambda$4686 EW & -0.56; -0.61 & 32\\
\fcov & 0.56 & 20\\
$r_{50}$ & 0.54; 0.18 & 11\\
\sitwo~$\lambda$1260 \vmax & -0.53 & 13\\
\cii~$\lambda$1334 \vmax & -0.51 & 14 \\
\sitwo~$\lambda$1190 EW & -0.50 & 12\\
EW (Median of low-ionization lines) & -0.50 & 22 \\
\cii* $\lambda$1335 EW & 0.50 & 15\\

\hline
\end{tabular}
\end{center}
\tablenotetext{a}{See Table~\ref{table:cor_ew}.}
\label{table:cor_vpeaks}
\end{table*}

\begin{table*}
\begin{center}
\caption{\fesclya\ Correlations\tablenotemark{a}}
\begin{tabular}{lcc}
\hline
\hline
Parameter & \rcor& Number of GPs  \\
\hline
\lya\ EW & 0.88 & 54\\
\fcov & -0.82 & 22\\
\fmin/\fcont & 0.82; 0.80 & 50\\
\ewnet(\cii*1335+\cii1334) & 0.80 & 12 \\
\vpeaks & -0.71; -0.67 & 42 \\
\sitwo~$\lambda$1190 EW & 0.69 & 14 \\
\sitwo~$\lambda$1260 EW & 0.65 & 21 \\
\ewnet(\sitwo*1265+\sitwo1260) & 0.63 & 12 \\
\vchar\ (Median of low-ionization lines) & -0.54 & 25\\
\sitwo~$\lambda$1193 EW & 0.54 & 10\\
\cii~$\lambda$1334 EW & 0.53 & 17 \\
\oi~$\lambda$1302 EW & 0.53 & 17\\
\sithree~$\lambda$1206 \vchar & -0.53 & 19\\
\vchar\ (Median of high-ionization lines) & -0.52 & 23 \\

\hline
\end{tabular}
\end{center}
\tablenotetext{a}{See Table~\ref{table:cor_ew}.}
\label{table:cor_fesc}
\end{table*}

\begin{table*}
\begin{center}
\caption{\fmin/\fcont\ Correlations\tablenotemark{a}}
\begin{tabular}{lcc}
\hline
\hline
Parameter & \rcor& Number of GPs  \\
\hline
\lya\ EW & 0.83; 0.84 & 52 \\
\fesclya & 0.82; 0.80 & 50\\
\vpeaks & -0.82; -0.82 & 43 \\
\oi~$\lambda$1302 EW & 0.75 & 15\\
\siiv~$\lambda$1403 EW & 0.71 & 10 \\
\sitwo~$\lambda$1304 EW & 0.68 & 14 \\
\ewnet(\oi*1305+\oi1302) & 0.65 & 11\\
\vblue & 0.63; 0.62 & 52 \\
\ewnet(\sitwo*1309+\sitwo1304) & 0.62 & 11\\
\ewnet(\cii*1335+\cii1334) & 0.61 & 11\\
\sitwo~$\lambda$1190 EW & 0.59 & 12\\
\neiii~$\lambda$3869/\oii~$\lambda$3727 & 0.58; 0.57 & 52\\
\cii~$\lambda$1334 EW & 0.58 & 16 \\
\fcov & -0.56 & 21 \\
\sii~$\lambda$6716/H$\alpha$ & -0.55; -0.52 & 49 \\
\oi~$\lambda$1302 \vchar & -0.54 & 15 \\
$L_{6000}$ & -0.52; -0.50 & 52\\
\oifb~$\lambda$6300/\oiii~$\lambda$5007 & -0.51 & 47 \\
\oiii~$\lambda$5007/\oii~$\lambda$3727 & 0.51; 0.49 & 52\\
\oii~$\lambda$3727/H$\beta$ & -0.50; -0.48 & 52\\
\hline
\end{tabular}
\end{center}
\tablenotetext{a}{See Table~\ref{table:cor_ew}.}
\label{table:cor_fmin}
\end{table*}


\begin{thebibliography}{116}
\expandafter\ifx\csname natexlab\endcsname\relax\def\natexlab#1{#1}\fi

\bibitem[{{Ahn} {et~al.}(2014){Ahn}, {Alexandroff}, {Allende Prieto}, {Anders},
  {Anderson}, {Anderton}, {Andrews}, {Aubourg}, {Bailey}, {Bastien}, \&
  et~al.}]{ahn14}
{Ahn}, C.~P., {et~al.} 2014, \apjs, 211, 17

\bibitem[{{Amor{\'{\i}}n} {et~al.}(2010){Amor{\'{\i}}n}, {P{\'e}rez-Montero},
  \& {V{\'{\i}}lchez}}]{amorin10}
{Amor{\'{\i}}n}, R.~O., {P{\'e}rez-Montero}, E., \& {V{\'{\i}}lchez}, J.~M.
  2010, \apjl, 715, L128

\bibitem[{{Atek} {et~al.}(2009){Atek}, {Schaerer}, \& {Kunth}}]{atek09}
{Atek}, H., {Schaerer}, D., \& {Kunth}, D. 2009, \aap, 502, 791

\bibitem[{{Baldwin} {et~al.}(1981){Baldwin}, {Phillips}, \&
  {Terlevich}}]{baldwin81}
{Baldwin}, J.~A., {Phillips}, M.~M., \& {Terlevich}, R. 1981, \pasp, 93, 5

\bibitem[{{Behrens} {et~al.}(2014){Behrens}, {Dijkstra}, \&
  {Niemeyer}}]{behrens14}
{Behrens}, C., {Dijkstra}, M., \& {Niemeyer}, J.~C. 2014, \aap, 563, A77

\bibitem[{{Bergvall} {et~al.}(2006){Bergvall}, {Zackrisson}, {Andersson},
  {Arnberg}, {Masegosa}, \& {{\"O}stlin}}]{bergvall06}
{Bergvall}, N., {Zackrisson}, E., {Andersson}, B.~G., {Arnberg}, D.,
  {Masegosa}, J., \& {{\"O}stlin}, G. 2006, \aap, 448, 513

\bibitem[{{Borthakur} {et~al.}(2014){Borthakur}, {Heckman}, {Leitherer}, \&
  {Overzier}}]{borthakur14}
{Borthakur}, S., {Heckman}, T.~M., {Leitherer}, C., \& {Overzier}, R.~A. 2014,
  Science, 346, 216

\bibitem[{{Boselli} {et~al.}(2002){Boselli}, {Gavazzi}, {Lequeux}, \&
  {Pierini}}]{boselli02}
{Boselli}, A., {Gavazzi}, G., {Lequeux}, J., \& {Pierini}, D. 2002, \aap, 385,
  454

\bibitem[{{Bromm} {et~al.}(1999){Bromm}, {Coppi}, \& {Larson}}]{bromm99}
{Bromm}, V., {Coppi}, P.~S., \& {Larson}, R.~B. 1999, \apjl, 527, L5

\bibitem[{{Cardamone} {et~al.}(2009){Cardamone}, {Schawinski}, {Sarzi},
  {Bamford}, {Bennert}, {Urry}, {Lintott}, {Keel}, {Parejko}, {Nichol},
  {Thomas}, {Andreescu}, {Murray}, {Raddick}, {Slosar}, {Szalay}, \&
  {Vandenberg}}]{cardamone09}
{Cardamone}, C., {et~al.} 2009, \mnras, 399, 1191

\bibitem[{{Cardelli} {et~al.}(1989){Cardelli}, {Clayton}, \&
  {Mathis}}]{cardelli89}
{Cardelli}, J.~A., {Clayton}, G.~C., \& {Mathis}, J.~S. 1989, \apj, 345, 245

\bibitem[{{Carr} {et~al.}(2018){Carr}, {Scarlata}, {Panagia}, \&
  {Henry}}]{carr18}
{Carr}, C., {Scarlata}, C., {Panagia}, N., \& {Henry}, A. 2018, \apj, 860, 143

\bibitem[{{Cen} \& {Kimm}(2015)}]{cen15}
{Cen}, R., \& {Kimm}, T. 2015, \apjl, 801, L25

\bibitem[{{Chisholm} {et~al.}(2018){Chisholm}, {Gazagnes}, {Schaerer},
  {Verhamme}, {Rigby}, {Bayliss}, {Sharon}, {Gladders}, \&
  {Dahle}}]{chisholm18}
{Chisholm}, J., {et~al.} 2018, \aap, 616, A30

\bibitem[{{Chisholm} {et~al.}(2017){Chisholm}, {Orlitov{\'a}}, {Schaerer},
  {Verhamme}, {Worseck}, {Izotov}, {Thuan}, \& {Guseva}}]{chisholm17}
{Chisholm}, J., {Orlitov{\'a}}, I., {Schaerer}, D., {Verhamme}, A., {Worseck},
  G., {Izotov}, Y.~I., {Thuan}, T.~X., \& {Guseva}, N.~G. 2017, \aap, 605, A67

\bibitem[{{Crawford} {et~al.}(1985){Crawford}, {Genzel}, {Townes}, \&
  {Watson}}]{crawford85}
{Crawford}, M.~K., {Genzel}, R., {Townes}, C.~H., \& {Watson}, D.~M. 1985,
  \apj, 291, 755

\bibitem[{{Crowther} {et~al.}(2016){Crowther}, {Caballero-Nieves}, {Bostroem},
  {Ma{\'{\i}}z Apell{\'a}niz}, {Schneider}, {Walborn}, {Angus}, {Brott},
  {Bonanos}, {de Koter}, {de Mink}, {Evans}, {Gr{\"a}fener}, {Herrero},
  {Howarth}, {Langer}, {Lennon}, {Puls}, {Sana}, \& {Vink}}]{crowther16}
{Crowther}, P.~A., {et~al.} 2016, \mnras, 458, 624

\bibitem[{{Dijkstra} {et~al.}(2016){Dijkstra}, {Gronke}, \&
  {Venkatesan}}]{dijkstra16}
{Dijkstra}, M., {Gronke}, M., \& {Venkatesan}, A. 2016, \apj, 828, 71

\bibitem[{{Dopita} \& {Sutherland}(2003)}]{dopita03}
{Dopita}, M.~A., \& {Sutherland}, R.~S. 2003, {Astrophysics of the diffuse
  universe}

\bibitem[{{Eldridge} {et~al.}(2017){Eldridge}, {Stanway}, {Xiao}, {McClelland},
  {Taylor}, {Ng}, {Greis}, \& {Bray}}]{eldridge17}
{Eldridge}, J.~J., {Stanway}, E.~R., {Xiao}, L., {McClelland}, L.~A.~S.,
  {Taylor}, G., {Ng}, M., {Greis}, S.~M.~L., \& {Bray}, J.~C. 2017, \pasa, 34,
  e058

\bibitem[{{Fernandez} \& {Shull}(2011)}]{fernandez11}
{Fernandez}, E.~R., \& {Shull}, J.~M. 2011, \apj, 731, 20

\bibitem[{{Finkelstein} {et~al.}(2015){Finkelstein}, {Ryan}, {Papovich},
  {Dickinson}, {Song}, {Somerville}, {Ferguson}, {Salmon}, {Giavalisco},
  {Koekemoer}, {Ashby}, {Behroozi}, {Castellano}, {Dunlop}, {Faber}, {Fazio},
  {Fontana}, {Grogin}, {Hathi}, {Jaacks}, {Kocevski}, {Livermore}, {McLure},
  {Merlin}, {Mobasher}, {Newman}, {Rafelski}, {Tilvi}, \&
  {Willner}}]{finkelstein15}
{Finkelstein}, S.~L., {et~al.} 2015, \apj, 810, 71

\bibitem[{{Fitzpatrick}(1999)}]{fitzpatrick99}
{Fitzpatrick}, E.~L. 1999, \pasp, 111, 63

\bibitem[{{Fletcher} {et~al.}(2018){Fletcher}, {Robertson}, {Nakajima},
  {Ellis}, {Stark}, \& {Inoue}}]{fletcher18}
{Fletcher}, T.~J., {Robertson}, B.~E., {Nakajima}, K., {Ellis}, R.~S., {Stark},
  D.~P., \& {Inoue}, A. 2018, ArXiv e-prints

\bibitem[{{Gazagnes} {et~al.}(2018){Gazagnes}, {Chisholm}, {Schaerer},
  {Verhamme}, {Rigby}, \& {Bayliss}}]{gazagnes18}
{Gazagnes}, S., {Chisholm}, J., {Schaerer}, D., {Verhamme}, A., {Rigby}, J.~R.,
  \& {Bayliss}, M. 2018, \aap, 616, A29

\bibitem[{{Giavalisco} {et~al.}(1996){Giavalisco}, {Koratkar}, \&
  {Calzetti}}]{giavalisco96}
{Giavalisco}, M., {Koratkar}, A., \& {Calzetti}, D. 1996, \apj, 466, 831

\bibitem[{{Gronke} {et~al.}(2016){Gronke}, {Dijkstra}, {McCourt}, \&
  {Oh}}]{gronke16b}
{Gronke}, M., {Dijkstra}, M., {McCourt}, M., \& {Oh}, S.~P. 2016, \apjl, 833,
  L26

\bibitem[{{Gronke} {et~al.}(2017){Gronke}, {Dijkstra}, {McCourt}, \& {Peng
  Oh}}]{gronke17}
{Gronke}, M., {Dijkstra}, M., {McCourt}, M., \& {Peng Oh}, S. 2017, \aap, 607,
  A71

\bibitem[{{Hanish} {et~al.}(2010){Hanish}, {Oey}, {Rigby}, {de Mello}, \&
  {Lee}}]{hanish10}
{Hanish}, D.~J., {Oey}, M.~S., {Rigby}, J.~R., {de Mello}, D.~F., \& {Lee},
  J.~C. 2010, \apj, 725, 2029

\bibitem[{{Hansen} \& {Oh}(2006)}]{hansen06}
{Hansen}, M., \& {Oh}, S.~P. 2006, \mnras, 367, 979

\bibitem[{{Hashimoto} {et~al.}(2015){Hashimoto}, {Verhamme}, {Ouchi},
  {Shimasaku}, {Schaerer}, {Nakajima}, {Shibuya}, {Rauch}, {Ono}, \&
  {Goto}}]{hashimoto15}
{Hashimoto}, T., {et~al.} 2015, \apj, 812, 157

\bibitem[{{Hayes} {et~al.}(2013){Hayes}, {{\"O}stlin}, {Schaerer}, {Verhamme},
  {Mas-Hesse}, {Adamo}, {Atek}, {Cannon}, {Duval}, {Guaita}, {Herenz}, {Kunth},
  {Laursen}, {Melinder}, {Orlitov{\'a}}, {Ot{\'{\i}}-Floranes}, \&
  {Sandberg}}]{hayes13}
{Hayes}, M., {et~al.} 2013, \apjl, 765, L27

\bibitem[{{Heckman} {et~al.}(2011){Heckman}, {Borthakur}, {Overzier},
  {Kauffmann}, {Basu-Zych}, {Leitherer}, {Sembach}, {Martin}, {Rich},
  {Schiminovich}, \& {Seibert}}]{heckman11}
{Heckman}, T.~M., {et~al.} 2011, \apj, 730, 5

\bibitem[{{Heckman} {et~al.}(2001){Heckman}, {Sembach}, {Meurer}, {Leitherer},
  {Calzetti}, \& {Martin}}]{heckman01}
{Heckman}, T.~M., {Sembach}, K.~R., {Meurer}, G.~R., {Leitherer}, C.,
  {Calzetti}, D., \& {Martin}, C.~L. 2001, \apj, 558, 56

\bibitem[{{Heger} {et~al.}(2003){Heger}, {Fryer}, {Woosley}, {Langer}, \&
  {Hartmann}}]{heger03}
{Heger}, A., {Fryer}, C.~L., {Woosley}, S.~E., {Langer}, N., \& {Hartmann},
  D.~H. 2003, \apj, 591, 288

\bibitem[{{Henry} {et~al.}(2015){Henry}, {Scarlata}, {Martin}, \&
  {Erb}}]{henry15}
{Henry}, A., {Scarlata}, C., {Martin}, C.~L., \& {Erb}, D. 2015, \apj, 809, 19

\bibitem[{{Iglesias-P{\'a}ramo} \&
  {Mu{\~n}oz-Tu{\~n}{\'o}n}(2002)}]{iglesias02}
{Iglesias-P{\'a}ramo}, J., \& {Mu{\~n}oz-Tu{\~n}{\'o}n}, C. 2002, \mnras, 336,
  33

\bibitem[{{Inoue} {et~al.}(2006){Inoue}, {Iwata}, \& {Deharveng}}]{inoue06}
{Inoue}, A.~K., {Iwata}, I., \& {Deharveng}, J.-M. 2006, \mnras, 371, L1

\bibitem[{{Izotov} {et~al.}(2017{\natexlab{a}}){Izotov}, {Guseva}, {Fricke},
  {Henkel}, \& {Schaerer}}]{izotov17a}
{Izotov}, Y.~I., {Guseva}, N.~G., {Fricke}, K.~J., {Henkel}, C., \& {Schaerer},
  D. 2017{\natexlab{a}}, \mnras, 467, 4118

\bibitem[{{Izotov} {et~al.}(2011){Izotov}, {Guseva}, \& {Thuan}}]{izotov11}
{Izotov}, Y.~I., {Guseva}, N.~G., \& {Thuan}, T.~X. 2011, \apj, 728, 161

\bibitem[{{Izotov} {et~al.}(2016{\natexlab{a}}){Izotov}, {Orlitov{\'a}},
  {Schaerer}, {Thuan}, {Verhamme}, {Guseva}, \& {Worseck}}]{izotov16a}
{Izotov}, Y.~I., {Orlitov{\'a}}, I., {Schaerer}, D., {Thuan}, T.~X.,
  {Verhamme}, A., {Guseva}, N.~G., \& {Worseck}, G. 2016{\natexlab{a}}, \nat,
  529, 178

\bibitem[{{Izotov} {et~al.}(2016{\natexlab{b}}){Izotov}, {Schaerer}, {Thuan},
  {Worseck}, {Guseva}, {Orlitov{\'a}}, \& {Verhamme}}]{izotov16b}
{Izotov}, Y.~I., {Schaerer}, D., {Thuan}, T.~X., {Worseck}, G., {Guseva},
  N.~G., {Orlitov{\'a}}, I., \& {Verhamme}, A. 2016{\natexlab{b}}, \mnras, 461,
  3683

\bibitem[{{Izotov} {et~al.}(2018{\natexlab{a}}){Izotov}, {Schaerer}, {Worseck},
  {Guseva}, {Thuan}, {Verhamme}, {Orlitov{\'a}}, \& {Fricke}}]{izotov18a}
{Izotov}, Y.~I., {Schaerer}, D., {Worseck}, G., {Guseva}, N.~G., {Thuan},
  T.~X., {Verhamme}, A., {Orlitov{\'a}}, I., \& {Fricke}, K.~J.
  2018{\natexlab{a}}, \mnras, 474, 4514

\bibitem[{{Izotov} {et~al.}(2017{\natexlab{b}}){Izotov}, {Thuan}, \&
  {Guseva}}]{izotov17b}
{Izotov}, Y.~I., {Thuan}, T.~X., \& {Guseva}, N.~G. 2017{\natexlab{b}}, \mnras,
  471, 548

\bibitem[{{Izotov} {et~al.}(2018{\natexlab{b}}){Izotov}, {Worseck}, {Schaerer},
  {Guseva}, {Thuan}, {Fricke}, \& {Orlitov{\'a}}}]{izotov18b}
{Izotov}, Y.~I., {Worseck}, G., {Schaerer}, D., {Guseva}, N.~G., {Thuan},
  T.~X., {Fricke}, A., V., \& {Orlitov{\'a}}, I. 2018{\natexlab{b}}, \mnras,
  478, 4851

\bibitem[{{Jaskot} \& {Oey}(2013)}]{jaskot13}
{Jaskot}, A.~E., \& {Oey}, M.~S. 2013, \apj, 766, 91

\bibitem[{{Jaskot} \& {Oey}(2014)}]{jaskot14}
---. 2014, \apjl, 791, L19

\bibitem[{{Jaskot} {et~al.}(2017){Jaskot}, {Oey}, {Scarlata}, \&
  {Dowd}}]{jaskot17}
{Jaskot}, A.~E., {Oey}, M.~S., {Scarlata}, C., \& {Dowd}, T. 2017, \apjl, 851,
  L9

\bibitem[{{Jones} {et~al.}(2013){Jones}, {Ellis}, {Schenker}, \&
  {Stark}}]{jones13}
{Jones}, T.~A., {Ellis}, R.~S., {Schenker}, M.~A., \& {Stark}, D.~P. 2013,
  \apj, 779, 52

\bibitem[{{Kakiichi} \& {Gronke}(2019)}]{kakiichi19}
{Kakiichi}, K., \& {Gronke}, M. 2019, arXiv e-prints, arXiv:1905.02480

\bibitem[{{Keenan} {et~al.}(2017){Keenan}, {Oey}, {Jaskot}, \&
  {James}}]{keenan17}
{Keenan}, R.~P., {Oey}, M.~S., {Jaskot}, A.~E., \& {James}, B.~L. 2017, \apj,
  848, 12

\bibitem[{{Kennicutt} \& {Evans}(2012)}]{kennicutt12}
{Kennicutt}, R.~C., \& {Evans}, N.~J. 2012, \araa, 50, 531

\bibitem[{{Khostovan} {et~al.}(2016){Khostovan}, {Sobral}, {Mobasher}, {Smail},
  {Darvish}, {Nayyeri}, {Hemmati}, \& {Stott}}]{khostovan16}
{Khostovan}, A.~A., {Sobral}, D., {Mobasher}, B., {Smail}, I., {Darvish}, B.,
  {Nayyeri}, H., {Hemmati}, S., \& {Stott}, J.~P. 2016, \mnras, 463, 2363

\bibitem[{{Krumholz} \& {Thompson}(2012)}]{krumholz_thompson12}
{Krumholz}, M.~R., \& {Thompson}, T.~A. 2012, \apj, 760, 155

\bibitem[{{Kunth} {et~al.}(1998){Kunth}, {Mas-Hesse}, {Terlevich}, {Terlevich},
  {Lequeux}, \& {Fall}}]{kunth98}
{Kunth}, D., {Mas-Hesse}, J.~M., {Terlevich}, E., {Terlevich}, R., {Lequeux},
  J., \& {Fall}, S.~M. 1998, \aap, 334, 11

\bibitem[{{Larson}(2005)}]{larson05}
{Larson}, R.~B. 2005, \mnras, 359, 211

\bibitem[{{Lebouteiller} {et~al.}(2013){Lebouteiller}, {Heap}, {Hubeny}, \&
  {Kunth}}]{lebouteiller13}
{Lebouteiller}, V., {Heap}, S., {Hubeny}, I., \& {Kunth}, D. 2013, \aap, 553,
  A16

\bibitem[{{Leitet} {et~al.}(2013){Leitet}, {Bergvall}, {Hayes}, {Linn{\'e}}, \&
  {Zackrisson}}]{leitet13}
{Leitet}, E., {Bergvall}, N., {Hayes}, M., {Linn{\'e}}, S., \& {Zackrisson}, E.
  2013, \aap, 553, A106

\bibitem[{{Leitherer} {et~al.}(1995){Leitherer}, {Ferguson}, {Heckman}, \&
  {Lowenthal}}]{leitherer95}
{Leitherer}, C., {Ferguson}, H.~C., {Heckman}, T.~M., \& {Lowenthal}, J.~D.
  1995, \apjl, 454, L19

\bibitem[{{Leitherer} {et~al.}(2016){Leitherer}, {Hernandez}, {Lee}, \&
  {Oey}}]{leitherer16}
{Leitherer}, C., {Hernandez}, S., {Lee}, J.~C., \& {Oey}, M.~S. 2016, \apj,
  823, 64

\bibitem[{{Luridiana} {et~al.}(2015){Luridiana}, {Morisset}, \&
  {Shaw}}]{luridiana15}
{Luridiana}, V., {Morisset}, C., \& {Shaw}, R.~A. 2015, \aap, 573, A42

\bibitem[{{Marks} {et~al.}(2012){Marks}, {Kroupa}, {Dabringhausen}, \&
  {Pawlowski}}]{marks12}
{Marks}, M., {Kroupa}, P., {Dabringhausen}, J., \& {Pawlowski}, M.~S. 2012,
  \mnras, 422, 2246

\bibitem[{{Mas-Hesse} {et~al.}(2003){Mas-Hesse}, {Kunth}, {Tenorio-Tagle},
  {Leitherer}, {Terlevich}, \& {Terlevich}}]{mashesse03}
{Mas-Hesse}, J.~M., {Kunth}, D., {Tenorio-Tagle}, G., {Leitherer}, C.,
  {Terlevich}, R.~J., \& {Terlevich}, E. 2003, \apj, 598, 858

\bibitem[{{McGaugh}(1991)}]{mcgaugh91}
{McGaugh}, S.~S. 1991, \apj, 380, 140

\bibitem[{{McKinney} {et~al.}(2019){McKinney}, {Jaskot}, {Oey}, {Yun}, {Dowd},
  \& {Lowenthal}}]{mckinney19}
{McKinney}, J., {Jaskot}, A., {Oey}, M.~S., {Yun}, M., {Dowd}, T., \&
  {Lowenthal}, J. 2019, ApJ Submitted

\bibitem[{{Micheva} {et~al.}(2017){Micheva}, {Oey}, {Jaskot}, \&
  {James}}]{micheva17}
{Micheva}, G., {Oey}, M.~S., {Jaskot}, A.~E., \& {James}, B.~L. 2017, \apj,
  845, 165

\bibitem[{{Micheva} {et~al.}(2018){Micheva}, {Oey}, {Keenan}, {Jaskot}, \&
  {James}}]{micheva18}
{Micheva}, G., {Oey}, M.~S., {Keenan}, R.~P., {Jaskot}, A.~E., \& {James},
  B.~L. 2018, \apj, 867, 2

\bibitem[{{Mostardi} {et~al.}(2015){Mostardi}, {Shapley}, {Steidel}, {Trainor},
  {Reddy}, \& {Siana}}]{mostardi15}
{Mostardi}, R.~E., {Shapley}, A.~E., {Steidel}, C.~C., {Trainor}, R.~F.,
  {Reddy}, N.~A., \& {Siana}, B. 2015, \apj, 810, 107

\bibitem[{{Naidu} {et~al.}(2018){Naidu}, {Forrest}, {Oesch}, {Tran}, \&
  {Holden}}]{naidu18}
{Naidu}, R.~P., {Forrest}, B., {Oesch}, P.~A., {Tran}, K.-V.~H., \& {Holden},
  B.~P. 2018, \mnras, 478, 791

\bibitem[{{Nakajima} {et~al.}(2016){Nakajima}, {Ellis}, {Iwata}, {Inoue},
  {Kusakabe}, {Ouchi}, \& {Robertson}}]{nakajima16}
{Nakajima}, K., {Ellis}, R.~S., {Iwata}, I., {Inoue}, A.~K., {Kusakabe}, H.,
  {Ouchi}, M., \& {Robertson}, B.~E. 2016, \apjl, 831, L9

\bibitem[{{Nakajima} \& {Ouchi}(2014)}]{nakajima14}
{Nakajima}, K., \& {Ouchi}, M. 2014, \mnras, 442, 900

\bibitem[{{Oey} {et~al.}(2017){Oey}, {Herrera}, {Silich}, {Reiter}, {James},
  {Jaskot}, \& {Micheva}}]{oey17}
{Oey}, M.~S., {Herrera}, C.~N., {Silich}, S., {Reiter}, M., {James}, B.~L.,
  {Jaskot}, A.~E., \& {Micheva}, G. 2017, \apjl, 849, L1

\bibitem[{{Orlitov{\'a}} {et~al.}(2018){Orlitov{\'a}}, {Verhamme}, {Henry},
  {Scarlata}, {Jaskot}, {Oey}, \& {Schaerer}}]{orlitova18}
{Orlitov{\'a}}, I., {Verhamme}, A., {Henry}, A., {Scarlata}, C., {Jaskot}, A.,
  {Oey}, M.~S., \& {Schaerer}, D. 2018, \aap, 616, A60

\bibitem[{{Osterbrock} \& {Ferland}(2006)}]{osterbrock06}
{Osterbrock}, D.~E., \& {Ferland}, G.~J. 2006, {Astrophysics of gaseous nebulae
  and active galactic nuclei}

\bibitem[{{Paardekooper} {et~al.}(2015){Paardekooper}, {Khochfar}, \& {Dalla
  Vecchia}}]{paardekooper15}
{Paardekooper}, J.-P., {Khochfar}, S., \& {Dalla Vecchia}, C. 2015, \mnras,
  451, 2544

\bibitem[{{P{\'e}rez-Montero}(2017)}]{perezmontero17}
{P{\'e}rez-Montero}, E. 2017, \pasp, 129, 043001

\bibitem[{{Prochaska} {et~al.}(2011){Prochaska}, {Kasen}, \&
  {Rubin}}]{prochaska11}
{Prochaska}, J.~X., {Kasen}, D., \& {Rubin}, K. 2011, \apj, 734, 24

\bibitem[{{Ramachandran} {et~al.}(2019){Ramachandran}, {Hamann}, {Oskinova},
  {Gallagher}, {Hainich}, {Shenar}, {Sander}, {Todt}, \&
  {Fulmer}}]{ramachandran19}
{Ramachandran}, V., {et~al.} 2019, arXiv e-prints

\bibitem[{{Rivera-Thorsen} {et~al.}(2017{\natexlab{a}}){Rivera-Thorsen},
  {Dahle}, {Gronke}, {Bayliss}, {Rigby}, {Simcoe}, {Bordoloi}, {Turner}, \&
  {Furesz}}]{riverathorsen17b}
{Rivera-Thorsen}, T.~E., {et~al.} 2017{\natexlab{a}}, \aap, 608, L4

\bibitem[{{Rivera-Thorsen} {et~al.}(2015){Rivera-Thorsen}, {Hayes},
  {{\"O}stlin}, {Duval}, {Orlitov{\'a}}, {Verhamme}, {Mas-Hesse}, {Schaerer},
  {Cannon}, {Ot{\'{\i}}-Floranes}, {Sandberg}, {Guaita}, {Adamo}, {Atek},
  {Herenz}, {Kunth}, {Laursen}, \& {Melinder}}]{riverathorsen15}
---. 2015, \apj, 805, 14

\bibitem[{{Rivera-Thorsen} {et~al.}(2017{\natexlab{b}}){Rivera-Thorsen},
  {{\"O}stlin}, {Hayes}, \& {Puschnig}}]{riverathorsen17a}
{Rivera-Thorsen}, T.~E., {{\"O}stlin}, G., {Hayes}, M., \& {Puschnig}, J.
  2017{\natexlab{b}}, \apj, 837, 29

\bibitem[{{Robertson} {et~al.}(2015){Robertson}, {Ellis}, {Furlanetto}, \&
  {Dunlop}}]{robertson15}
{Robertson}, B.~E., {Ellis}, R.~S., {Furlanetto}, S.~R., \& {Dunlop}, J.~S.
  2015, \apjl, 802, L19

\bibitem[{{Roman-Duval} {et~al.}(2016){Roman-Duval}, {Ely}, {Debes}, {Massa},
  {Oliveira}, {Penton}, {Proffitt}, {Sahnow}, \&
  {Sonnentrucker}}]{romanduval16}
{Roman-Duval}, J., {et~al.} 2016, {Optimization of Lifetime Position 3 of the
  COS/FUV Detector}, Tech. rep.

\bibitem[{{Rutkowski} {et~al.}(2016){Rutkowski}, {Scarlata}, {Haardt}, {Siana},
  {Henry}, {Rafelski}, {Hayes}, {Salvato}, {Pahl}, {Mehta}, {Beck}, {Malkan},
  \& {Teplitz}}]{rutkowski16}
{Rutkowski}, M.~J., {et~al.} 2016, \apj, 819, 81

\bibitem[{{Scarlata} \& {Panagia}(2015)}]{scarlata15}
{Scarlata}, C., \& {Panagia}, N. 2015, \apj, 801, 43

\bibitem[{{Schaerer} {et~al.}(2016){Schaerer}, {Izotov}, {Verhamme},
  {Orlitov{\'a}}, {Thuan}, {Worseck}, \& {Guseva}}]{schaerer16}
{Schaerer}, D., {Izotov}, Y.~I., {Verhamme}, A., {Orlitov{\'a}}, I., {Thuan},
  T.~X., {Worseck}, G., \& {Guseva}, N.~G. 2016, \aap, 591, L8

\bibitem[{{Schenker} {et~al.}(2014){Schenker}, {Ellis}, {Konidaris}, \&
  {Stark}}]{schenker14}
{Schenker}, M.~A., {Ellis}, R.~S., {Konidaris}, N.~P., \& {Stark}, D.~P. 2014,
  \apj, 795, 20

\bibitem[{{Schlafly} \& {Finkbeiner}(2011)}]{schlafly11}
{Schlafly}, E.~F., \& {Finkbeiner}, D.~P. 2011, \apj, 737, 103

\bibitem[{{Schneider} {et~al.}(2018){Schneider}, {Sana}, {Evans},
  {Bestenlehner}, {Castro}, {Fossati}, {Gr{\"a}fener}, {Langer},
  {Ram{\'{\i}}rez-Agudelo}, {Sab{\'{\i}}n-Sanjuli{\'a}n},
  {Sim{\'o}n-D{\'{\i}}az}, {Tramper}, {Crowther}, {de Koter}, {de Mink},
  {Dufton}, {Garcia}, {Gieles}, {H{\'e}nault-Brunet}, {Herrero}, {Izzard},
  {Kalari}, {Lennon}, {Ma{\'{\i}}z Apell{\'a}niz}, {Markova}, {Najarro},
  {Podsiadlowski}, {Puls}, {Taylor}, {van Loon}, {Vink}, \&
  {Norman}}]{schneider18}
{Schneider}, F.~R.~N., {et~al.} 2018, Science, 359, 69

\bibitem[{{Shapley} {et~al.}(2003){Shapley}, {Steidel}, {Pettini}, \&
  {Adelberger}}]{shapley03}
{Shapley}, A.~E., {Steidel}, C.~C., {Pettini}, M., \& {Adelberger}, K.~L. 2003,
  \apj, 588, 65

\bibitem[{{Shibuya} {et~al.}(2014){Shibuya}, {Ouchi}, {Nakajima}, {Hashimoto},
  {Ono}, {Rauch}, {Gauthier}, {Shimasaku}, {Goto}, {Mori}, \&
  {Umemura.}}]{shibuya14}
{Shibuya}, T., {et~al.} 2014, \apj, 788, 74

\bibitem[{{Shull} {et~al.}(2015){Shull}, {Moloney}, {Danforth}, \&
  {Tilton}}]{shull15}
{Shull}, J.~M., {Moloney}, J., {Danforth}, C.~W., \& {Tilton}, E.~M. 2015,
  \apj, 811, 3

\bibitem[{{Silich} \& {Tenorio-Tagle}(2017)}]{silich17}
{Silich}, S., \& {Tenorio-Tagle}, G. 2017, \mnras, 465, 1375

\bibitem[{{Silich} {et~al.}(2004){Silich}, {Tenorio-Tagle}, \&
  {Rodr{\'{\i}}guez-Gonz{\'a}lez}}]{silich04}
{Silich}, S., {Tenorio-Tagle}, G., \& {Rodr{\'{\i}}guez-Gonz{\'a}lez}, A. 2004,
  \apj, 610, 226

\bibitem[{{Smith} {et~al.}(2016){Smith}, {Crowther}, {Calzetti}, \&
  {Sidoli}}]{smith16}
{Smith}, L.~J., {Crowther}, P.~A., {Calzetti}, D., \& {Sidoli}, F. 2016, \apj,
  823, 38

\bibitem[{{Stark} {et~al.}(2011){Stark}, {Ellis}, \& {Ouchi}}]{stark11}
{Stark}, D.~P., {Ellis}, R.~S., \& {Ouchi}, M. 2011, \apjl, 728, L2

\bibitem[{{Stasi{\'n}ska} {et~al.}(2015){Stasi{\'n}ska}, {Izotov}, {Morisset},
  \& {Guseva}}]{stasinska15}
{Stasi{\'n}ska}, G., {Izotov}, Y., {Morisset}, C., \& {Guseva}, N. 2015, \aap,
  576, A83

\bibitem[{{Steidel} {et~al.}(2018){Steidel}, {Bogosavlevic}, {Shapley},
  {Reddy}, {Rudie}, {Pettini}, {Trainor}, \& {Strom}}]{steidel18}
{Steidel}, C.~C., {Bogosavlevic}, M., {Shapley}, A.~E., {Reddy}, N.~A.,
  {Rudie}, G.~C., {Pettini}, M., {Trainor}, R.~F., \& {Strom}, A.~L. 2018,
  ArXiv e-prints

\bibitem[{{Storey} \& {Hummer}(1995)}]{storey95}
{Storey}, P.~J., \& {Hummer}, D.~G. 1995, \mnras, 272, 41

\bibitem[{{St{\"o}rzer} \& {Hollenbach}(1998)}]{storzer98}
{St{\"o}rzer}, H., \& {Hollenbach}, D. 1998, \apjl, 502, L71

\bibitem[{{Stoughton} {et~al.}(2002){Stoughton}, {Lupton}, {Bernardi},
  {Blanton}, {Burles}, {Castand er}, {Connolly}, {Eisenstein}, {Frieman},
  {Hennessy}, {Hindsley}, {Ivezi{\'c}}, {Kent}, {Kunszt}, {Lee}, {Meiksin},
  {Munn}, {Newberg}, {Nichol}, {Nicinski}, {Pier}, {Richards}, {Richmond},
  {Schlegel}, {Smith}, {Strauss}, {SubbaRao}, {Szalay}, {Thakar}, {Tucker},
  {Vand en Berk}, {Yanny}, {Adelman}, {Anderson}, {Anderson}, {Annis},
  {Bahcall}, {Bakken}, {Bartelmann}, {Bastian}, {Bauer}, {Berman},
  {B{\"o}hringer}, {Boroski}, {Bracker}, {Briegel}, {Briggs}, {Brinkmann},
  {Brunner}, {Carey}, {Carr}, {Chen}, {Christian}, {Colestock}, {Crocker},
  {Csabai}, {Czarapata}, {Dalcanton}, {Davidsen}, {Davis}, {Dehnen},
  {Dodelson}, {Doi}, {Dombeck}, {Donahue}, {Ellman}, {Elms}, {Evans}, {Eyer},
  {Fan}, {Federwitz}, {Friedman}, {Fukugita}, {Gal}, {Gillespie}, {Glazebrook},
  {Gray}, {Grebel}, {Greenawalt}, {Greene}, {Gunn}, {de Haas}, {Haiman},
  {Haldeman}, {Hall}, {Hamabe}, {Hansen}, {Harris}, {Harris}, {Harvanek},
  {Hawley}, {Hayes}, {Heckman}, {Helmi}, {Henden}, {Hogan}, {Hogg}, {Holmgren},
  {Holtzman}, {Huang}, {Hull}, {Ichikawa}, {Ichikawa}, {Johnston}, {Kauffmann},
  {Kim}, {Kimball}, {Kinney}, {Klaene}, {Kleinman}, {Klypin}, {Knapp},
  {Korienek}, {Krolik}, {Kron}, {Krzesi{\'n}ski}, {Lamb}, {Leger},
  {Limmongkol}, {Lindenmeyer}, {Long}, {Loomis}, {Loveday}, {MacKinnon},
  {Mannery}, {Mantsch}, {Margon}, {McGehee}, {McKay}, {McLean}, {Menou},
  {Merelli}, {Mo}, {Monet}, {Nakamura}, {Narayanan}, {Nash}, {Neilsen},
  {Newman}, {Nitta}, {Odenkirchen}, {Okada}, {Okamura}, {Ostriker}, {Owen},
  {Pauls}, {Peoples}, {Peterson}, {Petravick}, {Pope}, {Pordes}, {Postman},
  {Prosapio}, {Quinn}, {Rechenmacher}, {Rivetta}, {Rix}, {Rockosi}, {Rosner},
  {Ruthmansdorfer}, {Sandford}, {Schneider}, {Scranton}, {Sekiguchi}, {Sergey},
  {Sheth}, {Shimasaku}, {Smee}, {Snedden}, {Stebbins}, {Stubbs}, {Szapudi},
  {Szkody}, {Szokoly}, {Tabachnik}, {Tsvetanov}, {Uomoto}, {Vogeley}, {Voges},
  {Waddell}, {Walterbos}, {Wang}, {Watanabe}, {Weinberg}, {White}, {White},
  {Wilhite}, {Wolfe}, {Yasuda}, {York}, {Zehavi}, \& {Zheng}}]{stoughton02}
{Stoughton}, C., {et~al.} 2002, \aj, 123, 485

\bibitem[{{Sukhbold} {et~al.}(2016){Sukhbold}, {Ertl}, {Woosley}, {Brown}, \&
  {Janka}}]{sukhbold16}
{Sukhbold}, T., {Ertl}, T., {Woosley}, S.~E., {Brown}, J.~M., \& {Janka}, H.~T.
  2016, \apj, 821, 38

\bibitem[{{Tanvir} {et~al.}(2019){Tanvir}, {Fynbo}, {de Ugarte Postigo},
  {Japelj}, {Wiersema}, {Malesani}, {Perley}, {Levan}, {Selsing}, {Cenko},
  {Kann}, {Milvang-Jensen}, {Berger}, {Cano}, {Chornock}, {Covino},
  {Cucchiara}, {D'Elia}, {Gargiulo}, {Goldoni}, {Gomboc}, {Heintz}, {Hjorth},
  {Izzo}, {Jakobsson}, {Kaper}, {Kr{\"u}hler}, {Laskar}, {Myers},
  {Piranomonte}, {Pugliese}, {Rossi}, {S{\'a}nchez-Ram{\'\i}rez}, {Schulze},
  {Sparre}, {Stanway}, {Tagliaferri}, {Th{\"o}ne}, {Vergani}, {Vreeswijk},
  {Wijers}, {Watson}, \& {Xu}}]{tanvir19}
{Tanvir}, N.~R., {et~al.} 2019, \mnras, 483, 5380

\bibitem[{{Trainor} {et~al.}(2016){Trainor}, {Strom}, {Steidel}, \&
  {Rudie}}]{trainor16}
{Trainor}, R.~F., {Strom}, A.~L., {Steidel}, C.~C., \& {Rudie}, G.~C. 2016,
  \apj, 832, 171

\bibitem[{{Vanzella} {et~al.}(2016){Vanzella}, {de Barros}, {Vasei}, {Alavi},
  {Giavalisco}, {Siana}, {Grazian}, {Hasinger}, {Suh}, {Cappelluti}, {Vito},
  {Amorin}, {Balestra}, {Brusa}, {Calura}, {Castellano}, {Comastri}, {Fontana},
  {Gilli}, {Mignoli}, {Pentericci}, {Vignali}, \& {Zamorani}}]{vanzella16}
{Vanzella}, E., {et~al.} 2016, \apj, 825, 41

\bibitem[{{Vasei} {et~al.}(2016){Vasei}, {Siana}, {Shapley}, {Quider}, {Alavi},
  {Rafelski}, {Steidel}, {Pettini}, \& {Lewis}}]{vasei16}
{Vasei}, K., {et~al.} 2016, \apj, 831, 38

\bibitem[{{Verhamme} {et~al.}(2015){Verhamme}, {Orlitov{\'a}}, {Schaerer}, \&
  {Hayes}}]{verhamme15}
{Verhamme}, A., {Orlitov{\'a}}, I., {Schaerer}, D., \& {Hayes}, M. 2015, \aap,
  578, A7

\bibitem[{{Verhamme} {et~al.}(2017){Verhamme}, {Orlitov{\'a}}, {Schaerer},
  {Izotov}, {Worseck}, {Thuan}, \& {Guseva}}]{verhamme17}
{Verhamme}, A., {Orlitov{\'a}}, I., {Schaerer}, D., {Izotov}, Y., {Worseck},
  G., {Thuan}, T.~X., \& {Guseva}, N. 2017, \aap, 597, A13

\bibitem[{{Vink} {et~al.}(2001){Vink}, {de Koter}, \& {Lamers}}]{vink01}
{Vink}, J.~S., {de Koter}, A., \& {Lamers}, H.~J.~G.~L.~M. 2001, \aap, 369, 574

\bibitem[{{Voges} \& {Walterbos}(2006)}]{voges06}
{Voges}, E.~S., \& {Walterbos}, R.~A.~M. 2006, \apjl, 644, L29

\bibitem[{{Wise} \& {Cen}(2009)}]{wise09}
{Wise}, J.~H., \& {Cen}, R. 2009, \apj, 693, 984

\bibitem[{{Wolfe} {et~al.}(2003){Wolfe}, {Prochaska}, \& {Gawiser}}]{wolfe03}
{Wolfe}, A.~M., {Prochaska}, J.~X., \& {Gawiser}, E. 2003, \apj, 593, 215

\bibitem[{{Wright} {et~al.}(1991){Wright}, {Mather}, {Bennett}, {Cheng},
  {Shafer}, {Fixsen}, {Eplee}, {Isaacman}, {Read}, {Boggess}, {Gulkis},
  {Hauser}, {Janssen}, {Kelsall}, {Lubin}, {Meyer}, {Moseley}, {Murdock},
  {Silverberg}, {Smoot}, {Weiss}, \& {Wilkinson}}]{wright91}
{Wright}, E.~L., {et~al.} 1991, \apj, 381, 200

\bibitem[{{Yang} {et~al.}(2017{\natexlab{a}}){Yang}, {Malhotra}, {Gronke},
  {Rhoads}, {Leitherer}, {Wofford}, {Jiang}, {Dijkstra}, {Tilvi}, \&
  {Wang}}]{yang17}
{Yang}, H., {et~al.} 2017{\natexlab{a}}, \apj, 844, 171

\bibitem[{{Yang} {et~al.}(2017{\natexlab{b}}){Yang}, {Malhotra}, {Rhoads},
  {Leitherer}, {Wofford}, {Jiang}, \& {Wang}}]{yang17a}
{Yang}, H., {Malhotra}, S., {Rhoads}, J.~E., {Leitherer}, C., {Wofford}, A.,
  {Jiang}, T., \& {Wang}, J. 2017{\natexlab{b}}, \apj, 838, 4

\bibitem[{{Zastrow} {et~al.}(2013){Zastrow}, {Oey}, {Veilleux}, \&
  {McDonald}}]{zastrow13}
{Zastrow}, J., {Oey}, M.~S., {Veilleux}, S., \& {McDonald}, M. 2013, \apj, 779,
  76

\end{thebibliography}
\end{document}